\documentclass[12pt,preprint,twocolumn]{aastex631}

\usepackage[utf8]{inputenc}

\usepackage{lipsum}
\newsavebox{\myimage}
\usepackage{graphics,graphicx}
\usepackage{helvet}
\usepackage{subfigure}
\usepackage[utf8]{inputenc}
\usepackage[T1]{fontenc}
\usepackage{soul} 
\hypersetup{citecolor=blue,colorlinks,pdfusetitle}
\usepackage{appendix}
\usepackage{amsmath, amssymb, gensymb}
\usepackage{breqn}
\usepackage{morefloats}
\usepackage{float}
\usepackage{ulem}
\usepackage{url}
\usepackage{natbib}
\usepackage{microtype}
\usepackage{multirow}
\usepackage{morefloats}
\usepackage{here}
\usepackage{xspace}
\usepackage{color}
\usepackage{lipsum}
\usepackage{xspace}
\usepackage{rotating}

\usepackage{float}
\usepackage{stmaryrd}

\usepackage{longtable}

\def\S{\mathcal{S}}

\def\eg {e.g.,\xspace} 
\def\ie {i.e.,\xspace} 

\def\etal {\textit{et al.}\xspace}

\def\cotwotoone {CO\,($J$\,=\,2\,$\rightarrow\,$1)\xspace}

\def\HtonetoOStwo {$\rm  H_{2}\,1\rightarrow0\,S(2) $\xspace}

\def\HttwotooneSthree {$\rm  H_{2}\,2\rightarrow1\,S(3) $\xspace}
\def\HtonetoOSone {$\rm  H_{2}\,1\rightarrow0\,S(1) $\xspace}

\def\HttwotooneStwo {$\rm  H_{2}\,2\rightarrow1\,S(2) $\xspace}

\def\HtonetoOSO {$\rm  H_{2}\,1\rightarrow0\,S(0) $\xspace}
\def\HttwotooneSone {$\rm  H_{2}\,2\rightarrow1\,S(1) $\xspace}

\def\HtOtoOSeight{$\rm  H_{2}\,0\rightarrow0\,S(8) $\xspace}
\def\HtOtoOSnine{$\rm  H_{2}\,0\rightarrow0\,S(9) $\xspace}
\def\HtOtoOSten{$\rm  H_{2}\,0\rightarrow0\,S(10) $\xspace}
\def\HtonetooneSnine{$\rm  H_{2}\,1\rightarrow1\,S(9) $\xspace}
\def\HtonetooneSten{$\rm  H_{2}\,1\rightarrow1\,S(10) $\xspace}
\def\HtonetooneSeleven{$\rm  H_{2}\,1\rightarrow1\,S(11) $\xspace}

\def\HttwotooneStwo{$\rm  H_{2}\,2\rightarrow1\,S(2) $\xspace}
\def\HttwotooneSthree{$\rm  H_{2}\,2\rightarrow1\,S(3) $\xspace}
\def\HttwotooneSone{$\rm  H_{2}\,2\rightarrow1\,S(1) $\xspace}
\def\HtonetoOSO{$\rm  H_{2}\,1\rightarrow0\,S(0) $\xspace}
\def\HtonetoOSone{$\rm  H_{2}\,1\rightarrow0\,S(1) $\xspace}

\def\HtonetoOSthree{$\rm  H_{2}\,1\rightarrow0\,S(3) $\xspace}

\def\HtonetoOOthree {$\rm  H_{2}\,1\rightarrow0\,O(3) $\xspace}

\def\COonetoOPeightneen {$\rm  CO\,1\rightarrow0\,P(18) $\xspace}

\def\Brgamma {Br$\gamma$\xspace}

\def\Ht {H$_2$\xspace}
\def\nH {$n_\textrm{\scriptsize{H}}$\xspace}
\def\b {$b$\xspace}
\def\vs {$v_\textrm{\scriptsize{s}}$\xspace}
\def\Go {$G_\textrm{\scriptsize{0}}$\xspace}
\def\CIR {$\zeta_{\small{\textrm{H}_2}}$\xspace}
\def\XPAH {$X(\textrm{\scriptsize{PAH}})$\xspace}

\begin{document}

\baselineskip=12pt 

\title{Unveiling two deeply embedded young protostars in the S68N Class 0 protostellar core with JWST/NIRSpec} 
\shorttitle{JWST/NIRSpec spectroscopy of the S68N Class 0 core}

\email{valentin.j.legouellec@nasa.gov}

\author[0000-0002-5714-799X]{Valentin J. M. Le Gouellec}
\affiliation{NASA Ames Research Center, Space Science and Astrobiology Division, M.S. 245-6 Moffett Field, CA 94035, USA}
\affiliation{NASA Postdoctoral Program Fellow}

\author[0000-0003-1487-6452]{Ben W. P. Lew}
\affiliation{Bay Area Environmental Research Institute, Moffett Field, CA 94035, USA}
\affiliation{NASA Ames Research Center, Space Science and Astrobiology Division, M.S. 245-6 Moffett Field, CA 94035, USA}

\author[0000-0002-8963-8056]{Thomas P. Greene}
\affiliation{NASA Ames Research Center, Space Science and Astrobiology Division M.S. 245-6 Moffett Field, CA 94035, USA}

\author[0000-0002-6773-459X]{Doug Johnstone}
\affiliation{NRC Herzberg Astronomy and Astrophysics, 5071 West Saanich Rd, Victoria, BC, V9E 2E7, Canada }
\affiliation{Department of Physics and Astronomy, University of Victoria, Victoria, BC, V8P 5C2, Canada}

\author[0000-0002-0354-1684]{Antoine Gusdorf}
\affiliation{Laboratoire de Physique de l’\'Ecole Normale Sup\'erieure, ENS, Universit\'e PSL, CNRS, Sorbonne Universit\'e, Universit\'e Paris Cit\'e, F-75005, Paris, France}
\affiliation{Observatoire de Paris, PSL University, Sorbonne Universit\'e, LERMA, 75014, Paris, France}

\author[0000-0001-8822-6327]{Logan Francis}
\affiliation{Leiden Observatory, Leiden University, PO Box 9513, 2300RA Leiden, The Netherlands}

\author[0000-0002-6528-3836]{Curtis DeWitt}
\affiliation{Space Science Institute, Boulder, CO 80301, USA}

\author[0000-0003-1227-3084]{Michael Meyer}
\affiliation{Department of Astronomy, University of Michigan, Ann Arbor, MI 48109, USA}

\author[0000-0002-9470-2358]{\L ukasz Tychoniec}
\affiliation{Leiden Observatory, Leiden University, PO Box 9513, 2300RA Leiden, The Netherlands}

\author[0000-0001-7591-1907]{Ewine F. van Dishoeck} 
\affiliation{Leiden Observatory, Leiden University, PO Box 9513, 2300RA Leiden, The Netherlands}

\author[0009-0003-2041-7911]{Mary Barsony} 
\affiliation{13115 Dupont Road, Sebastopol, CA 95472}

\author[0000-0003-0786-2140]{Klaus W. Hodapp}
\affiliation{University of Hawaii, Institute for Astronomy, 640 N. Aohoku Place, Hilo, HI 96720, USA}

\author[0000-0002-5236-3896]{Peter G. Martin}
\affiliation{Canadian Institute for Theoretical Astrophysics, University of Toronto, McLennan Physical Laboratories, 60 St. George Street, Toronto, Ontario M5S 3H8, Canada}

\author[0000-0002-9573-3199]{Massimo Robberto}
\affiliation{Space Telescope Science Institute, 3700 San Martin Drive, Baltimore, MD 21218, USA}
\affiliation{Department of Physics \& Astronomy, Johns Hopkins University, 3400 N. Charles Street, Baltimore, MD 21218, USA}




\shortauthors{Le Gouellec \etal}


\begin{abstract}
The near-infrared (NIR) emission of the youngest protostars still needs to be characterized to better understand the evolution of their accretion and ejection activity.
We analyze James Webb Space Telescope NIRSpec 1.7---5.3 $\mu$m observations of two deeply embedded sources in the S68N protostellar core in Serpens.
The North Central (NC) source exhibits a highly obscured spectrum ($A_K \sim 4.8$ mag) that is modeled with a pre-main-sequence photosphere and a hot disk component. The photospheric parameters are consistent with a young, low-mass photosphere, as suggested by the low surface gravity, log $g$ of 1.95 $\pm$ 0.15 cm s$^{-2}$. The hot disk suggests that accretion onto the central protostellar embryo is ongoing, although prototypical accretion-tracing emission lines \ion{H}{1} are not detected. 
The South Central (SC) source, which is even more embedded ($A_K \sim 8$ mag; no continuum is detected shortward of 3.6 $\mu$m) appears to be driving the large-scale S68N protostellar outflow, and launches a collimated hot molecular jet detected in \Ht and CO ro-vibrational lines. Shock modeling of the \Ht (ro)vibrational lines establishes that fast $C$-type shocks ($\geq$ 30 km s$^{-1}$), with high pre-shock density ($\geq$ $10^7$ cm$^{-3}$), and strong magnetic field (\b $\sim$ 3--10, where $B = b\,\times\,\sqrt{\textrm{n}_{\scriptsize{\textrm{H}}} (\textrm{cm}^{-3})}\,\mu\textrm{G}$) best match the data. The bright CO fundamental line forest suggests energetic excitation, with the contribution of non-LTE effects, \ie irradiation pumping. Detected OH and CH$^{+}$ ro-vibrational lines support this hypothesis. 
These two Class 0 protostars seem to be in very young evolutionary stages and still have to acquire the bulk of their final stellar masses. These results demonstrate that JWST enables unprecedented diagnostics of these first stages of the protostellar evolutionary phase.
\end{abstract}

\section{Introduction}
\label{sec:intro}

Stars accrete most of their mass during the protostellar stage, where their progenitors are still embedded in a dense and cold collapsing protostellar envelope. The Class 0 protostellar phase designates the initial short-lifetime main accretion phase that follows the formation of the second Larson core \citep{Larson1969,Andre1993,Andre2000,Andre2002}. The accretion onto the central protostellar embryo triggers the vigorous ejection of material in the form of (bipolar) outflows \citep{Frank2014,Bally2016}. The accretion mechanism(s) at play in the Class 0 stage is still poorly constrained. These young systems concurrently form their disks while mass infalling from the envelope accumulates in the inner regions. As they are still deeply embedded and dynamically connected to the surrounding collapsing envelope, the properties of these young disks are still poorly understood. 
Young disks are subject to rapid mass infall, which can drive them closer to gravitational instability, and even to fragmentation \citep{Adams1989,Bonnell1994,Stamatellos2009,Kratter2010}. 
How the angular momentum is extracted and carried away in order to allow the accretion to occur is the subject of different fields of study, including the redistribution of angular momentum in the inner envelope via the magnetic field (\eg \citealt{AnezLopez2024}), and the extraction of angular momentum via the launching of outflows. The current paradigm is that outflows are magneto-hydrodynamically (MHD) driven, with a wide-angle wind launched from the disk and a collimated jet launched at smaller radii \citep{Pascucci2023}. The ``X-wind'' model proposes an alternative framework based on magnetized star–disk interaction, where the wind and jet are both launched magnetocentrifugally from the inner edge of the disk \citep{Shang2020,Shang2023a}.

While significant progress has been realized in the last decade with the advent of ALMA and NOEMA observations of protostellar embedded disks \citep{Maury2019,Sheehan2022,Tobin2024}, infalling envelopes (\eg \citealt{Cabedo2021,Gaudel2020,Sai2022,Yen2023}), and outflows (\eg \citealt{Podio2021,HsiehCH2023}), infrared (IR) spectroscopy allows for highly complementary diagnostics of protostellar evolution, via the study of the photosphere, inner disk, and atomic and molecular gas tracing the shocked hot gas within outflow cavities. Indeed, IR vibrational spectroscopy traces the hot gas in cavities and disk inner regions while millimeter observations trace the colder phases of the protostellar systems (\eg entrained outflow, outer disk). Ground based observations of Class 0s in the near-IR (NIR) \citep{Laos2021,LeGouellec2024a} revealed that the accretion onto the central regions is more vigorous during the Class 0 phase than for more evolved Class I objects, and potentially of different nature than the adopted magnetospheric accretion paradigm for T-Tauri stars. Notably, Class 0 protostars exhibit a net higher detection rate of CO overtone bands in emission than in Class I objects (see \citealt{Doppmann2005,Connelley2010,Fiorellino2021}), suggesting a more active inner disk at the more embedded phases. 
These latter results are particularly interesting in view of the difficulty of reproducing the observed distribution of accretion-driven protostellar luminosities (\eg see the surveys of \citealt{Kenyon1990b,Kenyon1994,Evans2009,Dunham2010a,Dunham2013} and the models of \citealt{Myers2009,Myers2010,Offner2011b,Fischer2017}). Several observational campaigns have investigated whether protostellar accretion is a variable process (see the recent review by \citealt{Fischer2023}), which can be related to the fact that Class 0 accretion disks are more active than the further-evolved Class I disks \citep{LeGouellec2024a}.
Only a few photospheres of Class 0 protostellar embryos have been detected thus far \citep{Greene2018,LeGouellec2024a}, within sources that, surprisingly, do not show 
evidence of strong accretion activity (sources with no detections of CO overtone emission or \ion{H}{1} lines, and with faint \Ht lines), although detailed modeling will be required to quantify the accretion activity of these objects.

Outflows have been studied in the IR with ISO, \textit{Spitzer} and ground based spectroscopic observations of [\ion{Fe}{2}], \Ht, and  OH lines \citep{Rosenthal2000,CarattiOGaratti2006,Antoniucci2008,Tappe2008,Maret2009,Neufeld2009,Greene2010,Dionatos2014}, for which the driving source was often too embedded to be directly detected. While [\ion{Fe}{2}] line studies reveal the properties of embedded atomic jets, \Ht is more suitable to probe shock excitation and gas cooling given that it is the major NIR coolant. In addition to useful quantification of the outflow mass ejection rate which is intrinsically related to the accretion activity of protostars \citep{Bontemps1996,Ellerbroek2013,Yildiz2015,LeeCF2020}, studying outflows and the shocks propagating in cavities provide crucial insights on the outflow energetics, chemical composition of the ejected and shocked gas, UV illumination of cavities, and properties of the pre-shock gas (\eg density and magnetic field strength of the shocked cavity walls). JWST now enables mapping the excitation of IR lines tracing the different layers of shocked gas and ionized phases. It is also sensitive enough in the NIR to detect the spectra of embedded continuum sources and reveal their nature.


S68N\footnote{This source is also called Serpens Emb 8 in the literature \citep{Enoch2009}.} is a bona fide protostellar core located in the Serpens Main star forming region, for which we adopt a distance of 445 pc \citep{OrtizLeon2017b,Zucker2019,Herczeg2019,Zucker2020}. Initially characterized in several far-IR (FIR) and sub-millimeter studies \citep{McMullin1994,Hurt1996a,Hurt1996,WolfChase1998,Enoch2009,Enoch2011}, it has a bolometric luminosity and bolometric temperature of $L_{\textrm{bol}}\,=\,15.9\,L_\odot$ and $T_{\textrm{bol}}\,=\,33\,K$ \citep{Dunham2015,Pokhrel2023}.
The SED modeling of \citet{Pokhrel2023} confirmed the embedded nature of this object, and constrained the envelope mass to be 6.5 M$_\odot$.
The protostellar envelope has a systemic velocity ($v_{\textrm{lsr}}$) of 8.45 km s$^{-1}$ \citep{Lee2014}.
Recent higher angular resolution ALMA 870 $\mu$m dust continuum observations resolved four different fragments and dense filament-like substructures embedded within the protostellar envelope, alongside dust polarization studies revealing a complex magnetic field morphology \citep{Hull2017a,LeGouellec2019a}.
No disk was detected in the molecular gas kinematics or dust continuum emission in the IRAM Plateau de Bure Interferometer survey CALYPSO \citep{Maury2019,Maret2020}.
Among the four fragments in the S68N core, at least one of them harbors a Class 0 protostar, which launches a low-velocity molecular outflow identified by \citet{Tychoniec2019,Podio2021}, and has obvious sub-millimeter envelope emission in the ALMA data \citep{LeGouellec2019a}. This source also opens a clear outflow cavity in the inner envelope on the blueshifted side, proposed to be substantially irradiated, as suggested by the bright CCH emission observed by \citet{LeGouellec2023a}. CCH emission has been found to be a good tracer of intense irradiation \citep{Jansen1994,Sternberg1995,Aikawa1999,Stauber2004,Benz2016}, and/or a signpost for a high C/O gas ratio \citep{Bergner2020}.

\begin{figure*}[!tbh]
\centering
\includegraphics[scale=1.2,clip,trim= 0cm 0.0cm 0cm 0cm]{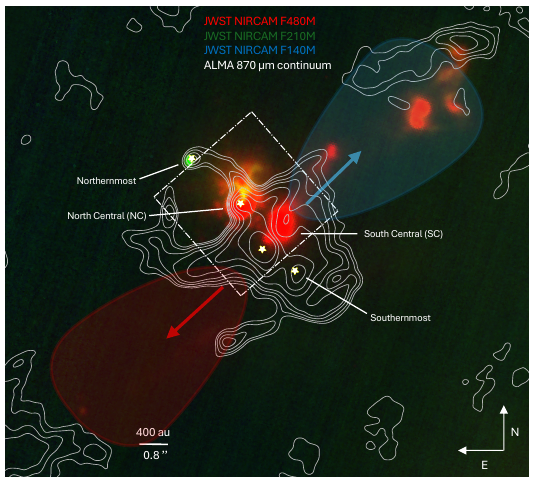}

\vspace{-0.1cm} 
\caption{\small Multi-wavelength view of the S68N protostellar core. The color scale is a JWST NIRCAM 3 color image using F480M (red), F210M (green), and F140M (blue) (full data are presented by \citealt{Green2024}). The gray contours, which are tracing the 870 $\mu$m dust continuum emission from \citet{LeGouellec2019a}, are 5, 9, 16, 24, 44, 74, 128 $\times$ the rms noise $\sigma_I$ of the Stokes $I$ map, where $\sigma_I$ = 60 $\mu$Jy beam$^{-1}$. The four white stars  correspond to the four fragments identified in the ALMA continuum map.
The Northernmost, NC and SC IR sources are labeled.
The red (blue) arrow and lobe outline the redshifted (blueshifted) outflow cavity identified with ALMA observations of the CO (2$\rightarrow$1) line (more details on the ALMA data are presented in Section \ref{sec:obs_alma}), which is associated with the SC fragment, the known Class 0 source in the S68N system.
The dashed-line white square corresponds to the coverage of the JWST NIRSpec IFU data presented here.
Table \ref{t.sources_coord} presents the coordinates of the millimeter and associated IR sources.
Note that the different NIRCAM images have been plotted on different flux scales to enhance their visibility; as a result, relative fluxes should not be inferred from this image.}
\label{fig:RGB_alma_image}
\end{figure*}

The second brightest sub-millimeter fragment has been characterized in the $K$-band with Keck NIRSPEC spectroscopy by \citet{Greene2018,LeGouellec2024a}. These investigations found that different absorption features (CO, Na, and Ca) were consistent with a low-temperature, low-gravity photosphere with significant extinction and a modest near-infrared (NIR) continuum veiling. In the present study, we aim to investigate further the nature of the driving sources within the S68N protostellar core.
 
This paper is structured as follows. In Section \ref{sec:obs}, we present the observations used in this study, \ie new JWST NIRSpec IFU observations from 1.7 to 5.3 $\mu$m alongside archival ALMA observations. One-dimensional (1D) extracted spectra, emission line and continuum emission maps are presented in Section \ref{sec:results}. Section \ref{sec:mol_em} focuses on the excitation of the molecular gas seen in emission, especially \Ht and CO lines. In Section \ref{sec:photosphere}, we present our results of the modeling of the photospheric absorption features and hot disk emission of the source initially characterized by \citet{Greene2018}. We present a principal component analysis of the NIRSpec IFU data in Section \ref{sec:PCA}. We discuss our results in Section \ref{sec:discussion} in the context of the NIR properties of Class 0 protostars. We draw our conclusions in Section \ref{sec:ccl}.

\section{Observations}
\label{sec:obs}

\subsection{Multi-wavelength view of the S68N core}

Figure \ref{fig:RGB_alma_image} presents a multi-wavelength view of the S68N protostellar core. A JWST NIRCam three-color image is overlaid with contours tracing ALMA 870 $\mu$m dust continuum emission from \citet{LeGouellec2019a}. 
The JWST NIRCam data are from the Cycle 1 GO program 1611 (PI: Klaus M. Pontoppidan) and are presented by \citet{Green2024}. The S68N core is shown in their Figure 2, where they present F140M, F210M, F360M, and F480M NIRCam filter observations.
Here, three $\lambda / \delta \lambda \simeq 10$ NIRCam filters were combined to produce Figure \ref{fig:RGB_alma_image}, F480M (covering several CO lines, H$_{2}$\,1-1\,S\,10, and H$_{2}$\,1-1\,S\,9), F210M (covering H$_{2}$\,1-0\,S\,1), and F140M.

All four submillimeter fragments hosted by the S68N core are very embedded, and the envelope obscures most of their potential NIR emission. Among the four fragments identified in the ALMA 870~$\mu$m dust continuum observations (roughly aligned in the NE - SW direction, see the four star symbols in Figure \ref{fig:RGB_alma_image}), the two Northern sources are directly seen in the NIR. The JWST NIRSpec observations presented in Section \ref{sec:obs_jwst} only covers the two central sources, which we will call the North Central (NC) and the South Central (SC) sources, throughout this paper (outlined in Figure \ref{fig:RGB_alma_image}). The NC source is clearly detected in the F480M and F210M filters (and detected at a low signal-to-noise ratio in the F140M filter). As for the SC source, whose sub-millimeter fragment is the brightest source in the ALMA data, its blueshifted cavity is clearly detected in the F360M and F480M filters. No NIR emission is seen on the peak of the submillimeter fragment, suggesting that the SC infrared source consists of scattered light and outflow emission.
The Northernmost source (outside of the NIRSpec coverage, yet labeled in Figure \ref{fig:RGB_alma_image}) is the bluest, as it is the only source with a clear detected contribution in the F140M NIRCam filter. The Southernmost source (also outside of the NIRSpec coverage and labeled in Figure \ref{fig:RGB_alma_image}), is totally extincted and no infrared contributions are seen in the NIRCam images. We outline the coordinates and names of the different millimeter-identified fragments in Table \ref{t.sources_coord}, alongside the coordinates and names of the corresponding IR sources. 

\begin{table*}
\centering
\small
\caption[]{Fragments of the S68N core}
\label{t.sources_coord}
\setlength{\tabcolsep}{0.3em} 
\begin{tabular}{p{0.2\linewidth}cccccc}
\hline \hline \noalign{\smallskip}
IR source name$^a$ & $\alpha_{\textrm{IR},\,\textrm{J2000}}$ & $\delta_{\textrm{IR},\,\textrm{J2000}}$ & mm source name$^b$ & $\alpha_{\textrm{mm},\,\textrm{J2000}}$ & $\delta_{\textrm{mm},\,\textrm{J2000}}$ \\
& $\rm {h}$\,:\,$\rm {m}$\,:\,$\rm {s}$ & $\rm {\circ}$\,:\,$\rm {\prime}$\,:\,${\prime\prime}$ & & $\rm {h}$\,:\,$\rm {m}$\,:\,$\rm {s}$ & $\rm {\circ}$\,:\,$\rm {\prime}$\,:\,${\prime\prime}$ \\
\hline
North Central (NC) & 18:29:48.13 & +01:16:44.69 & Serpens Emb 8-b & 18:29:48.13 & +01:16:44.57 \\
\noalign{\smallskip}
South Central (SC) & 18:29:48.06 & +01:16:43.98 & Serpens Emb 8 & 18:29:48.09 & +01:16:43.30 \\
\noalign{\smallskip}
Northernmost source & 18:29:48.22 & +01:16:45.77 & --- & 18:29:48.22 & +01:16:45.83  \\
\noalign{\smallskip}
Southernmost source$^c$ & --- & --- & Serpens Emb 8-c & 18:29:48.03 & +01:16:42.70  \\
\noalign{\smallskip}
\hline
\end{tabular}
\tablecomments{\small 
$^a$ Names of the IR sources identified in the JWST data. \\
$^b$ Names of the millimeter (mm) sources identified by \citet{LeGouellec2019a}.\\
$^c$ Undetected in the NIR.}
\end{table*}

\subsection{JWST NIRSpec}
\label{sec:obs_jwst}

Our JWST spectroscopic observations of S68N were executed on 2023 April 16 UT under GTO Program 1186 (P.I Thomas Greene). The NIRSpec/IFU \citep{Boker2022,Jakobsen2022} observation visit includes F170LP/G235H ($R \simeq 2700$), F170LP/G235M ($R \simeq 1000$), and G395H/F290LP ($R \simeq 2700$) gratings.
The target position was 18$\rm ^{h}$\,29$\rm ^{m}$\,48$\fs$12, $+$01$\rm ^{\circ}$\,16$\rm ^{\prime}$\,44$\farcs$60 (J2000) centered on a field of view of $3^{\prime\prime} \times 3^{\prime\prime}$.
The G235H and G395H spectral observations include 6 dither pointings with the small cycling pattern to avoid target spectra falling on bad pixels and MSA failed open and closed slits.    
At each dither, two integrations of G235H spectra are recorded with the NRSIRS2 readout mode. Each integration contains 13 groups with a total exposure time of 11554s, or around 3.2 hours. A ``leakcal'' exposure was taken for the G235H grating, consisting of one dither and one integration for a total 1926s exposure time.
For the G395H spectral observation, one integration is executed at each dither, and each integration has 16 groups with the NRS2IRSRAPID readout mode. The total exposure time for the G395H spectroscopy is 1488s, or around 25 minutes. 
In addition, a 904s G235M/F190LP observation composed of two integrations was obtained to fill in the wavelength gap in the G395H spectra because of the spacing between the NIRSpec NRS1 and NRS2 detectors.

\label{sec:obs_jwst_reduc}

The data reduction was performed by following the standard JWST data reduction pipeline of version 1.13.0 and a reference file version of pmap 1174 with modifications as described below. 
At the jump detection step in the first stage of the data reduction, "sat\_required\_snowball" was set to False, and "expand\_large\_events" was set to True for detecting large snowball events. 
We then applied the \texttt{nsclean} algorithm \citep{Rauscher2024} to remove faint correlated noise, that appears in the form of vertical banding in the NRS1 and NRS2 detector plane maps. For each exposure, it uses areas where the dark is known to fit a background model in Fourier space.
In the second stage, the reduced data ("cal. fits") were processed through a customized module to check if all pixels with unreliable flats or lacking saturation checks in the data quality array are flagged as "DO\_NOT\_USE".
This module also checks the spatial flux difference between adjacent pixels.
If the flux difference between a pixel and the adjacent ones is at least eight sigmas or higher among pixels within a 5x7 pixel region, it is considered an artifact and flagged as "DO\_NOT\_USE" (the same procedure is performed by \citealt{Nisini2024b}).
The flagged pixels were then replaced with linearly interpolated adjacent good pixels.
Finally, via the third and final stage of the pipeline, we apply an ultimate outlier detection step (\ie the "outlier\_detection" step, with threshold\_percent = 99.8 and a kernel\_size = 33, comparing the threshold with the normalized minimum pixel difference inside the kernel, used to identify bad pixels) and build the spectral cube in the "cube\_build" step with the "emsm" weighting method.

\section{Spectral analysis}
\label{sec:results}

\subsection{Continuum}
\label{sec:obs_jwst_cont}

\begin{figure*}[!tbh]
\centering
\includegraphics[scale=0.55,clip,trim= 0cm 0.0cm 0cm 0.0cm]{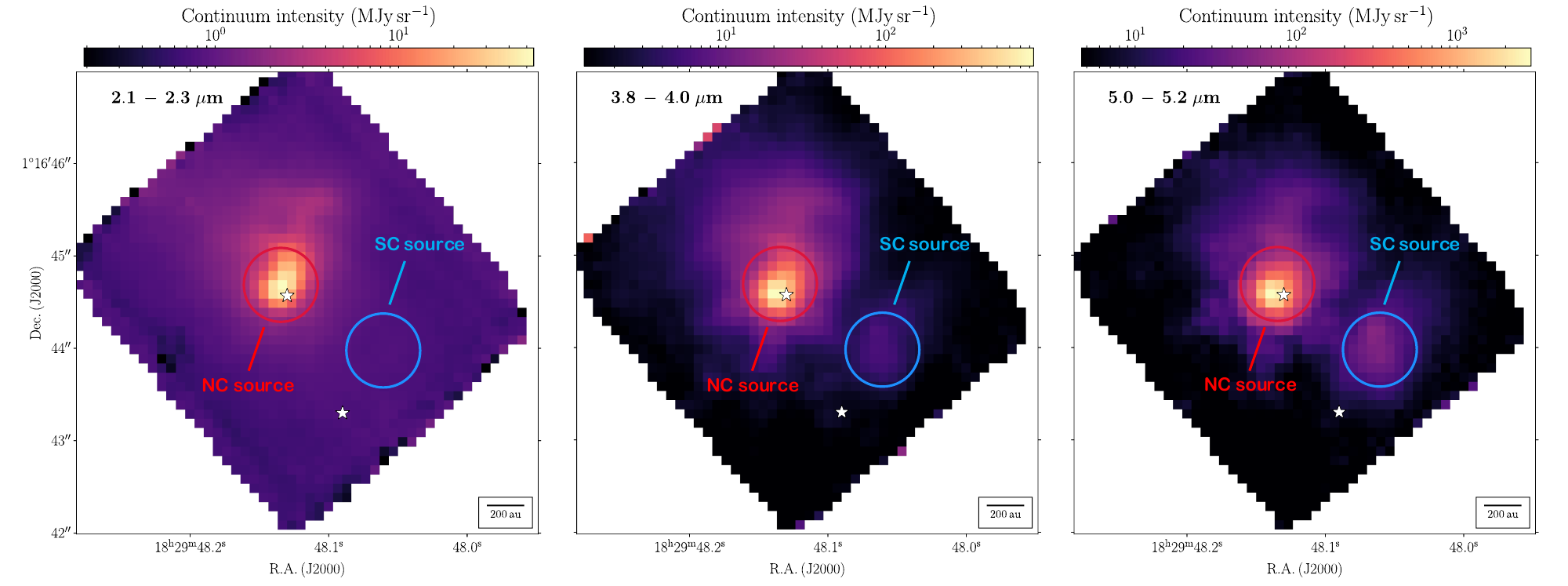}
\vspace{-0.3cm}
\caption{\small Continuum emission extracted from the NIRSpec IFU G235H and G395H cubes, between 2.1 and 2.3 $\mu$m (\textit{left panel}), 3.8 and 4.0 $\mu$m(\textit{middle panel}), and 5.0 and 5.2 $\mu$m (\textit{right panel}). The white stars correspond to the location of the submillimeter emission peaks identified by \citet{LeGouellec2019a}. The red and blue circles show the chosen apertures for spectral extraction of the NC and SC sources (see Section \ref{sec:obs_jwst_spec}), respectively. The aperture diameters in units of $\lambda/D$ (where $\lambda$ is the central wavelength of the corresponding spectral window, and $D$ the effective mirror diameter of JWST) are equal to 11.5, 6.5, and 5, at 2.2 $\mu$m, 3.9 $\mu$m, and 5.1 $\mu$m, respectively}.
\label{fig:cont_images}
\end{figure*}

The continuum emission extracted from the NIRSpec IFU G235H and G395H cubes between 2.1 and 2.3 $\mu$m, 3.8 and 4.0 $\mu$m, and 5.0 and 5.2 $\mu$m, are shown in Figure \ref{fig:cont_images}. To derive the continuum in each pixel within the above spectral regions, we compute the local median of the sigma-clipped spectra (using a 5 standard deviation clipping threshold) in a moving spectral region of 0.03 $\mu$m in size. We then fit the resulting spectra with third-order polynomials and compute their means to estimate the continuum level in each pixel in each spectral region. We note that the CO line forest may contaminate the determination of the continuum between 5.0 and 5.2 $\mu$m, as we are limited by the faint continuum level of the SC source (see the spectra extracted in Section \ref{sec:obs_jwst_spec}).

The NIRSpec IFU cubes reveal two continuum point sources. In the 2.1---2.3 $\mu$m continuum image, we recover the extended emission of the NC source seen in the NIRCam F210M data presented in Figure \ref{fig:RGB_alma_image}.
Since the continuum map we build with our NIRSpec data is devoid of emission line contributions, this suggests that the NIRCam F210M data are more sensitive to NIR continuum, which can be in-situ NIR emission or scattered light.
If coming from a young disk, the NIR continuum would produce an axisymmetric emission. Given the morphology of the NIR light to the North of the NC source, scattered light must contribute to the NIR continuum.
This structure is also seen in the 3.8---4.0 $\mu$m and 5.0---5.2 $\mu$m images, but the JWST PSF shape starts to dominate at longer wavelengths.
The FWHM of the continuum emission (obtained with a 2D Gaussian fit in the image plane) is $\sim0\farcs22$ at 3.9 and 5.1 $\mu$m, while $\lambda/D$ equals (where is $D$ is 6.5 meters, the effective mirror diameter of JWST) to $0\farcs12$ and $0\farcs16$ at 3.9 and 5.1 $\mu$m, respectively. Again, this suggests that scattered light contributes to the spatial extent of the NIR emission.
We note that the infrared location of the NC source coincides well with the corresponding submillimeter peak emission. 

The continuum of the SC source is totally extincted at 2.1---2.3 $\mu$m, but a faint continuum is detected at 3.8---4.0 $\mu$m and 5.0---5.2 $\mu$m. However, the infrared location of the SC source is offset by $\sim0\farcs75$ northwest of the submillimeter peak emission toward the blueshifted outflow cavity identified by \citet{LeGouellec2019a} (see the white stars in our Figure \ref{fig:RGB_alma_image} and Figure \ref{fig:cont_images}, and Section \ref{sec:obs_alma}). The $\sim 0\farcs5$ region centered on the submillimeter peak is itself totally obscured in the NIR data. 
The nature of the 3.8---4.0 $\mu$m and 5.0---5.2 $\mu$m SC source continuum is thus most likely scattered light coming from the inner regions that eject the bipolar outflow.

\subsection{Spectral extraction and line detection}
\label{sec:obs_jwst_spec}

\begin{figure*}[!tbh]
\centering
\includegraphics[scale=0.565,clip,trim= 2.5cm 1cm 3.5cm 1.3cm]{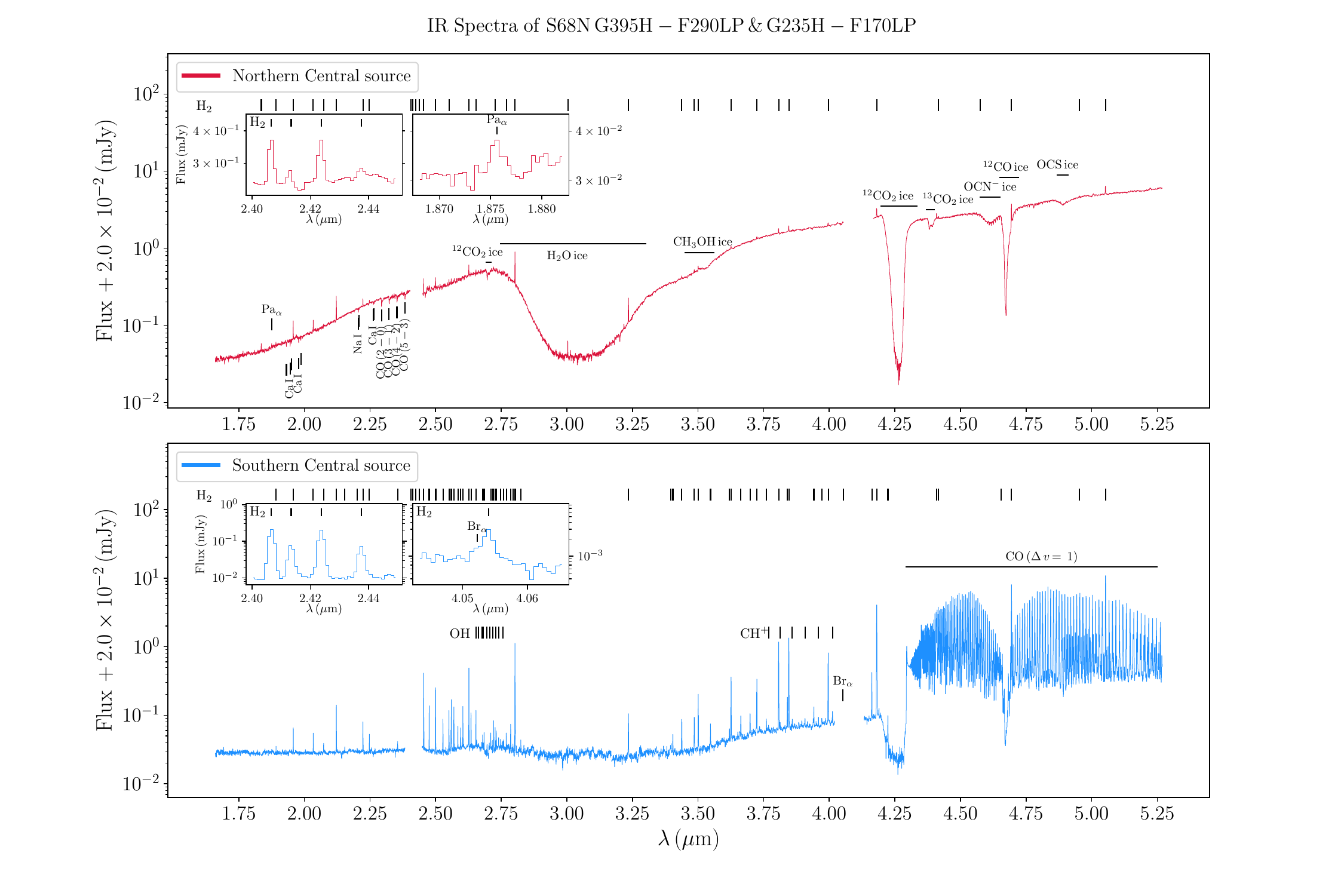}

\vspace{-0.3cm}
\caption{\small Spectra extracted toward the NC (\textit{top panel}) and SC (\textit{bottom panel}) sources from the F170LP/G235H, F170LP/G235M, and G395H/F290LP NIRSpec IFU observations. In each panel the main spectra have been extracted from the G235H and G395H observations, using a fixed aperture radius of 0.4$^{\prime\prime}$, and offset by 0.02 mJy. The gaps around 2.4 $\mu$m and 4.1 $\mu$m are caused by the gaps between the NRS1 and NRS2 detectors' wavelength coverage. In each panel the subpanel on the left represents the spectra extracted from the G235M grating between 2.4 $\mu$m and 2.45 $\mu$m filling the gap of the G235H grating. The subpanel on the right shows faint H\small{I} lines seen in the NC and SC sources. One subpanel shows HI Br$\alpha$ toward the SC source, but this is only a tentative detection.
As the NC source has strong continuum emission, we outline the main ice absorption features in the top panel. The absorption lines (CO overtone, Ca \small{I}, Na \small{I}) coming from the photosphere of the NC source are identified below the spectra. All other spectral emission features correspond to \Ht, OH, CH$^{+}$ or CO fundamental ($\Delta v = 1$) emission lines. We note faint CO fundamental absorption in the spectra extracted toward the NC source. The spectra shown in this figure are not extinction corrected.}
\label{fig:spec_2_sources}
\end{figure*}

The position of the apertures used for the NC and SC sources are (18$\rm ^{h}$\,29$\rm ^{m}$,  $+$01$\rm ^{\circ}$\,16$\rm ^{\prime}$) 48$\rm ^{s}$.14, 44$\farcs$69, and 48$\rm ^{s}$.06, 43$\farcs$.98 (J2000), respectively. The apertures were chosen to be centered on the respective extracted source's peak NIR continuum emission. We use a fixed aperture radius of $0\farcs4$ to extract the 1D spectra presented in Figure \ref{fig:spec_2_sources}. 
We propagate the uncertainty values of every pixel from the pipeline during the circular aperture-based signal extraction.

The NC source exhibits a bona fide YSO spectrum, with strong extinction and deep ice absorption bands from the cold surrounding protostellar envelope (namely CO$_2$, CH$_3$OH+NH$_3$$\cdot$H$_2$O, H$_2$O stretch, $^{12}$CO$_2$, $^{12}$CO, $^{13}$CO$_2$, OCS, OCN$^{-}$; see \citealt{McClure2023,Brunken2024,Brunken2024b,Slavicinska2024}). We also note a faint CO$_2$ ice feature at 2.77 $\mu$m, and faint ice absorption lines from H$_2$O-dangling bonds between 2.72 and 2.76 $\mu$m. The OH-dangling feature (due to water molecules not fully bound to neighboring water molecules; \citealt{Noble2024}) in this latter spectral window is not detected. Shortward of 2.5 $\mu$m, several photospheric features, \ie Na, Ca, and CO overtone ($\Delta v = 2$) bands, are seen in absorption. Only one faint H\small{I} recombination line is seen in emission, the Pa$\alpha$ line at 1.876 $\mu$m. Around 40 \Ht emission lines are detected.

The spectra extracted toward the SC source exhibit only emission lines and no continuum shortward of 3.25 $\mu$m, suggesting strong extinction. A faint continuum is detected longward of 3.25 $\mu$m. The spectra are very rich in molecular emission lines, with $\sim$70 \Ht lines detected and a bright CO fundamental ($\Delta v = 1$) line forest longward of 4.25 $\mu$m. We note the clear detection of the CO ($v = 1\rightarrow0$), ($v = 2\rightarrow1$), and ($v = 3\rightarrow2$) bandheads at 4.29, 4.35, and 4.41 $\mu$m, respectively. \ion{H}{1} Br$\alpha$ is only tentatively detected; the line falls in between NRS1 and NRS2, whose separation in the 2D detector coincides with the location of the SC source. It is also blended with $\rm  H_{2}\,2\rightarrow 1\,O(7)$ (see Figure \ref{fig:spec_2_sources}). 

In addition to CO and \Ht, we report detections of OH, CH$^{+}$, and candidate detection of H$_{3}^{+}$ in Section \ref{sec:mol_em_other_species}. Finally, no [\ion{Fe}{2}] lines are detected in either of the two sources.


\subsection{NIRSpec emission line maps}
\label{sec:obs_jwst_maps}

\begin{figure*}[!tbh]
\centering
\subfigure{\includegraphics[scale=0.34,clip,trim= 3cm 0.5cm 0cm 1.0cm]{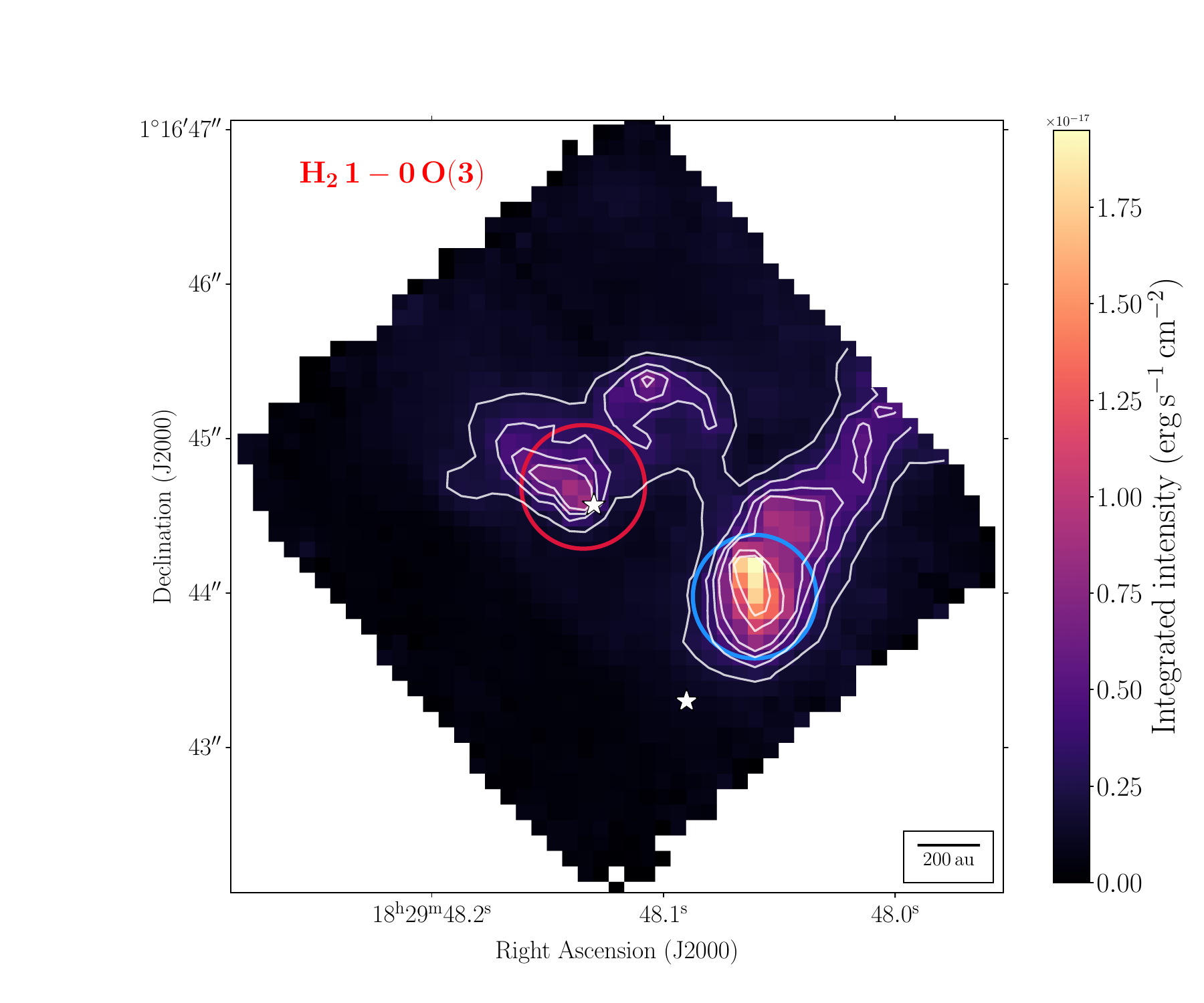}}
\subfigure{\includegraphics[scale=0.34,clip,trim= 5.7cm 0.5cm 0cm 1.0cm]{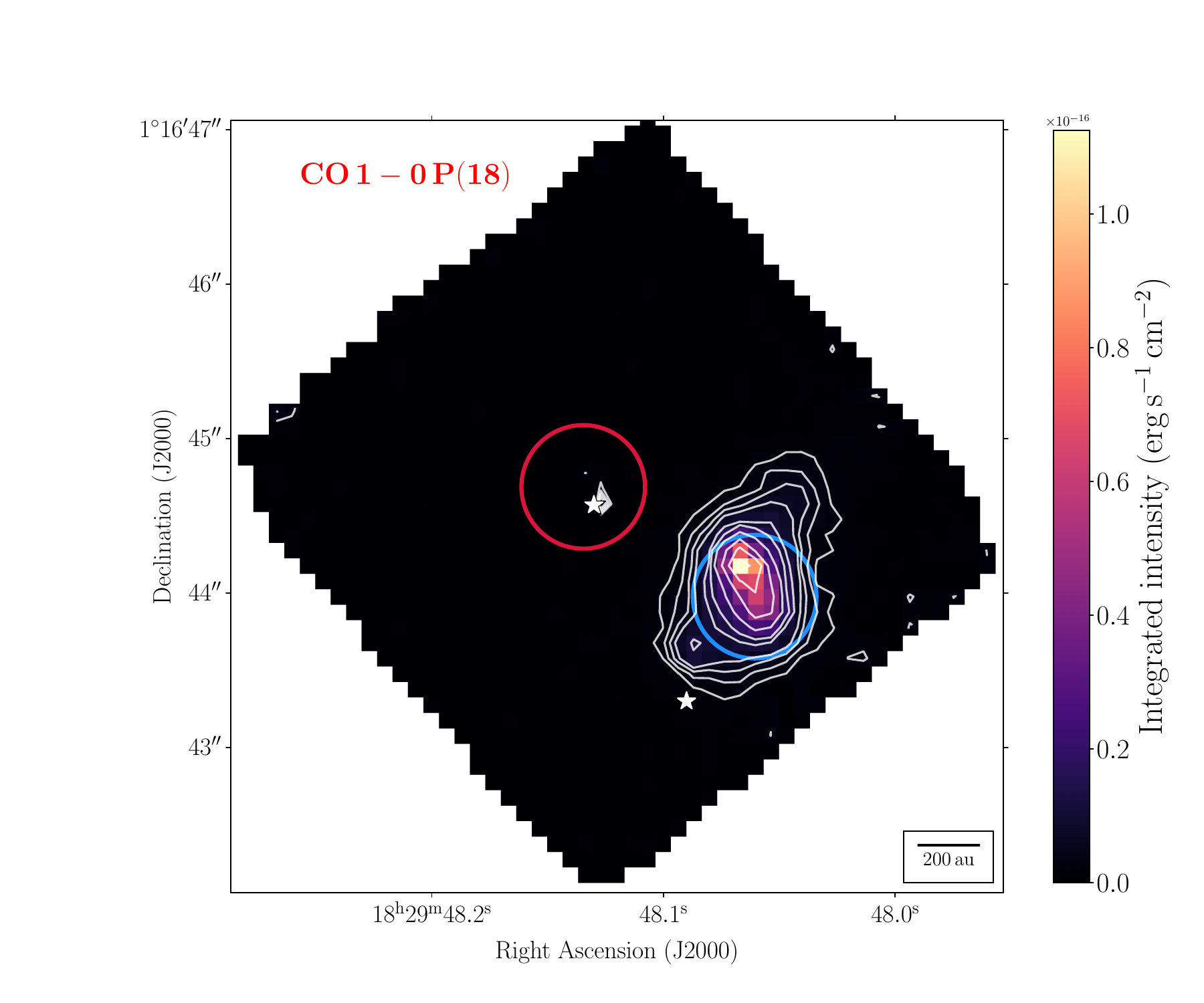}}

\vspace{-0.2cm}
\caption{\small Examples of 2D extracted emission line maps in the NIRSpec IFU data of S68N. The left panel shows the $\rm H_{2}\,1-0\,O(3)$ emission line map at 2.80 $\mu$m extracted from the G235H data. The right panel shows the $\rm CO\,1-0\,P(18)$ emission line map at 4.83 $\mu$m extracted from the G395H data. The color scale and the white contours trace the integrated line intensity. The white stars correspond to the locations of the submillimeter emission peaks identified by \citet{LeGouellec2019a}. The red and blue circles show the chosen apertures for spectral extraction of the NC and SC sources (see Section \ref{sec:obs_jwst_spec}), respectively.
}
\label{fig:line_H2_CO_int_flux}
\end{figure*}

\begin{figure*}[!tbh]
\centering
\subfigure{\includegraphics[scale=0.34,clip,trim= 3cm 0.5cm 0cm 1.0cm]{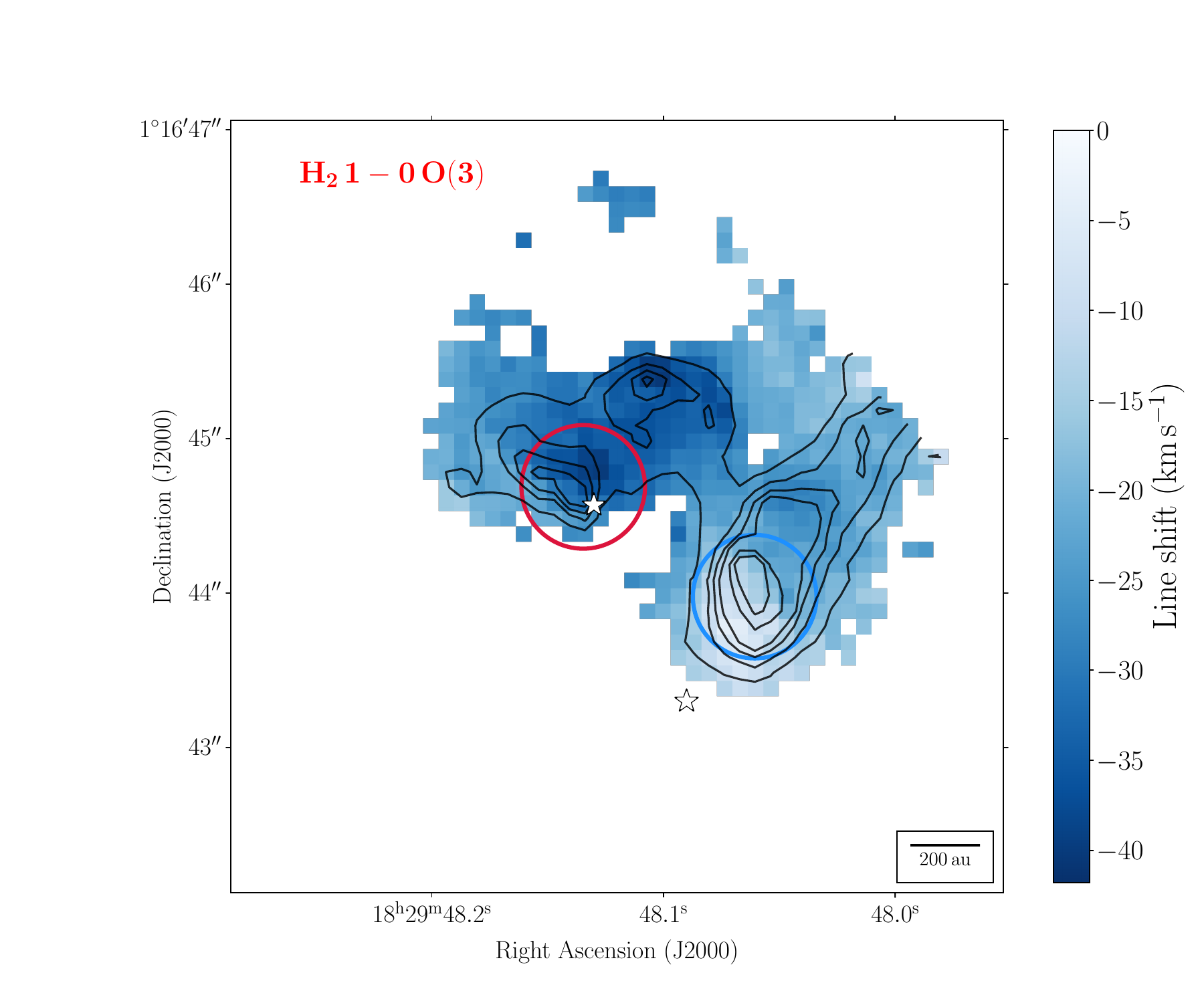}}
\subfigure{\includegraphics[scale=0.34,clip,trim= 5.7cm 0.5cm 0cm 1.0cm]{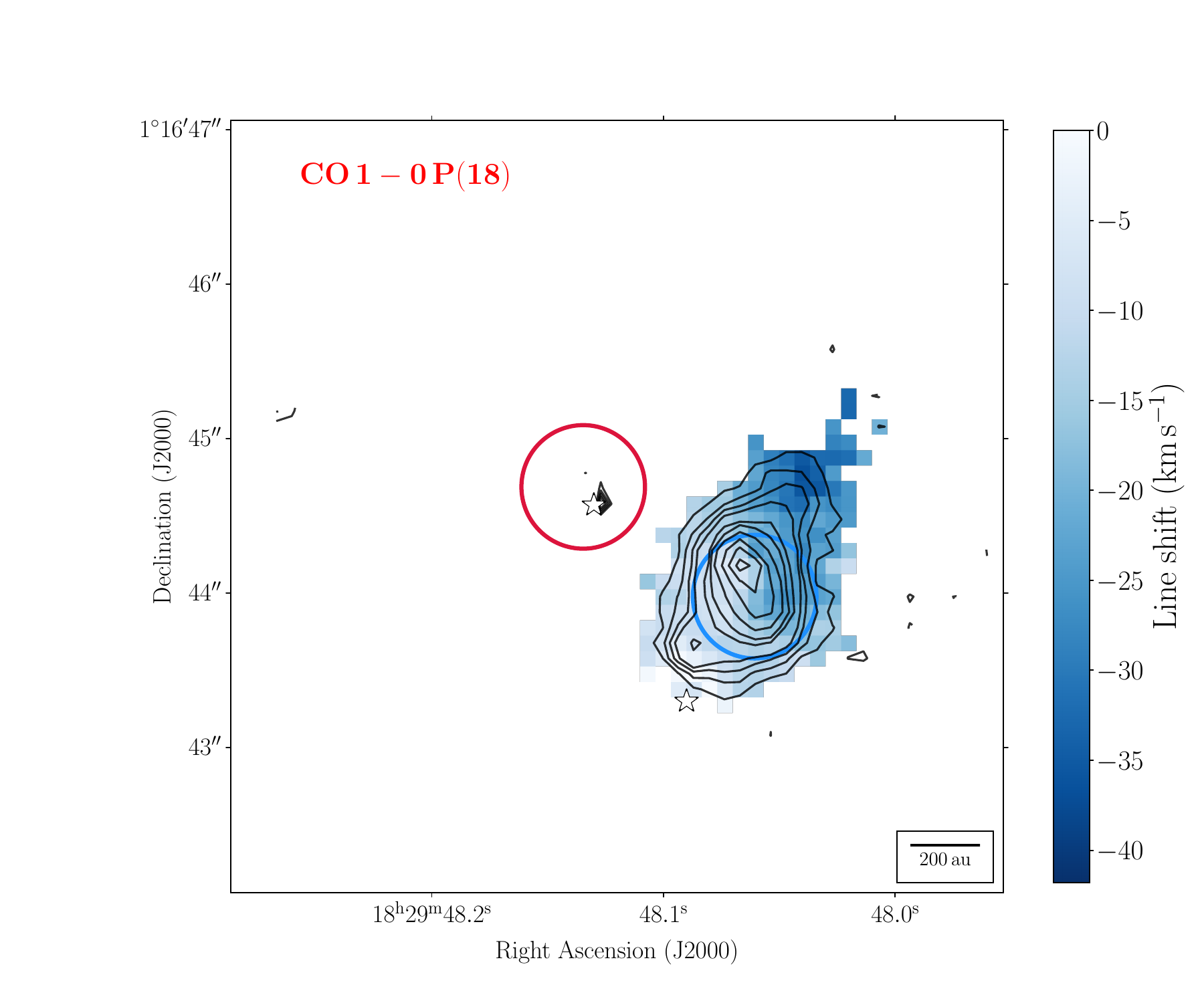}}

\vspace{-0.2cm}
\caption{\small Same as Figure \ref{fig:line_H2_CO_int_flux} but showing the velocity shifts of the $\rm H_{2} 1-0 \,O(3)$ and $\rm CO 1-0\,P(18)$ emission lines (extracted in each pixel). The velocities have been corrected for the LSR velocity of the S68N core (\ie 8.45 km s$^{-1}$). The \Ht and CO lines are thus clearly blueshifted.}
\label{fig:line_H2_CO_vel_shift}
\end{figure*}

In order to extract the emission lines, we subtract the 1D continuum spectra extracted using the method outlined in Section \ref{sec:obs_jwst_cont}. The spectra are then fitted with 1D Gaussian profiles. We propagate the uncertainty values from the pipeline into the \texttt{curve\_fit} routine of the \texttt{scipy} python package to derive the uncertainty of the fitted line profile parameters, \ie amplitude, full-width-half-maximum (FWHM), and velocity shift. To determine whether a line is considered detected or not, we require the signal-to-noise ratio (S/N) of the line emission peak (obtained as an output of the 1D Gaussian profile fitting attempt with the curve\_fit method of scipy) and integrated flux (obtained integrating the continuum subtracted spectra over [-3/2 $\times$ FWHM; 3/2 $\times$ FWHM]) to be $\geq$ 10. If a line is not detected, its upper limit is set to $10 \times \sigma  \times \lambda_{\textrm{line}}/$R, where R is the spectral resolution, and $\sigma$ is the average uncertainty around $\lambda_{\textrm{line}}$ (computed within a window of 2 $\times$ R/$\lambda_{\textrm{line}}$ in size, centred on $\lambda_{\textrm{line}}$).

Figure \ref{fig:line_H2_CO_int_flux} presents the molecular emission line maps of \HtonetoOOthree and \COonetoOPeightneen, where each spaxel was treated independently as a 1D spectrum. The integrated flux is shown where the line is detected. 
Attempting a line fit of \Ht and CO for each spaxel allows for a rigorous signal extraction, as the line flux and uncertainty are correctly propagated and detections accurately determined. 
This also allows us to build maps of emission line parameters, \ie FWHM and velocity shift, as shown in Figure \ref{fig:line_H2_CO_vel_shift}. Our molecular emission lines are typically unresolved in velocity width in our NIRSpec IFU data ($\delta$v $\sim\,$90---140 km s$^{-1}$, the spectral resolution of NIRSpec IFU gratings). However, relative systemic velocities are measured with much higher precision. The uncertainties we derive for line velocity shifts ($\lesssim 0.5$ km s$^{-1}$) are most likely underestimates, and the absolute uncertainties must account for systematic errors in the wavelength calibration. Any such errors are not precisely constrained so far, and the JWST pipeline documentation only refers to the fact that the wavelength scale shall be determined with an accuracy of much better than 1/8 of a resolution element after calibration \citep{Boker2023}, \ie corresponding to an accuracy $\lesssim\,14$ km\,s$^{-1}$ for the high spectral resolution gratings of NIRSpec.

The \Ht emission line map exhibits different morphology between the NC and the SC source. The \HtonetoOOthree line of the NC source exhibits two extended patches of emission whose morphology matches well the extended 2 $\mu$m continuum emission seen in the NIRCam data. The SC source presents a clear jet-like collimated structure in the \Ht emission. In both sources, there is an offset between the submillimeter continuum emission and the \Ht emission, suggesting that only one side of the cavity, \ie the blueshifted side given the velocity shift map of Figure \ref{fig:line_H2_CO_vel_shift}, is seen. The \Ht emission is blueshifted by $\sim\,$30--40 km s$^{-1}$ in the NC source, while it is blueshifted by $\sim\,$15--30 km s$^{-1}$ in the SC source with a small gradient in the direction of the collimation (see Section \ref{sec:mol_em_jet_south}; these measured velocities have been corrected to the S68N's $v_{\textrm{LSR}}$). We note that in the jet-like structure of the SC source, the morphology of the \Ht emission is clearly different than the CO emission. The CO emission is less spatially extended and confined toward the base of the blueshifted cavity, where the 3.5--5$\,\mu$m continuum is detected (see Section \ref{sec:obs_jwst_cont}). This suggests that the CO and the \Ht emitting gases may be coming from different physical components of the jet system. 
However, the relative emission of CO and \Ht can intrinsically be sensitive to different environmental conditions (\ie irradiation, density, temperature, and chemistry).
The CO emission is slightly more blueshifted than \Ht in the SC source (\ie blueshifted by $\sim\,$15--35 km s$^{-1}$; the difference in velocity shift with respect to \Ht is more than three times larger than the velocity uncertainty), but the small gradient in velocity shift is consistent in the two molecules. This gradient suggests that this molecular emission is locally produced, but we cannot rule out that scattered molecular emission contributes to the total line emission (we discuss the excitation conditions of the CO and the \Ht gas in Section \ref{sec:mol_em_jet_south}).

\subsection{ALMA molecular emission}
\label{sec:obs_alma}

\begin{figure*}[!tbh]
\centering
\subfigure{\includegraphics[scale=0.42,clip,trim= 3.0cm 1.0cm 4.8cm 0.2cm]{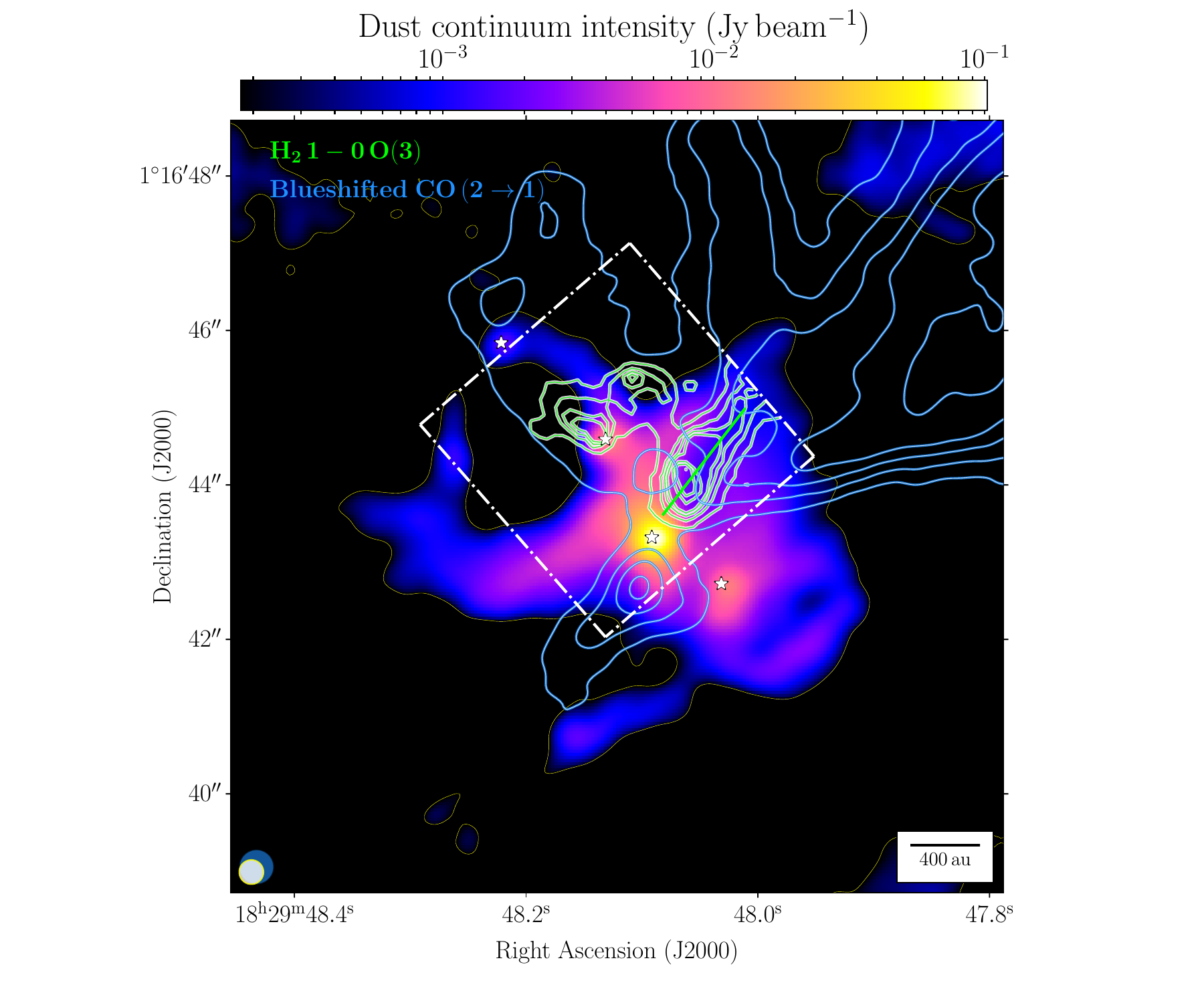}}
\subfigure{\includegraphics[scale=0.42,clip,trim= 5.7cm 1.0cm 4.8cm 0.2cm]{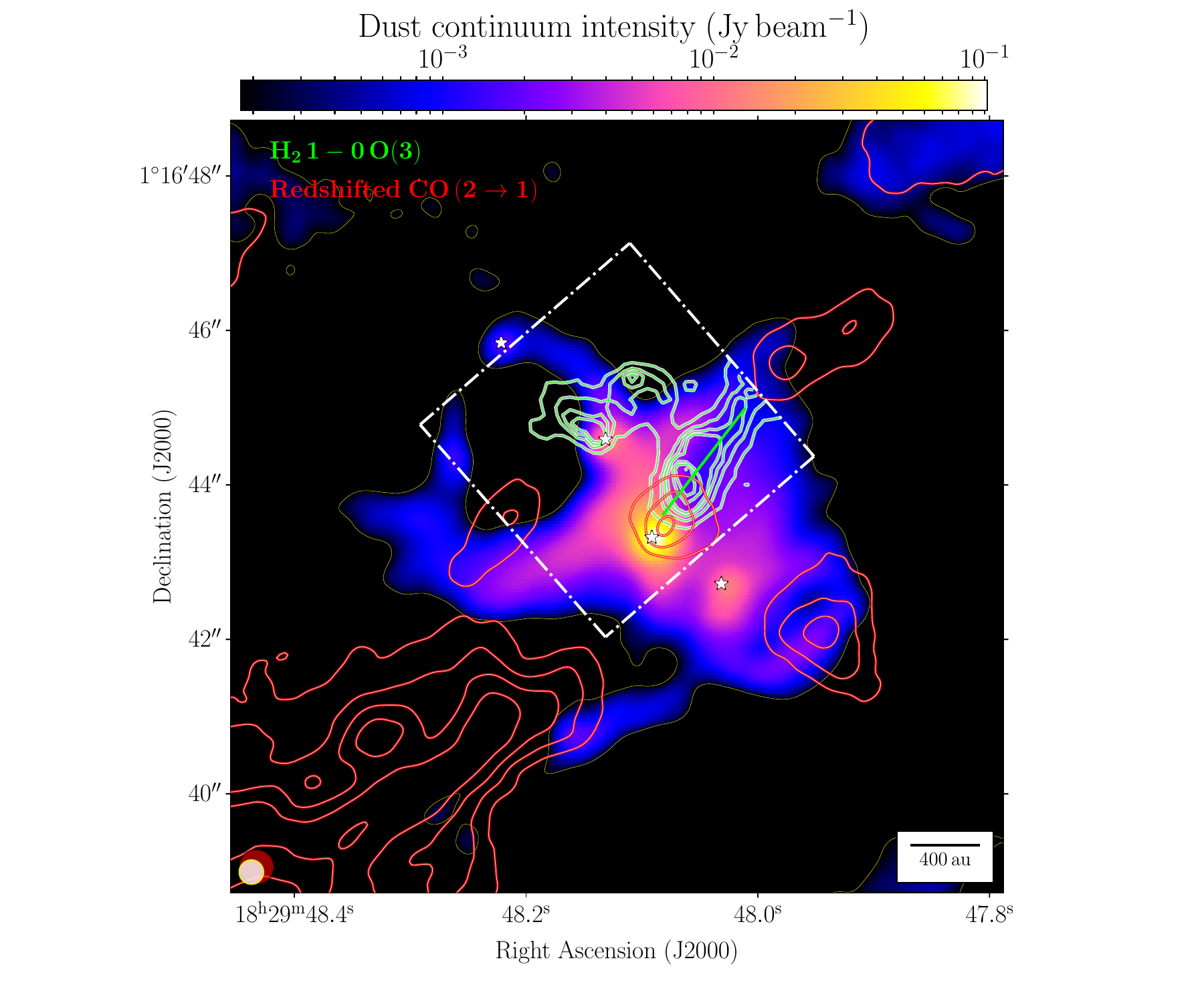}}

\vspace{-0.2cm}
\caption{\small Overlay of JWST NIRSpec IFU \Ht observations of S68N with ALMA observations of \cotwotoone and 870 $\mu$m dust continuum emission. In both panels, the color scale is the 870 $\mu$m dust continuum emission from \citet{LeGouellec2019a}, with the single yellow contours tracing the 3 $\sigma_I$ level, where $\sigma_I$ = 60 $\mu$Jy beam $^{-1}$. In the \textit{left} and \textit{right} panels, the blue and red contours trace the velocity-integrated blueshifted and redshifted \cotwotoone emission, integrated between $-53$ and $6$ km s$^{-1}$, and $11$ and $40$ km s$^{-1}$, respectively. The green contours trace the $\rm H_{2}$ $1-0$ O(3) emission line map at 2.80 $\mu$m extracted from the G235H data.
The four white stars correspond to the four fragments identified in the ALMA continuum map. The white square corresponds to the coverage of the JWST NIRSpec IFU data presented here. The ellipses in the bottom left corner of each plot show the beamsize of the ALMA observations, white being for the dust continuum, and blue/red being for the CO integrated emission. The green line indicates the axis the of molecular jet, along which we compute the \Ht excitation in Section \ref{sec:mol_em}.}
\label{fig:JWST_ALMA_images}
\end{figure*}

We now compare the morphology of the hot component of the outflows associated with \Ht emission that traces the hot shocked gas (\ie $\sim\,1000-3000$ K), with the millimeter CO rotational emission line, which traces colder (\ie $\simeq$ 100 K) molecular gas. Figure \ref{fig:JWST_ALMA_images} presents an overlay of the $\rm H_{2}$ $1\rightarrow 0$ O(3) integrated emission line with the ALMA blueshifted and redshifted components of the \cotwotoone emission line maps, and the ALMA 870 $\mu$m dust continuum observations from \citet{LeGouellec2019a} and \citet{Tychoniec2019} (ALMA program 2013.1.00726.S, PI: C. L. H. Hull). The ALMA CO and dust continuum data have a spatial resolutions of 170 and 240 AU, respectively.
The main ALMA source, corresponding to the SC source in our JWST observations, seems to be the source launching the large conical bipolar outflow. The \Ht jet-like collimated structure coincides well with the ``V''-shape blueshifted emission of the CO outflow. Faint blueshifted (redshifted) CO emission is seen toward the redshifted (blueshifted) outflow cavity, suggesting that the outflow directions are close to the plane-of-the-sky (POS). 
The low velocity SiO/SO emission seen by \citet{Tychoniec2019,Podio2021} in each outflow cavity suggests fast ejection (these molecules are generally shock tracers and associated with fast, or extremely high velocity jet emission in embedded protostars). The extended low velocity CO emission is also accompanied by low velocity HCN and H$_2$CO emission.

The \Ht emission of the NC source seems to be associated with a more extended faint blueshifted \cotwotoone emission. The Northernmost IR source (labeled as such in Figure \ref{fig:RGB_alma_image}; its millimeter counterpart is the Northernmost white star in Figure \ref{fig:JWST_ALMA_images}) lying outside the NIRSpec IFU coverage but detected in the NIRCam data does not seem to be associated with \cotwotoone emission. We also note the lack of extended $2\mu$m emission in the NIRCam data (covering the \HtonetoOSone line) and thus suggest that no outflow structures are associated with this source. Finally, the southernmost millimeter source (outside the NIRSpec IFU coverage; identified as Serpens Emb 8-c by \citealt{LeGouellec2019a}) could be associated with a faint patch of redshifted and blueshifted \cotwotoone emission, but no NIRCam emission is detected, likely due to the high extinction of this part of the protostellar envelope. This suggests that this latter source could be either a deeply embedded protostar or a prestellar object.

\section{IR molecular emission line analysis}
\label{sec:mol_em}

\begin{figure*}[!tbh]
\centering
\subfigure{\includegraphics[scale=0.50,clip,trim= 0cm 0.0cm 0cm 0cm]{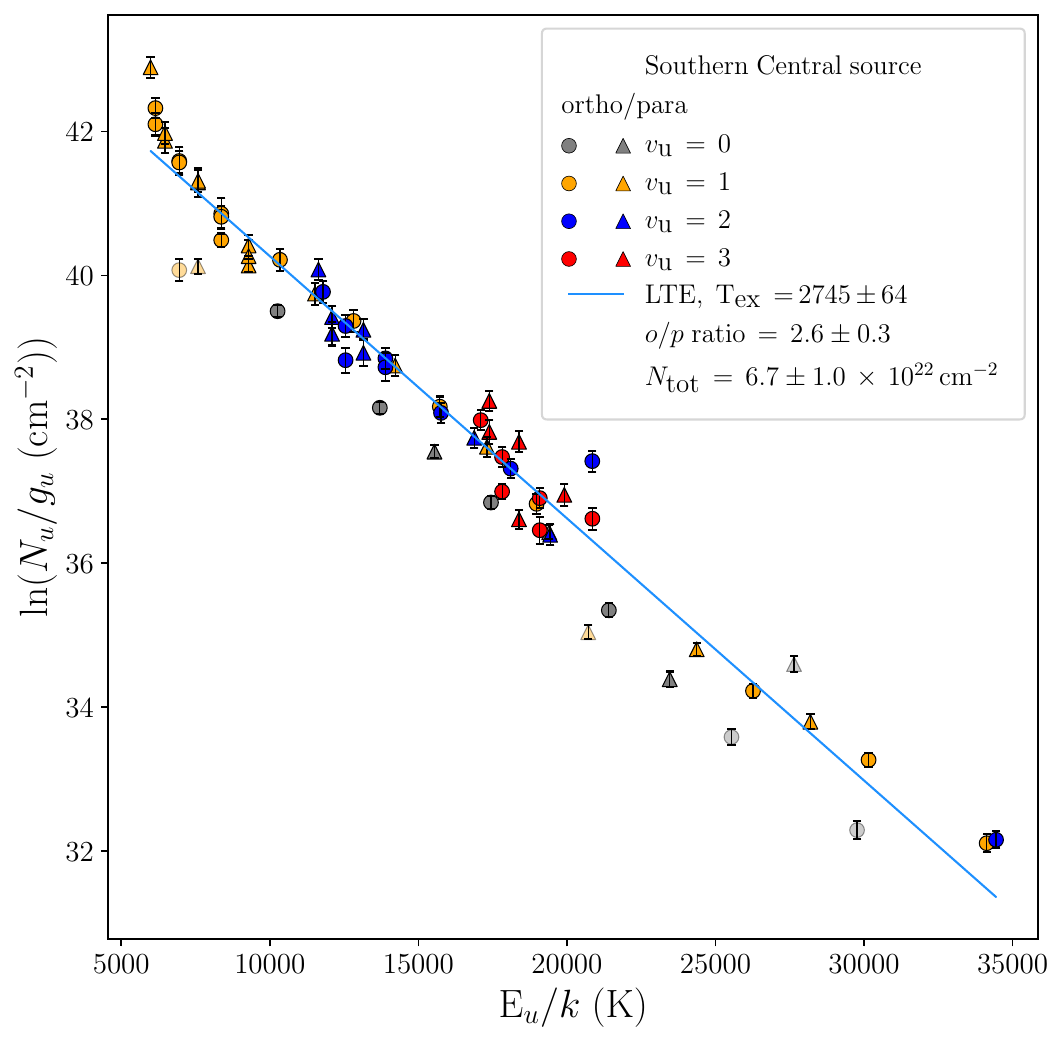}}
\subfigure{\includegraphics[scale=0.50,clip,trim= 0cm 0.0cm 0cm 0cm]{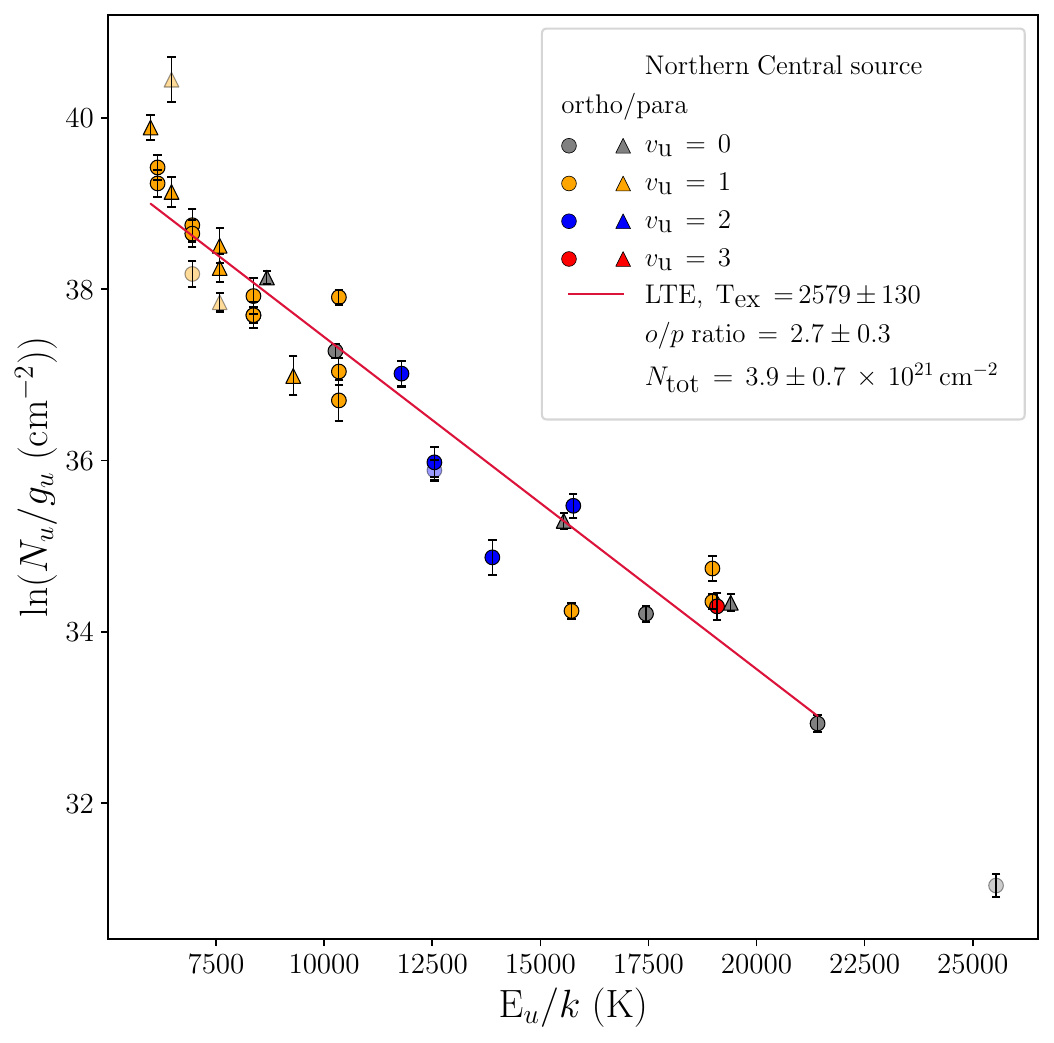}}

\vspace{-0.3cm}
\caption{\small \Ht excitation diagrams computed with extinction-corrected line fluxes obtained toward the SC source (\textit{left panel}), and the NC source (\textit{right panel}), extracting spectra at the two aperture locations indicated in Figures \ref{fig:line_H2_CO_int_flux} and \ref{fig:line_H2_CO_vel_shift}. Each \Ht line is represented by a color point corresponding to the rotational ladder of its upper state (gray for $v_{\textrm{u}}\,=\,0$, orange for $v_{\textrm{u}}\,=\,1$, blue for $v_{\textrm{u}}\,=\,2$, and red for $v_{\textrm{u}}\,=\,3$). Ortho-\Ht lines are represented by circles, and para-\Ht by triangles. 
The lightly shaded points correspond to \Ht lines located in strong ice absorption bands and are not considered in the excitation analysis (\ie temperature, column density, and $o/p$ ratio).
In each panel, the solid line represents the best single-component fit through all the ro-vibrational transitions, and the corresponding excitation temperature and total column density are indicated, alongside the derived $o/p$ ratio.}
\label{fig:H2_diagram_2sources}
\end{figure*}

These JWST NIRSpec IFU observations are rich in \Ht and CO molecular emission lines. Going back to the two spectra extracted toward the NC and SC sources (red and blue apertures in Figures \ref{fig:cont_images}, \ref{fig:line_H2_CO_int_flux}, and \ref{fig:line_H2_CO_vel_shift}; see also the spectra in Figure \ref{fig:spec_2_sources}), we detect up to 70 different \Ht lines toward the SC source, pure-rotational lines ($v_{\textrm{u}}\,=\,0$) up to vibrational of $v_{\textrm{u}}\,=\,3$. A bright forest of CO fundamental ($\Delta\,v\,=\,1$) lines is also detected toward the SC source. In this Section, we constrain the excitation conditions of the molecular gas responsible for these emission lines. We start by presenting an analysis of the \Ht ro-vibrational level populations of both sources in Section \ref{sec:mol_em_H2}. We then focus in Section \ref{sec:mol_em_jet_south} on the molecular jet of the SC source, where we invoke shocks as the origin of the \Ht emission and discuss whether the CO excitation conditions are consistent with the same shock structures.

\subsection{Molecular Hydrogen emission lines}
\label{sec:mol_em_H2}

The ro-vibrational levels of the \Ht molecule ground electronic state \citep{Black1976,Black1987,Sternberg1989a} can be excited by either UV excitation or collisional thermal excitation. The ro-vibrationally excited molecules decay via collisions or radiative cascade of ro-vibrational transitions, whose wavelengths range from the near- to the mid-infrared. Since \Ht is a homonuclear diatomic molecule without a permanent electronic dipole moment, only electric quadrupole transitions ($\Delta J\,=\,0,\,\pm2$) are allowed, which can usually be considered as optically thin. The line flux from each \Ht line is thus proportional to the number of molecules in a given upper level, and we can in turn infer the emitting gas conditions from the ro-vibrational level populations.

To extract all the observed \Ht lines, we use the \Ht line catalog by \citet{Roueff2019}. For a given aperture, the \Ht lines are fitted and then subtracted from the spectra until they no longer meet our detection criteria. The \Ht lines located among the CO fundamental line forest are not considered because our CO model is not accurate enough to produce high-quality CO-subtracted spectra. In addition, the \Ht lines located in the strongest ice absorption bands are extracted but not considered in the \Ht excitation analysis. Especially in the SC source, the optical depths of the ice absorption features are hard to determine, which affects the reliability of our \Ht line flux extraction.

\subsubsection{Extinction measurements}
\label{sec:mol_em_H2_extinction}

We measure the foreground extinction of the \Ht emitting regions by comparing the observed to theoretical line flux ratio of different pairs of lines arising from the same upper vibrational and rotational level (using the computations of \citealt{Roueff2019}). 
The larger the separation between the wavelengths of the \Ht transition pair is, the smaller the uncertainty in the estimated extinction value. 
We adopt the extinction law from \citet{Pontoppidan2024a} throughout the article. 
This extinction law is well adapted to embedded protostellar envelopes, as it reproduces the dense clouds extinction curve \citep{Chapman2009a}. It uses the optical constants of \citet{Draine2003,Zubko1996}, and incorporates the optical constants for mantles of H$_2$O, CO$_2$, and CO ice.
Restricting ourselves to \Ht lines with $v_{\textrm{u}} \,\leq\,1$, we obtain the mean and standard deviation values of $A_K\,=\,4.8\,\pm\,1.4$ and $7.5\,\pm\,1.0$ mag, for the NC and SC source extracted spectra (see Figure \ref{fig:spec_2_sources}), respectively (with a total of 10 and 12 pairs of \Ht lines for the NC and SC sources, respectively). Using other extinction laws such as \citet{Indebetouw2005,Hensley2021} typically yields extinction values within 0.1 mag, suggesting our estimation of the extinction is robust.
As these \Ht lines are very bright, the obtained uncertainties on the individual extinction values are typically small (\ie $\lesssim\,0.1$ mag). For example, the mean (and standard deviation) of the distribution of uncertainties of the different extinction values $A_K$ we estimate for each corresponding \Ht line ratio is 0.05 mag (0.03 mag). The absolute uncertainties are thus governed by the accuracy of our adopted extinction law. Therefore, we choose to not propagate uncertainties from such extinction determinations because these uncertainties reflect systematic effects of the extinction law and wavelength-dependence of the scattered light, which are not modeled in this work.
the distribution of uncertainties of the different extinction values we estimate for each corresponding \Ht line ratio


\subsubsection{Ortho-to-para Ratio}
\label{sec:mol_em_H2_opr}

The spins of the two protons in \Ht molecules are either aligned or anti-aligned, corresponding to two distinct spin isomers called ortho-\Ht (spins aligned) and para-\Ht (spins anti-aligned). The nuclear spin statistical weights for \Ht molecules in thermal equilibrium give an ortho-to-para abundance ratio ($o/p$ ratio) of 3. Since the \Ht radiative transitions (only $\Delta J\,=\,0,\,\pm2$ are allowed) conserve the ortho and para state of the molecule, the $o/p$ ratio of an \Ht gas is set or modified by \Ht formation, collisions that exchange a nucleus (with \Ht, H, and ions such as H$^+$ and H$_{3}^{+}$), or selective dissociation. It follows that the $o/p$ ratio of an \Ht line emitting gas can be a relevant indicator of hot gaseous conditions, such as those encountered in the passage of a shock wave. 

We measure the $o/p$ ratio by minimizing the distance (in the ln($N_{\textrm{u}}/g_{\textrm{u}}$) space) between the \Ht-para and -ortho level populations over a range of $o/p$ ratio values, for \Ht lines with $v_{\textrm{u}} \leq 3$ and $J_{\textrm{u}} \leq 11$. At each step of the fitting routine, we apply an $o/p$ correction factor to the \Ht-para levels and measure the vertical distance with the \Ht-ortho levels in the ln($N_{\textrm{u}} \ g_{\textrm{u}}$) space by linearly interpolating them along the line upper energy levels, for each rotation ladder (method outlined by \citealt{Kaplan2021}, which yields statistical uncertainties obtained from the standard deviation of the best models around the optimum). We obtain mean $o/p$ ratios of 2.6 $\pm$ 0.3 and 2.7 $\pm$ 0.3 for the SC and the NC source, respectively. These values typically correspond to shocked gas, of hot temperature ($T\,\geq 2000$ K) and high pre-shock density (n$_{\textrm{H}}\,\geq 10^{5}$ cm$^{-3}$) following \citet{Neufeld1998,Wilgenbus2000,Neufeld2006}. We stress that such derived values of $o/p$ ratios will depend on the set of \Ht lines chosen to compute them. The comparisons with such numerical shock calculations thus must be done with care, as line upper levels of different excitation energies are populated differently across the shock wave (see Section 4.1 of \citealt{Wilgenbus2000}).

\subsubsection{\Ht excitation}
\label{sec:mol_em_H2_excitation}

As the \Ht lines are optically thin, their line flux is proportional to the column density $N_u$ of \Ht molecules in the upper states of the transitions. We derive the column density of \Ht in the upper state from the following equation:
\begin{equation}
N_u\,=\,\frac{F_{ul}}{\Delta E_{ul} h c A_{ul}}\,\,,
\end{equation}
where $F_{ul}$ is the line flux of the radiative transition from upper ($u$) to lower ($l$) ro-vibrational states, $\Delta E_{ul}$ is the energy difference of the states (in cm$^{-1}$), $A_{ul}$ is the transition probability (from \citet{Roueff2019}), $h$ is Planck’s constant, and $c$ is the speed of light. We fit a single temperature component and extract the total column density of molecules $N_{\textrm{tot}}$ assuming the ro-vibrational level populations follow the Boltzmann distribution, \ie $\textrm{ln}\left(N_u / N_{\textrm{tot}}\right)\,=\, \textrm{ln}\left(g_u / Z(T)\right)\,+\,E_u/kT$, where $g_u$ is the degeneracy of a given upper level and $Z(T)$ the partition function of \Ht at the temperature $T$. 

The fits for the SC and NC source extracted spectra are shown in Figure \ref{fig:H2_diagram_2sources}. The single temperature component fit ($2745 \pm 64$ K and $2579 \pm 130$ K for the SC and NC sources, respectively), expected in energetic shock conditions, appears to be a good approximation, especially for the SC source molecular outflow. A colder component is seen around E$_{u}\,\lesssim\,5000$ K, for only the $\sim$ 3 lowest excitation energy lines among the \Ht lines we cover. 
Measurements of pure rotational \Ht lines located within the JWST/MIRI MRS spectral range would thus be necessary to accurately constrain this cold component. The $N_{\textrm{tot}}$ we derive (\ie $6.7 \pm 1.0 \times 10^{22}$ cm $^{-2}$ and $3.9 \pm 0.7 \times 10^{21}$ cm $^{-2}$ for the SC and NC sources, respectively) is thus a lower limit and only represents the column density of the hot \Ht gas. We detect up to $v\,=\,3\rightarrow2$ in the SC source, and up to $v\,=\,2\rightarrow1$ in the NC source.

\subsection{The molecular wind of the SC source}
\label{sec:mol_em_jet_south}

\begin{figure}[!tbh]
\centering
\includegraphics[scale=0.45,clip,trim= 0cm 0.0cm 0cm 0cm]{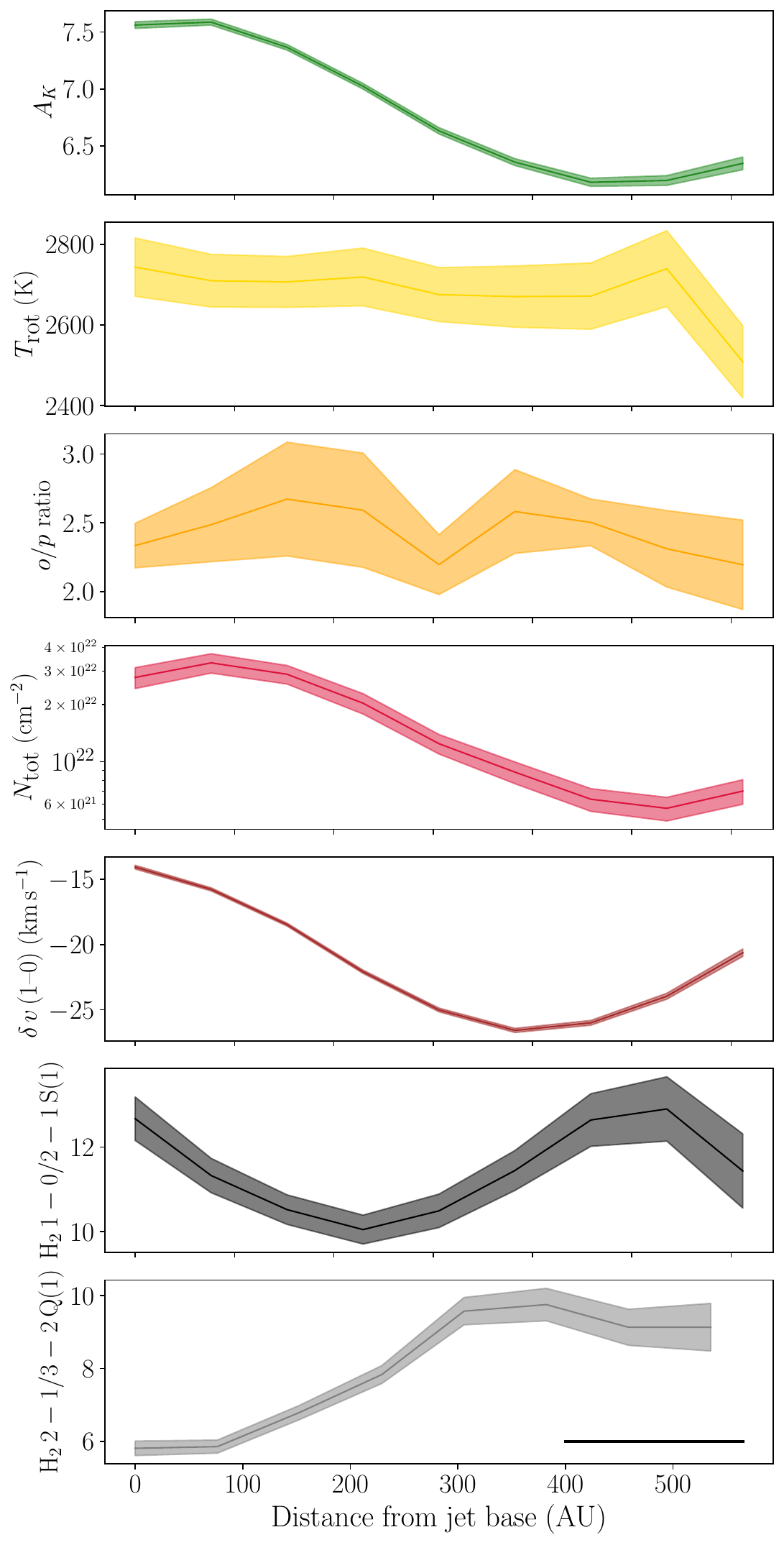}

\vspace{-0.2cm}
\caption{\small Evolution of the \Ht excitation parameters as a function of position along the major axis of the molecular jet of the SC source (whose axis is outlined in Figure \ref{fig:JWST_ALMA_images} with a green line). Shown top to bottom are the \Ht based $K$-band extinction, the fitted excitation temperature, $o/p$ ratio, the \Ht column density from the rotational diagram, the mean velocity shift of \Ht lines (corrected for $v_{\textrm{lsr}}$), and the two \Ht line extinction-corrected flux ratios $\rm H_{2}\,1-0/2-1\,S(1)$ and $\rm H_{2}\,2-1/3-2\,Q(1)$. These two ratios are between \Ht lines having the two same rotational level but different vibrational levels. The shaded area indicates the corresponding uncertainties propagated in the calculation of each parameter. The horizontal black line in the bottom plot shows the aperture size.}
\label{fig:H2_param_jet}
\end{figure}

\begin{figure*}[!tbh]
\centering
\includegraphics[scale=0.6,clip,trim= 0cm 0.0cm 0cm 0cm]{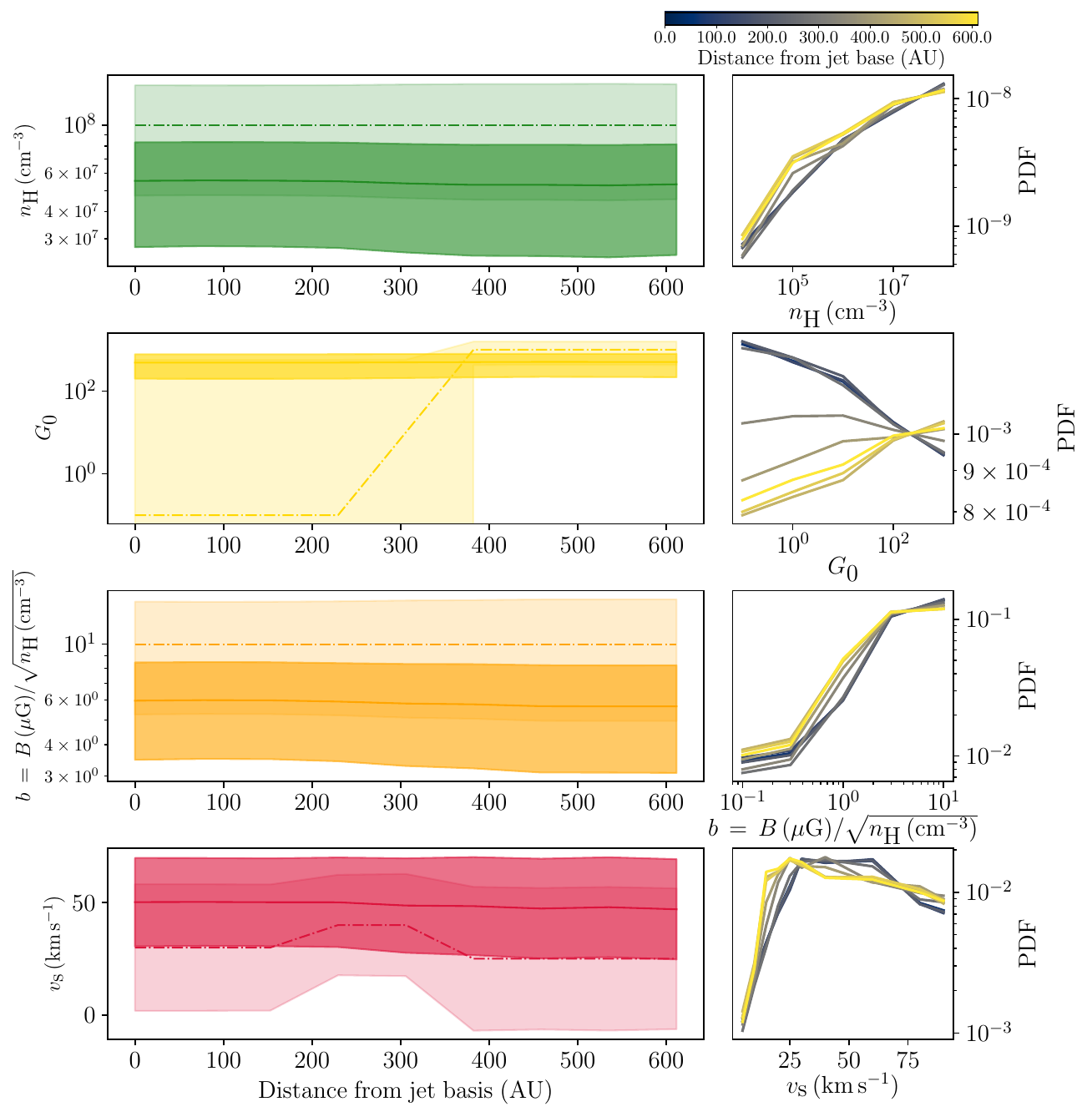}

\vspace{-0.3cm}
\caption{\small Shock model parameter exploration using the grid of \citet{Kristensen2023}. 
This grid varies the volume density \nH, the transverse magnetic field strength \b such that $B = b\,\times\,\sqrt{\textrm{n}_{\small{\textrm{H}}} (\textrm{cm}^{-3})}\,\mu\textrm{G}$, the shock velocity \vs, and the external UV irradiation field strength \Go. For each spatial location (\ie aperture radius of 0.4$^{\prime\prime}$) along the molecular jet of the SC source, the extinction-corrected flux of the detected \Ht lines is compared for every shock model of the grid. A likelihood function is then calculated (see Section \ref{sec:mol_em_jet_south_H2shocks}).
Every row corresponds to one parameter of the model grid. 
The left panels show the evolution of the mean (solid line) and the maximum (dashed line) of the 1D probability function (see appendix \ref{app:shock_models}) along the jet. The shaded areas show the respective standard deviations. The right panel shows the 1D probability function for each aperture along the jet, color-coded as a function of distance from the jet base.}
\label{fig:H2_shock_modeling}
\end{figure*}

We focus in this Section on the molecular excitation analysis of the SC source outflow. So far we have performed our \Ht excitation analysis only for a 1D extracted spectrum  (left panel of Fig.\  \ref{fig:H2_diagram_2sources}). Figure \ref{fig:H2_param_jet} presents the evolution of the \Ht excitation parameters as a function of position along the major axis of the blueshifted molecular jet. For each position, the spectrum is extracted with the same aperture every time (0$\farcs$4 radius), and the same \Ht analysis as in Section \ref{sec:mol_em_H2_excitation} is performed. We oversample the number of pointings along the jet axis for plotting purposes; the diagrams of Figure \ref{fig:H2_param_jet} have $\sim\,$4 independent measurements along the jet sampled by 9 locations. The \Ht based $K$-band extinction, the fitted excitation temperature, $o/p$ ratio, and \Ht column density are from rotational diagrams built at each position along the jet. The mean velocity shift of \Ht lines and the extinction-corrected line flux ratios $\rm H_{2}\,1-0/2-1\,S(1)$ and $\rm H_{2}\,2-1/3-2\,Q(1)$ are shown. 

We note a clear decrease of extinction and \Ht column density as we move away from the jet base. The excitation temperature and $o/p$ ratio remain constant along the jet, within our computed uncertainties. A clear gradient of \Ht line velocity shifts of $\sim 12.5$ km\,s$^{-1}$ per arcsecond is seen, with the \Ht line centroid becoming bluer moving away from the source. The uncertainties shown in this diagram are from the 1D gaussian fitting and are of the order of $~0.1$ km\,s$^{-1}$. These uncertainties are relatively low because the high SNR of the \Ht lines results in very precise line parameter fitting (see our related comments in Section \ref{sec:obs_jwst_maps}). The \Ht velocity shift values are relatively small compared to bona-fide Class 0 protostellar jet systems, whose velocities easily exceed 50 km s$^{-1}$ (\eg \citealt{Lefloch2015}). As mentioned in Section \ref{sec:mol_em}, this suggests that the outflowing gas seen in \Ht emission lies close to the POS. 
This hypothesis was also proposed by \citet{Podio2021} in their analysis of CO, SO, and SiO millimeter emission lines of this protostellar outflow.
Finally, the two \Ht line flux ratios suggest that the \Ht excitation conditions evolve along the jet, with an increase of the both $\rm H_{2}\,1-0/2-1\,S(1)$ and $\rm H_{2}\,2-1/3-2\,Q(1)$ beyond 100 AU from the jet base.

\subsubsection{Shock modeling of the \Ht emission}
\label{sec:mol_em_jet_south_H2shocks}

Because the energy levels of \Ht are widely spaced, \Ht predominantly traces warm gas $\geq$ 100 K, making its emission sensitive to shocks and irradiated regions like PDRs (\eg see \citealt{Kaplan2017,Kaplan2021}). Both \Ht excitation mechanisms can be at play in protostellar outflows. Outflow cavities of Class 0 protostars are dense enough that collisional excitation will play a role \citep{Gusdorf2008a,Gusdorf2008b}. However, UV excitation could contribute to setting the \Ht ro-vibrational level populations in such regions in two cases: (1) if the accretion luminosity emanating from the accretion shock at the stellar surface produces enough UV to propagate in the base of outflow cavities before getting attenuated by the ambient medium (\eg see \citealt{vanKempen2009,Visser2012,Yildiz2012,Benz2016}), and (2) if the shocks occurring in the cavities are fast enough, and with a small enough transverse magnetic field strength, to become self-irradiated \citep{Lehmann2020,Lehmann2022}. Because of the clear gradient in local conditions (ionization temperature, density, etc.) across the shock front, multiple components are responsible for setting the total \Ht excitation diagram integrated across the shock. The hotter gas ultimately dominates the excitation of states at the highest energies, while the cooler \Ht molecules dominate the excitation of states at the lowest energies. We now attempt to reproduce the \Ht excitation observed along the jet of the SC source by using shock models that self-consistently compute the \Ht excitation mechanisms.

\paragraph{The shock model grid} We explore the recent shock model grid by \citet{Kristensen2023}, that uses the Paris-Durham shock code over an extensive grid of $\sim$\,14 000 plane-parallel stationary shock models. The Paris-Durham
shock code (\citealt{Flower2003,Lesaffre2013,Godard2019}; and references therein) is a public numerical tool\footnote{Available on the ISM platform \href{https://ism.obspm.fr}{https://ism.obspm.fr}.} that computes the dynamical, thermal, and chemical evolution of interstellar matter in a steady-state plane (1D) parallel shock wave. 

Before computing each model, equilibrium pre-shock initial conditions are determined via chemical steady-state calculations, computed with the given density and radiation field. If the shock is irradiated, a photon-dominated region (PDR) calculation is also performed (see \citealt{Godard2019}). Depending on the initial conditions, the code either implements a Jump (J-, or CJ-type; discontinuous) solution or a Continuous (C-, or C$^\star$-type; continuous) solution. Depending on the transverse magnetic field strength and ratio of the neutral mass density to the charged fluid mass density, the shock velocity is below the magnetosonic speed of the charged fluid, in which case the charged fluid decouples from the neutral fluid. This stream of charged particles preheats and decelerates the gas ahead via collisions in a magnetic precursor. C-, C$^\star$-,and CJ-type shocks are thus solved with a multifluid code. This makes the interstellar shock wave sensitive to the ambient magnetic field strength. In turn, interstellar shocks can be used to constrain the pre-shock transverse magnetic field strength (\eg see \citealt{Gustafsson2010}). Otherwise, if the shock velocity exceeds the ion magnetosonic speed, J-type shocks occur and generate higher temperatures and gradients.

The code self-consistently includes a 1D treatment of the UV continuum radiative transfer, heating/cooling charge exchange mechanisms from grains and PAHs physics, photodestruction processes of atoms and molecules, a chemical network \citep{Flower2015}, radiative pumping of \Ht level populations, chemical cooling via the excitation of the fine structure lines and metastable lines of atomic species and ro-vibrational lines of molecules (\eg see \citealt{Neufeld1993,LeBourlot2002}). For each model, the molecular abundances of the most abundant species, temperature and density parameters, shock width, \Ht integrated intensities and $o/p$ ratio are extracted.

The \citet{Kristensen2023} grid samples six input parameters, namely the preshock density \nH (10$^2$---10$^8$ cm$^{-3}$), transverse (to the shock propagation direction) magnetic field strength \b (0.1---10, such that $B = b\,\times\,\sqrt{\textrm{n}_{\small{\textrm{H}}} (\textrm{cm}^{-3})}\,\mu\textrm{G}$), shock velocity \vs (2---90 km s$^{-1}$, depending on \b), semi-isotropic external UV radiation field strength \Go (0---10$^{3}$, in units of the \citet{Mathis1983} field), the \Ht cosmic-ray ionization rate \CIR (10$^{-17}$---10$^{-15}$ s$^{-1}$), and the polycyclic aromatic hydrocarbons (PAH) abundance fraction \XPAH (10$^{-8}$---10$^{-6}$). The grid sampling step size of the \nH, \Go, \CIR and \XPAH parameters are factors of ten. The \b and \vs parameters are not uniformly sampled because they (with also \Go and \nH) ultimately set the models implementation routine \ie (C-, C$^\star$-, CJ-, or J-type; see the Figure 6 by \citealt{Kristensen2023}). This grid does not cover the parameter space corresponding to strongly dissociative self-irradiated J-type shocks, mostly occurring for J-type shocks with \vs\,$\geq\,$30 km s$^{-1}$ and \b\,$\geq\,$1. Also, the magnetic field orientation is assumed to be perpendicular to the direction of shock propagation (no oblique shocks), which is purely 1D (\ie with no implementation of shock sideways expansion). Shocks of this grid are also assumed to be stationary (for reference, see \citealt{Lesaffre2004,Gusdorf2008b}). Finally, ice mantles and grain-grain interaction physics are not included in this grid.

\paragraph{Shock parameters exploration} For each spatial location along the SC source molecular jet, the H$_2$ extracted fluxes are compared to the shock model grid. The results of these comparisons are shown in Figure \ref{fig:H2_shock_modeling}, where the most likely (the likelihood functions, and their maximum and mean values are shown) parameters of the shock grid are plotted as a function of the distance from the jet base (details on our likelihood function computation are presented in appendix \ref{app:shock_models}). 

In all the locations along the jet, the best model is a fast, strongly magnetized, C-type shock, with high pre-shock density (see as example, the model best matching the \Ht excitation diagram of the first location along the jet in Figure \ref{fig:H2_diagram_comp_model} of the Appendix \ref{app:shock_models}). J- and CJ-type shock models significantly differ from observed H$_2$ level population. Notably, in J- and CJ-type shock models the shape of the excitation diagram is more curved for $E_{\textrm{u}}\,\leq\,5000$\,K, and appears as
a single temperature component after 10,000 K compared to our data. The probability distribution functions shown in Figure \ref{fig:H2_shock_modeling} clearly indicate that strong constraints (relative to the corresponding range of values explored in the shock model grid for each parameter) are obtained for \nH, \b, and \vs. The highest probable pre-shock density values range between 3$\times10^{7}$ and $10^{8}$ cm$^{-3}$, the shock magnetization \b between 3 and 10, and the shock velocity between 20 and 70 km s$^{-1}$. Such high pre-shock densities are consistent with dense outflow cavity walls being compressed by the emerging outflow shocks. These density values are also consistent with the bright sub-millimeter dust emission toward the outflow cavity walls seen in the ALMA data surrounding the inner jet/outflow system (see discussion sections of \citealt{LeGouellec2019a,Hull2020a}), infalling envelope models of young protostars (\eg see \citealt{Kristensen2012, Visser2012}), and numerical simulations of protostellar collapse presented by \citealt{LeGouellec2023b,Basu2024}. This suggests that the jet is still nestled within the dense and young protostellar envelope. 
The high shock velocity also confirms the jet nature of this young protostellar system. 

This shock model comparison also suggests that the inner regions of this Class 0 molecular jet are significantly magnetized, with $B\,\sim\,10$---100 mG.
The external UV irradiation field is less well constrained, especially toward the base of the jet, but the results are not consistent with a total absence of UV irradiation. However, at $\geq\,300$ AU from the jet base, the shocks are consistent with a significant UV irradiation of \Go\,$\geq\,10$. This latter result is surprising, given that the main source of UV photons is likely to be located toward the inner regions where the accretion luminosity escapes. A possibility is that shocks located further away from the jet base are self-irradiated and produce their own UV field (the centroid of \Ht and CO lines are bluer further away from the jet base in Figures and \ref{fig:line_H2_CO_vel_shift} and \ref{fig:H2_param_jet}, which can suggest faster shocks). However, we note that our capability of constraining the UV field strength is most certainly affected by the fact that the shock model grid we use does not implement the shocks' self-irradiation (see \citealt{Lehmann2020,Lehmann2022}). The last possibility of an external UV irradiation field coming from outside the S68N core and outflow is less likely for two reasons: we did not identify a nearby star capable of
generating such a UV field, and the actual extinction from the surrounding Serpens Main cloud (\ie beyond the S68N core) is significant and would make the penetration of external UV photons difficult, especially to produce \Go\,$\geq\,10$.
The shock modeling has a low dependence on the \CIR and \XPAH parameters (especially since the mean and maximum of the probability distribution functions do not coincide). The low dependence on \CIR is expected for UV irradiated shocks (see appendix C of \citealt{Kristensen2023}). The grid exploration of these two parameters are presented in Appendix \ref{app:shock_models}.

Our H$_2$ observations are thus consistent with energetic C-type shocks. J-type shocks would actually populate more intensively the higher ro-vibrational levels, resulting in different H$_2$ excitation diagrams than those obtained via our data. C-type shocks deposit energy over a larger distance than J-type shocks. We have retrieved additional parameters of these shock models that best match our observations, and found they are consistent with our observations. Namely, the shock width (measured as the extent of the region, across the shock propagation direction, where 80\% of the emission is radiated away) of the best shock models varies within $\sim$1---200 au for the brightest H$_2$ lines along the jet, while the radius of our apertures for spectral extraction is 0.4$^{\prime\prime}$, \ie $\sim$ 180 au at the distance of Serpens Main. We note that the width of the molecular jet (perpendicular to the direction of propagation) is $\sim$200---250 au. The $o/p$ ratio integrated throughout the shock of the best shock models varies within $\sim$2---2.5 among the ro-vibrational level populations, which is consistent with the $o/p$ ratio we extract from the same ro-vibrational levels. 
We note that our estimates of H$_2$ column densities along the jet (see Figure \ref{fig:H2_param_jet}) overestimate the column density of the best models by a factor $\sim$2. This mismatch can be explained by our aperture size (of $\sim$ 180 AU in radius) encompassing various shock working surfaces.


Finally, we note that the observed small velocity shifts of the \Ht lines (\ie $-$15 --- $-$25 km s$^{-1}$) relative to the high shock velocities we constrained suggests that the direction of the fast molecular jet is close to the POS.
Indeed, we compute an inclination of 64$^\circ$ (with respect to the line-of-sight) using the average ratio of the velocity shift of \Ht lines (corrected for the $v_{\textrm{lsr}}$, see below) and the shock speed of the best models along the jet.
\citet{CarattioGaratti2024} recently showed with a MIRI mosaic of the HH211 protostellar outflow that the hot \Ht gas indeed originates from the molecular phase of the fast moving jet.
Also, as the low-$J$ CO line emission whose velocity is close to the $v_{\textrm{lsr}}$ is not recoverable in ALMA data due to the embedded nature of protostars in their filtered CO-rich dense molecular cloud, a low inclination can explain the faint ALMA CO (2$\rightarrow$1) emission toward the base of outflow cavity, where the bright H$_2$ molecular emission is observed. The apparent low-velocity CO (2$\rightarrow$1) emission is thus located toward the radially expanding outflow cavity, further away from the central jet. 

\begin{figure}[!tbh]
\centering
\includegraphics[scale=0.35,clip,trim= 0cm 0.0cm 0cm 0cm]{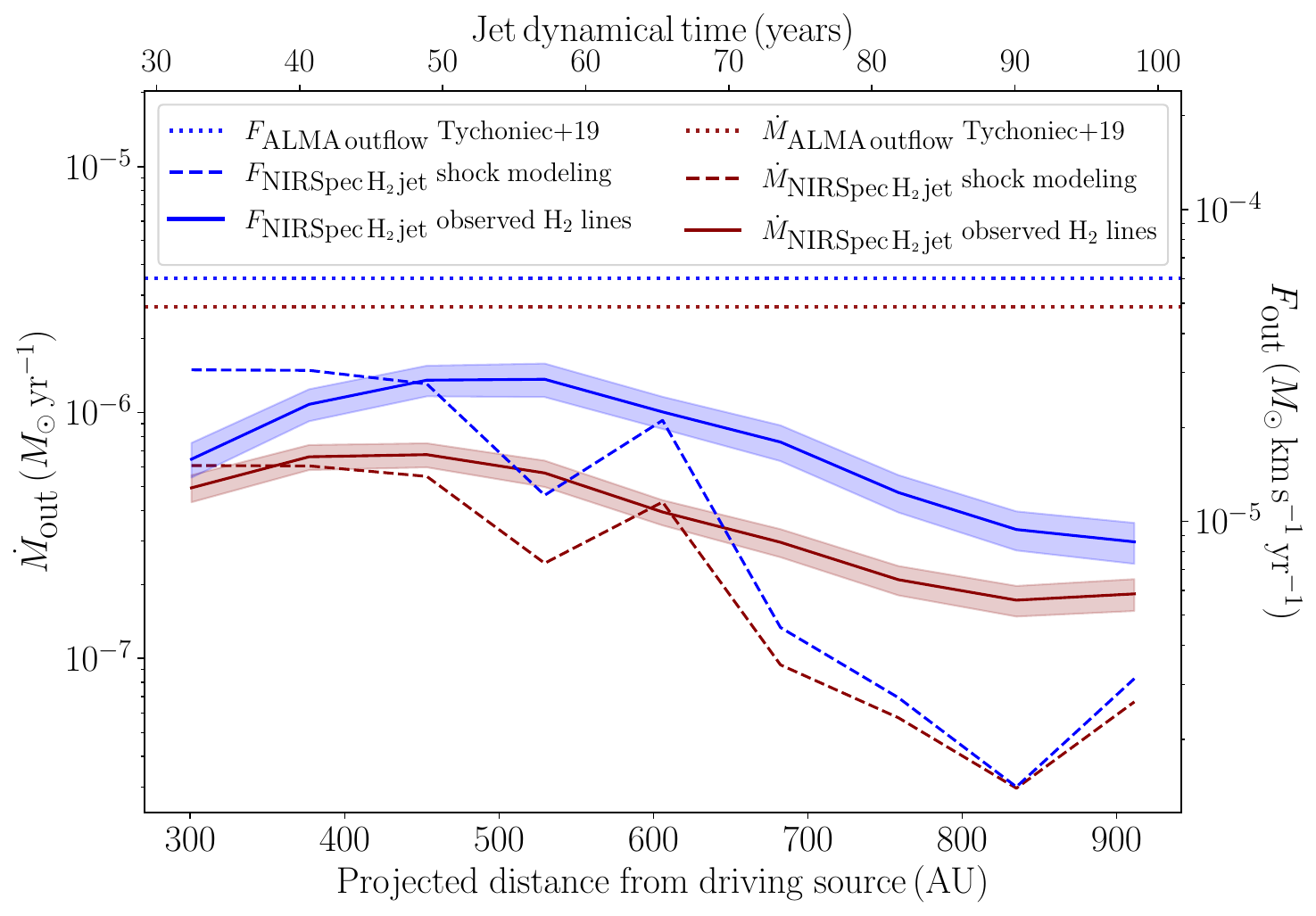}

\vspace{-0.2cm}
\caption{\small Evolution of the mass ejection rate $\dot{M}_{\textrm{out}}$ (blue lines) and force $F_{\textrm{out}}$ (red lines) of the blueshifted molecular jet/outflow of the SC source as a function of distance from the driving protostar. 
$\dot{M}_{\textrm{out}}$ and $F_{\textrm{out}}$ obtained from the observed \Ht lines are shown in solid lines (and the associated uncertainties are shown with the shaded areas).
$\dot{M}_{\textrm{out}}$ and $F_{\textrm{out}}$ obtained from the best shock models along the jet are shown in dashed lines. 
For reference, we show with dotted horizontal lines the values of the outflow mass ejection rate and force computed by \citet{Tychoniec2019} from the ALMA \cotwotoone data for the entire blueshifted outflow lobe (integrated between -2 and -22 km s$^{-1}$, after $v_{\textrm{lsr}}$ correction), extending up to $\sim$ 6000 AU.
The dynamical time, computed from the ratio of the projected distance from the driving source over the average best models' shock speed along the jet (\ie 40 km s$^{-1}$), is shown at the top.
}
\label{fig:outflow_force_rate}
\end{figure}

\paragraph{Energetics of the molecular jet} We compute the mass-loss rate from the driving protostar of the SC outflow as follows:
\begin{equation}
    \dot{M}_{\textrm{out}}\,=\,\mu_{\textrm{H}_{2}} \times \textrm{m}_{\textrm{H}} \times N_{\textrm{H}_{\textrm{2}}} \times A / t_{\textrm{dyn}}\,\,,
\end{equation}
where $N_{\textrm{H}_{\textrm{2}}}$ is the column density of \Ht at a given location along the jet, $\mu_{\textrm{H}_{2}}$ is the mean molecular weight (we use a value of 2.8; \citealt{Kauffmann2008}), $A$ is the pixel area, $t_{\textrm{dyn}}$ is the dynamical time associated with a given location along the jet, and $\textrm{m}_{\textrm{H}}$ is the mass of the hydrogen atom. The dynamical time is computed with $\delta R/v_{\textrm{jet}}$, where $\delta R$ is the aperture diameter corrected for inclination, and $v_{\textrm{jet}}$ is the speed of the \Ht jet material. Then, the outflow force $F_{\textrm{out}}$ is obtained multiplying $\dot{M}_{\textrm{out}}$ by $v_{\textrm{jet}}$. We extract $\dot{M}_{\textrm{out}}$ and $F_{\textrm{out}}$ directly from the observations, using the \Ht column density of Figure \ref{fig:H2_param_jet} and $|\delta_v|$ (where $\delta_v$, shown in Figure \ref{fig:H2_param_jet}, is the mean velocity shift of \Ht lines \textrm{corrected for $v_{\textrm{lsr}}$}) for $v_{\textrm{jet}}$ after correction due to the inclination. We determine the inclination from the mean of the $|\delta_v|/$\vs ratio along the jet, and obtain a value of $64^{\circ}$ (with respect to the line-of-sight), which yields a mean and standard deviation for $v_{\textrm{jet}}$ of $49 \pm 10$ km s $^{-1}$.
We also extract $\dot{M}_{\textrm{out}}$ and $F_{\textrm{out}}$ from the best shock models along the jet, using at each location the integrated \Ht column density across the shock, and the shock speed \vs for jet velocity $v_{\textrm{jet}}$. 
These quantities are shown in Figure \ref{fig:outflow_force_rate}, as a function of the distance from the driving protostar (we use the sub-millimeter continuum peak). The modeled mass ejection rate and outflow force agree within a factor of 2 with the observations up to 600 AU from the source ($\sim 10^{-7}$---$10^{-6} M_\odot$ yr$^{-1}$ and $\sim 5 \times 10^{-6}$---$2 \times 10^{-5}$ for $\dot{M}_{\textrm{out}}$ and $F_{\textrm{out}}$, respectively), after which the model under-predicts them by a factor 5---10. 

These quantities tend to decrease with distance form the driving source, which also suggests a recent increase in the mass accretion rate. Such decrease happening on the scale of $\sim$ 1000 AU is consistent with recent numerical simulations implementing protostellar outflows (\eg see \citealt{Verliat2022,Lebreuilly2024b}).
The warm \Ht component traced by the pure-rotational lines covered by MIRI would yield higher \Ht column densities, but would also correspond to slower gas though (\eg, see \citealt{CarattioGaratti2024}), than the hot \Ht component traced by the lines covered by NIRSpec. This likely causes our estimation of $\dot{M}_{\textrm{out}}$ and $F_{\textrm{out}}$ to be underestimates.

We also compare these values with the outflow mass ejection rate and force computed by \citet{Tychoniec2019} using ALMA \cotwotoone data summed over the entire blueshifted outflow lobe extending out to $\sim$ 6000 AU. Interestingly, the inner regions of the SC molecular jet observed with \Ht lines in the NIRSpec data is lower by a factor of 3---10 compared to the mass ejection rate and force of the ALMA cold molecular outflow. 
This can suggest that the mass ejection rate and outflow force were higher in the past. The NIRCAM 480M data reveal bright shocked structures $\sim$ 5000 AU from the SC source.
Considering that for this comparison the entire blueshifted outflow lobe was integrated in the ALMA data, and that we map the \Ht jet only from 300 to 900 AU with NIRSpec, we posit that integrating the \Ht jet in its entire extent (up to 5000 AU from the central source in NIRCAM data of Figure \ref{fig:RGB_alma_image}) would probably yield a comparable energy budget compared to the one of the entire blueshifted outflow, obtained with ALMA observations of the cold/warm CO gas.


\subsubsection{CO molecular wind}
\label{sec:mol_em_jet_south_COwind}

\begin{figure*}[!tbh]
\centering
\includegraphics[scale=0.53,clip,trim= 0.3cm 0.0cm 0cm 0cm]{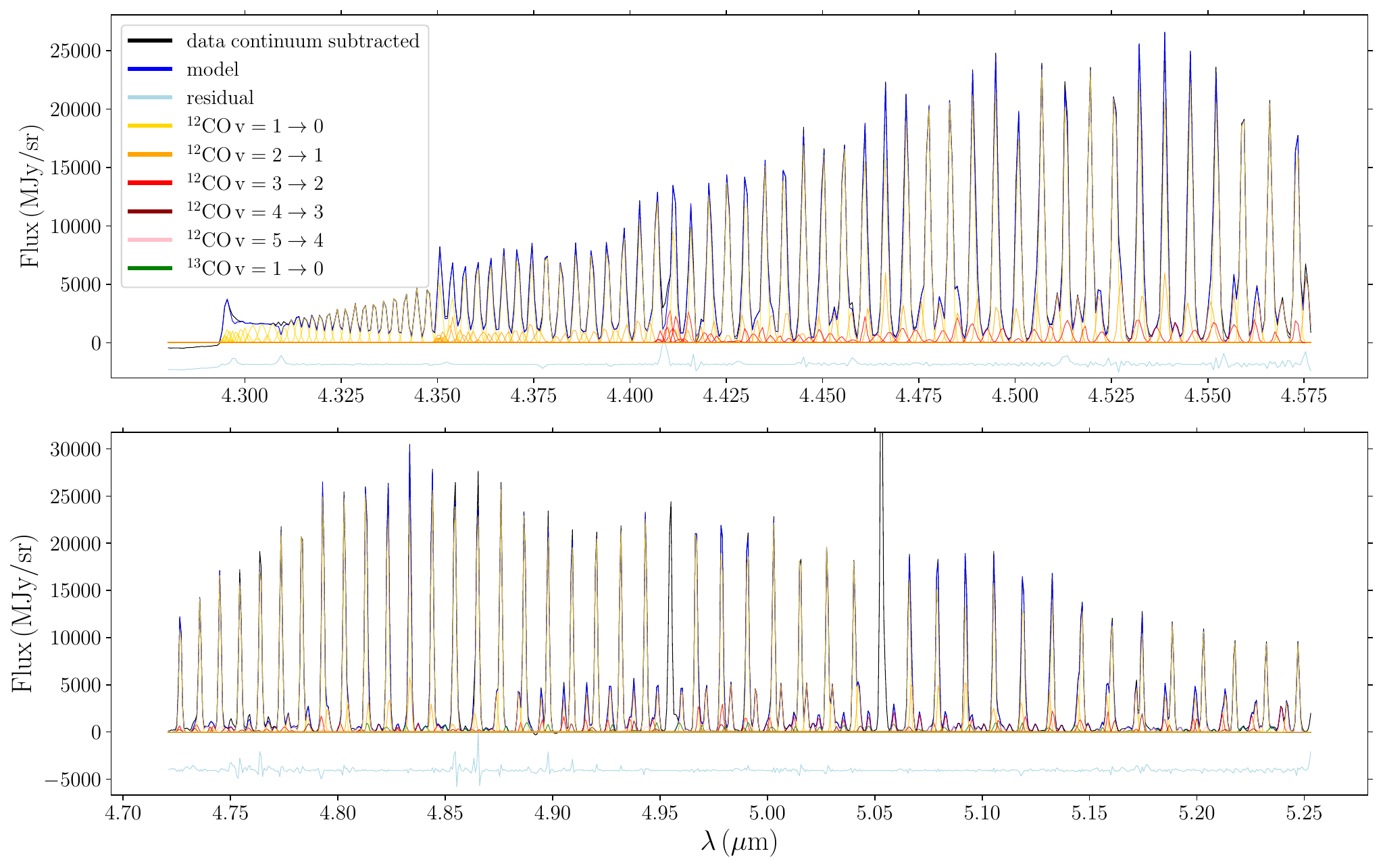}

\vspace{-0.2cm}
\caption{\small CO fundamental line forest of the SC source. The black line is the observed continuum-subtracted spectrum, the blue is the sum of Gaussian fits to all detected CO emission lines, and the light blue line is the residual (offset by the residual’s max in each window). All single-line fits are also shown, color coded by rotational ladder (\ie yellow for $^{12}$CO $v = 1-0$, orange for $^{12}$CO $v = 2-1$, red for $^{12}$CO $v = 3-2$, dark red for $^{12}$CO $v = 4-3$, pink for $^{12}$CO $v = 5-4$, and green for $^{13}$CO $v = 1-0$). We do not attempt to fit the spectra around the very bright \Ht lines \Ht\,0-0\,S(9), \Ht\,0-0\,S(8), and \Ht\,1-1\,S(9), and the saturated $^{12}$CO ice absorption band (between 4.58 and 4.72 $\mu$m).}
\label{fig:CO_spectra}
\end{figure*}

\begin{figure}[!tbh]
\centering
\includegraphics[scale=0.49,clip,trim= 0.2cm 0.0cm 0cm 0cm]{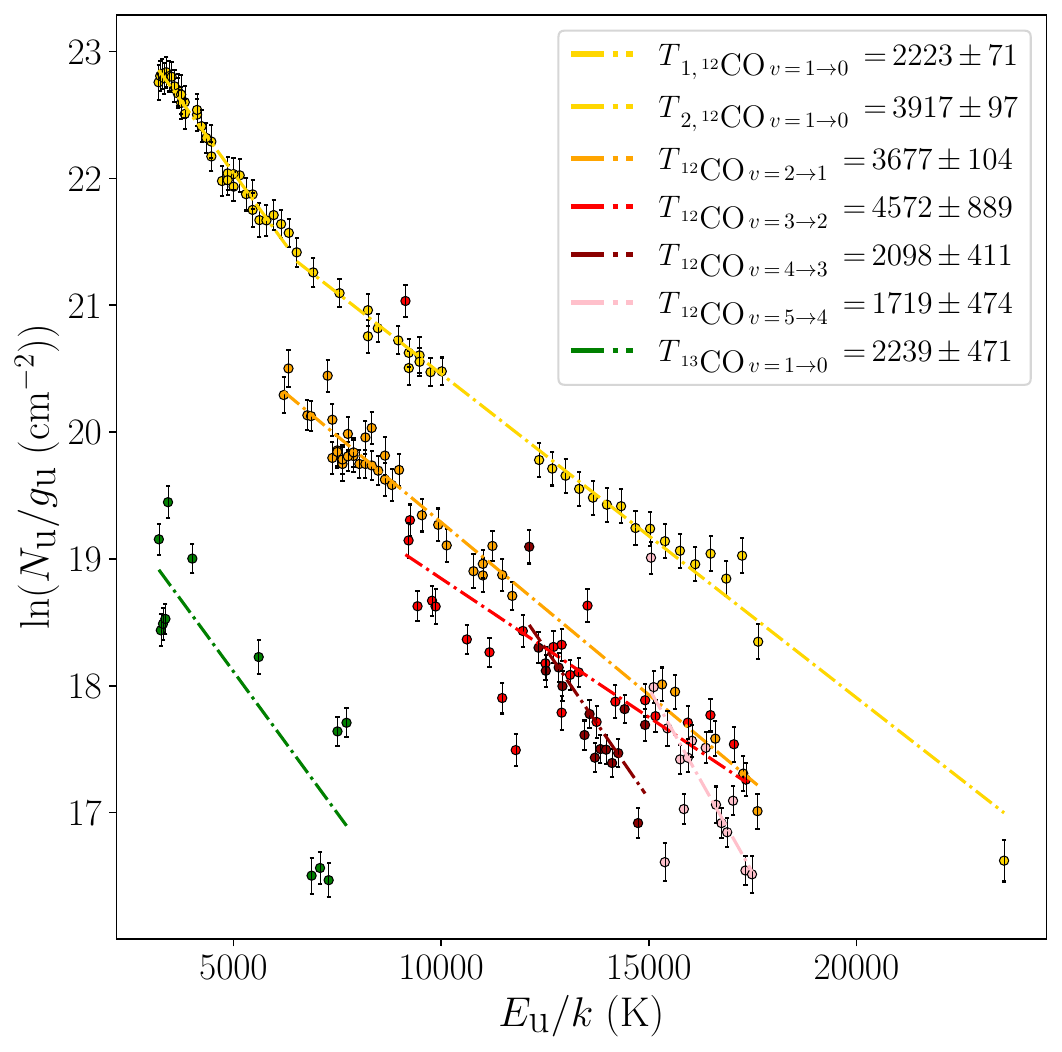}

\vspace{-0.2cm}
\caption{\small CO population diagrams built from the extraction of extinction corrected line fluxes from the CO fundamental line forest. The color code is the same as in Figure \ref{fig:CO_spectra}. The $^{12}$CO $v = 1-0$  rotational ladder is fit with two linear components, while the other rotational ladders are fit with a single linear regression. The corresponding temperatures are listed in the legend.}
\label{fig:CO_excitation_diagram}
\end{figure}

\begin{figure}[!tbh]
\centering
\includegraphics[scale=0.38,clip,trim= 0cm 0.0cm 0cm 0cm]{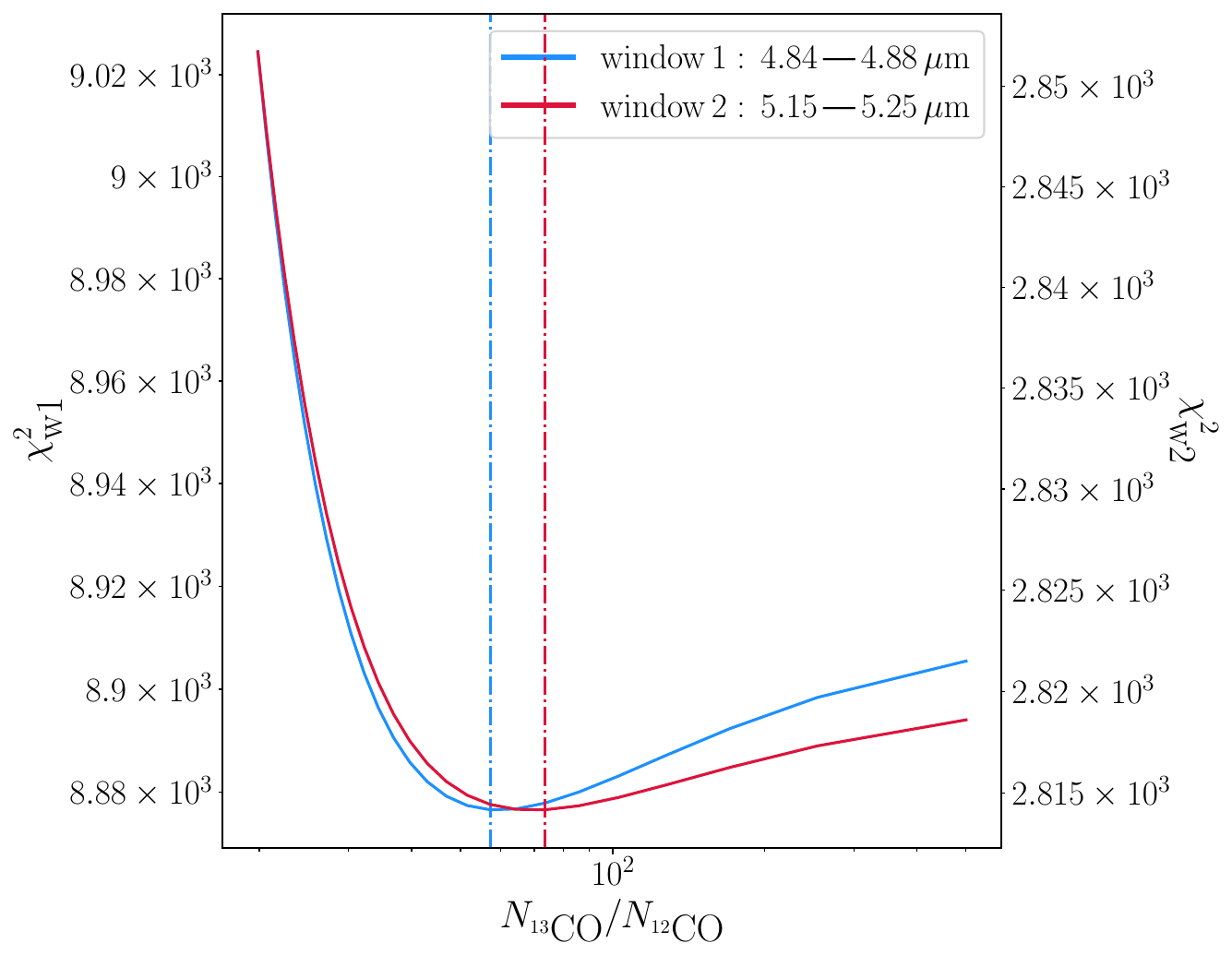}

\vspace{-0.2cm}
\caption{\small Reduced $\chi^2$ between the LTE model of Figure \ref{fig:CO_data_model_spec} and the observed spectra as a function of $N_{^{13}\textrm{CO}}/N_{^{12}\textrm{CO}}$, for two spectral windows: window \#1 4.84---4.88 $\mu$m (blue), and window \#2 5.15---5.25 $\mu$m (red). These spectral regions were chosen to only include regions where several $^{13}$CO lines are not blended with the $^{12}$CO lines. The most probable $N_{^{12}\textrm{CO}}/N_{^{13}\textrm{CO}}$ values are 57 and 73 for the 4.84---4.88 $\mu$m and 5.15---5.25 $\mu$m spectral regions, respectively.}
\label{fig:13CO_12CO}
\end{figure}

The SC source spectra exhibit a bright CO fundamental (\ie ro-vibrational lines with $\Delta\,v\,=\,$1-0) line forest ranging from 4.3 $\mu$m until the end of the G395H spectral coverage (these CO lines are also visible in the first $JWST$/MIRI MRS channel up to $\sim$ 6 $\mu$m, priv. communication with the JOYS MIRI GTO team). We cannot strictly attribute the spatial origin of CO emission. The shock conditions we constrained from \Ht must be responsible for one or more components of the CO emission. However, the velocity structure seen in Figure \ref{fig:line_H2_CO_vel_shift} shows slight differences between the H$_2$ ($v\,=\,1\rightarrow0$) and CO ($v\,=\,1\rightarrow0$) lines. The 3--5 $\mu$m continuum emission located in the blueshifted cavity is spatially shifted from the ALMA dust continuum peak, suggesting that the molecular emission may originate from the disk or a compact wind and be scattered into our line of sight by the outflow cavity.

We subtract the \HtonetooneSten, \HtonetooneSeleven, \HtOtoOSten line fluxes interpolating the H$_2$ excitation diagram of the corresponding rotational ladder (see Section \ref{sec:mol_em_H2_excitation}). However, we do not attempt to fit the spectra around the bright  \HtOtoOSnine, \HtOtoOSeight, and \HtonetooneSnine lines and the saturated $^{12}$CO ice absorption band (between 4.58 and 4.72 $\mu$m). The corresponding spectral regions are flagged and not used in this CO analysis. We employ the continuum subtraction method outlined in Section \ref{sec:obs_jwst_cont} from 3.6 $\mu$m (where the continuum of the SC source is clearly detected, see Figure \ref{fig:spec_2_sources}) to 5.3 $\mu$m. We note that we cannot differentiate the potential pseudo continuum created by the bright CO fundamental line forest from the actual line-free IR continuum given that the CO lines are still bright by the long-wavelength end of the G395H extracted spectra. The continuum detected between 3.6 and 4.2 $\mu$m offers a robust baseline on which to base the longer-wavelength continuum. The continuum level is shown in Figure \ref{fig:CO_cont} of Appendix \ref{app:CO_models}.

In this Section, we treat this SC source spectrum as having CO only in emission. The difficulty of determining a continuum level and the low continuum flux preclude us from determining if CO absorption is detected, which has been seen in several Classical T-Tauri star (CTTS) disks, and winds (\eg see \citealt{Herczeg2011,Pontoppidan2011,Federman2024,Rubinstein2024}). We use the HITEMP CO line list from the HITRAN database \citep{Rothman2010,Gordon2022} to retrieve the line transition data tables of the first five rotational ladders of $^{12}$CO, and the first one of $^{13}$CO. The CO lines are all fitted with multiple Gaussian profiles, and separated in successive groups of lines (typically $\sim$\,30 lines per group) within a given spectral region. While we fit the FWHM and peak intensity of each line, the same velocity shift is used for all lines (\ie $-$6.94 km s$^{-1}$, measured as the velocity shift that minimizes the mean offset of CO lines fitted beforehand; not considering the $v_{\textrm{lsr}}$ correction). This allows us to fit groups of blended lines together. 
The sum of all fitted lines are shown in Figure \ref{fig:CO_spectra}, alongside the observed spectra and the residuals. The $^{12}$CO $v = 1-0$ bandhead, just longward of the $^{12}$CO$_2$ ice absorption band, is detected, as well as the $^{12}$CO $v = 2-1$  and $^{12}$CO $v = 3-2$  bandheads. These corresponding highly energetic upper states suggest that a component of hot ($\geq\,3000$ \,K) gas contributes to the emission, or that UV photoexcitation is at play \citep{Krotkov1980,vanDishoeck1988,Brittain2007}. We note that we do not attempt to fit the $^{12}$CO $v = 4-3$, $^{12}$CO $v = 5-4$, and $^{13}$CO $v = 1-0$ lines shortward of the $^{12}$CO ice absorption band, given their corresponding R-branch lines are very faint and blended, such that no clear detections are expected in this wavelength range. The increase in spacing between P-branch lines with decreasing $J$ allows us to isolate these lines in the line forest longward of 4.7 $\mu$m. Several faint $^{13}$CO $v = 1-0$ lines are detected within the 4.84---4.87 $\mu$m and 5.15---5.25 $\mu$m spectral regions.


\begin{table*}
\small
\caption[]{CO fundamental excitation parameters toward the SC source}
\label{t.CO_excitation_param}
\setlength{\tabcolsep}{0.4em} 
\begin{tabular}{p{0.25\linewidth}ccccc}
\hline \hline \noalign{\smallskip}
Rotational ladder & $N_{\textrm{tot}}$$^{a}$ & $\mathcal{N}_{\textrm{tot}}$$^{b}$ & $T_{\textrm{rot}}$$^{c}$  & $T_{\textrm{vib}}$$^{d}$ \\
~ & cm$^{-2}$ & & K & K \\
\noalign{\smallskip}
\hline
\noalign{\smallskip}
1$^{\textrm{st}}$ component $^{12}$CO\,$v=1\rightarrow 0$ & $3.8\pm0.2\times10^{13}$ & $1.7\pm0.1\times10^{43}$ & $2223\pm71$ & \multirow{3}{*}{$^{12}$CO\,($v=1-2$) \; $1426\pm14$} \\
\noalign{\smallskip}
2$^{\textrm{nd}}$ component $^{12}$CO\,$v=1\rightarrow 0$ & $2.6\pm0.2\times10^{13}$  & $1.2\pm0.1\times10^{43}$ & $3917\pm97$ & \\
\noalign{\smallskip}
$^{12}$CO\,$v=2\rightarrow 1$ & $8.1\pm0.6\times10^{12}$  & $3.8\pm0.3\times10^{42}$ & $3677\pm104$ & \multirow{2}{*}{$^{12}$CO\,($v=2-3$) \; $2234\pm59$}\\
\noalign{\smallskip}
$^{12}$CO\,$v=3\rightarrow 2$ & $4.3\pm1.8\times10^{12}$  & $2.1\pm1.2\times10^{42}$ & $4572\pm889$ & \multirow{2}{*}{$^{12}$CO\,($v=3-4$) \; $4822\pm495$}\\
\noalign{\smallskip}
$^{12}$CO\,$v=4\rightarrow 3$ & $4.0\pm7.0\times10^{13}$  & $1.5\pm1.9\times10^{43}$ & $2098\pm411$ & \multirow{2}{*}{$^{12}$CO\,($v=4-5$) \; $4228\pm416$}\\
\noalign{\smallskip}
$^{12}$CO\,$v=5\rightarrow 4$ & $2.0\pm4.0\times10^{15}$  & $1\pm3\times10^{44}$ & $1719\pm474$ & \\
\noalign{\smallskip}
$^{13}$CO\,$v=1\rightarrow 0$ & $1.6\pm0.8\times10^{12}$  & $7\pm4\times10^{41}$ & $2239\pm471$ &  \\
\noalign{\smallskip}
\hline
\end{tabular}
\tablecomments{\small 
$^{a}$ Column densities of CO molecules assuming optically thin emission, obtained from the fit of the given rotational ladder (component) in the CO population diagram (see Figure \ref{fig:CO_excitation_diagram}).
$^{b}$ Total number of CO molecules derived from the column densities.
$^{c}$ Rotational temperatures obtained from the fit of the given rotational ladder (component) in the CO population diagram.
$^{d}$ Vibrational temperatures between each pair of neighboring vibrationally excited states, 
The determination of the $^{12}$CO\,$v=4\rightarrow 3$ and $^{12}$CO\,$v=5\rightarrow 4$ column densities and rotational temperatures suffer from the small number of detected lines and are thus highly uncertain.}
\end{table*}

\paragraph{CO line excitation} Among all lines fitted in Figure \ref{fig:CO_spectra}, we first select those that meet our S/N criterion of 10. Then, we select among those the lines whose closest neighboring detected line lies further than 1/2 spectral resolution element (assuming the theoretical NIRSpec/IFU spectral resolution, see \citealt{Jakobsen2022}). Similarly to the H$_2$ lines, we can derive the extinction of the CO emitting gas using the pairs of P and R transitions sharing the same upper state ro-vibrational level ($v_{\textrm{u}},\, J_{\textrm{u}}$), for which the intrinsic flux ratio is predicted. Using the \citet{Pontoppidan2024a} extinction law, we find mean and standard deviation values for $A_K$ of 7.6$\pm$1.0 mag for the $^{12}$CO $v = 1-0$ ladder and 8.0$\pm$0.7 mag for the $^{12}$CO $v = 2-1$ ladder (propagating the line flux uncertainties with the eleven, and nine pairs of lines for each ladder, respectively). 
These extinction values are surprisingly consistent with those obtained with the \Ht line ratio in Section \ref{sec:mol_em_H2_extinction}.
However, infrared radiation can contribute to the excitation of CO ro-vibrational lines and generate a P- versus R-branch asymmetry, with absorption against the infrared continuum and/or self-absorption affecting more the R- compared to the P-branch \citep{GonzalezAlfonso2002,Lacy2013}. 
The lack of P- versus R-branch asymmetry observed here and the consistency with the extinction computed with \Ht suggests that collisions govern the CO excitation.
Our observations lack the spectral resolution to investigate this effect, and we thus use the extinction computed in Section \ref{sec:mol_em_H2_extinction} to de-redden the CO fundamental lines (see \citealt{Evans1991,Rettig2005,Barentine2012,GonzalezAlfonso2024,Buiten2024,GarciaBernete2024,PereiraSantaella2024}, for such R-branch absorption seen in the CO ro-vibrational lines).
We placed the $^{12}$CO and $^{13}$CO extinction-corrected line fluxes into a CO population diagram in Figure \ref{fig:CO_excitation_diagram} using the HITRAN line transition parameters \citep{Rothman2010,Gordon2022}. We fit the $^{12}$CO $v = 1-0$ level population with two temperature components (for lines with $E_{\textrm{u}}/k$ smaller and larger than 6500 K). The rotational temperatures and associated column densities we obtain are listed in Table \ref{t.CO_excitation_param}.
Given the errors, the fit on the $^{12}$CO $v = 3-2$ gives consistent results with the $^{12}$CO $v = 2-1$ component. However, the $^{12}$CO $v = 4-3$ and $^{12}$CO $v = 5-4$ rotational ladders lack sufficient line detections to accurately determine their rotational temperatures. We note that the temperature $T_{^{13}\textrm{CO}\,v=1-0}$ is consistent with $T_{1,\,^{12}\textrm{CO}\,v=1-0}$.

The $v_{\textrm{u}}\,\geq\,2$ components  appear less abundant than the $^{12}$CO $v = 1-0$ level population in the excitation diagram, which argues toward non-LTE excitation effects, \ie not only collisional excitation in shocks, but also potentially radiatively excited CO gas by UV pumping near the protostar or produced locally by self-irradiated shocks, or IR pumping. To investigate this point, we follow the procedure by \citet{Rubinstein2024} (we refer to their Section 3.4) to compute the vibrational temperatures for the successive neighboring vibrationally excited states of $^{12}$CO (\ie $v = 1$ and $v = 2$, $v = 2$ and $v = 3$, $v = 3$ and $v = 4$, and $v = 4$ and $v = 5$). The vibrational temperature values for each pair of vibrational levels are listed in Table \ref{t.CO_excitation_param}. The $T_{\textrm{vib},\,v=1-2}$ value we obtain for the SC source of S68N is higher than that found for the protostars studied by \citet{Rubinstein2024}.
Having $T_{\textrm{vib},\,v=1-2}$ lower than $T_{\textrm{rot,1},\,v=1\rightarrow0}$ suggests that
the vibrational levels are not in equilibrium with the rotational levels, meaning
this ro-vibrational emission is not in LTE. This could set a constraint on the gas volume density and thus the origin of the hot CO gas responsible for this emission. However, we would also need to know the kinetic temperature of the gas to constrain the local CO critical density, and determine if optical thickness effects are affecting the temperatures extracted from the CO excitation analysis.
Additionally, the higher vibrational temperatures (\ie $\geq\,2000$ K) we obtain for the higher vibrational levels suggest that IR and/or UV pumping  contribute to the excitation of the CO molecules \citep{Krotkov1980,Scoville1980}, which is commonly seen toward protoplanetary disks exposed to UV irradiation stars \citep{Brittain2007,Bast2011,Brown2013,vanderPlas2015}.
The lack of P/R asymmetry found above comparing the \Ht- with the CO-derived extinction can suggest that radiative pumping is not the dominant mechanism.

\paragraph{Shock Excitation} The cavity shock properties constrained (see Figure \ref{fig:H2_shock_modeling}; \vs $\geq$ 30 km s$^{-1}$), \nH $\geq$ $10^7$ cm$^{-3}$, \b $\sim$ 3--10) with the analysis of the numerous H$_2$ emission lines seem not to be inconsistent with our CO analysis. While the extinction values computed from CO lines are consistent with the extinction values we found using the H$_2$ lines in Section \ref{sec:mol_em_H2_excitation}, suggesting the H$_2$ and CO emitting gas come from the same medium, the rotational temperatures extracted from the H$_2$ and CO population diagrams are not straightforward to compare given the complexity of the gas excitation. 
Comparisons with the shock model grid suggest external irradiation from the protostar can also be responsible for some amount of energetic CO excitation. The best shock model toward the location where the CO spectra is extracted predicts an integrated CO column density of $\sim\,5\,\times\,10^{18}\,\textrm{cm}^{-2}$, which is several orders of magnitude higher than the CO column densities computed from the CO ro-vibrational population diagram (see Figure \ref{fig:CO_excitation_diagram}, and Table \ref{t.CO_excitation_param}).
We have also extracted the total number of CO molecules from the shocks models (considering the typical shock size to and the total column density), and also found that they exceed the observed CO ro-vibrational population by several orders of magnitude.
The low amount of ro-vibrationally excited CO relative to \Ht lines can be explained by the fact that the bulk of the CO population lies at lower excitation energy than what we trace with the ro-vibrational lines. This effect is more pronounced for CO since its pure rotational lines lie at lower energy than those of \Ht (\ie in the (sub)-millimeter).

Unfortunately comparing with the CO abundance derived from the pure rotational lines traced by ALMA is not possible, as the ALMA CO emission is not co-spatial with the jet basis of the SC source, as explained above. \citet{Tychoniec2019} derived a column density of $5\times10^{15}$---$5\times10^{17}$ cm$^{-2}$ throughout the entire blueshifted outflow lobe.
In addition, $^{12}$CO can also be optical thick, and the non-LTE effects mentioned above potentially at play make our $^{12}$CO column density estimations to be lower limits.
Finally, the CO fundamental line forest we detect can also have a different spatial origin than that of \Ht.
CO fundamental emission has been proposed to originate from the inner disk upper layers and/or slow disk winds in Class I/II objects (\eg see \citet{Pontoppidan2011,Herczeg2011,Brown2013}), which in the case of S68N SC would be scattered into our line of sight via the outflow cavity. As discussed in \citet{Rubinstein2024}, the gas density of the shocked gas traced by \Ht may not be dense enough compared to the critical number density of these CO lines.

\paragraph{CO isotope ratio} Finally, we now attempt to constrain the $^{12}\textrm{C}/^{13}\textrm{C}$ abundance ratio using the $^{13}$CO lines we clearly detect around 4.85 $\mu$m and 5.2$\mu$m thanks to sufficient wavelength spacing from neighboring $^{12}$CO lines. From the column densities derived from the CO excitation diagram that fits two LTE components for $^{12}$CO $v = 1-0$, we obtain the ratio ($N_{^{12}\textrm{CO}_{v\,=\,1\rightarrow0} ,1}$\,+\,$N_{^{12}\textrm{CO}_{v\,=\,1\rightarrow0} ,2}$)/$N_{^{13}\textrm{CO}_{v\,=\,1\rightarrow0}}\,=\,42\pm$22. In addition, we also use the slab models of Figure \ref{fig:CO_data_model_spec} to constrain $N_{^{13}\textrm{CO}}/N_{^{12}\textrm{CO}}$ ratio. We use the slab model parameters of Figure \ref{fig:CO_data_model_spec} for $^{12}$CO, and add a $^{13}$CO$_{v\,=\,1\rightarrow0}$ component of 1800\,K and 100\,AU surface area radius to the model. We find the $N_{^{13}\textrm{CO}}$ value that minimizes the reduced $\chi^2$ for the two spectral regions (\ie 4.84---4.87 $\mu$m, and 5.15---5.25 $\mu$m) where $^{13}$CO lines are clearly identified (see Figure \ref{fig:13CO_12CO}). The most probable $N_{^{12}\textrm{CO}}/N_{^{13}\textrm{CO}}$ values are 57 and 73. These estimations rely on the accuracy of the continuum level, but most importantly on the assumption that the $^{12}$CO emission is optically thin. We note that our reduced $\chi^2$ values are relatively high given that the slab model poorly fits the observed CO spectrum which is of relatively high SNR (the CO line peaks are typically $\sim$ 100---200 times the noise derived from the pipeline). These elements make it difficult to estimate uncertainties for these two values of $N_{^{13}\textrm{CO}}/N_{^{12}\textrm{CO}}$.
The $N_{^{13}\textrm{CO}}/N_{^{12}\textrm{CO}}$ values we constrain are consistent with the interstellar value ($\sim$ 60---70; \citealt{Hawkins1987,Langer1993,Wilson1994,Milam2005}), but are higher by a factor of $\sim$ 1.5---2 with respect to the values found in the ice absorption bands of this source \citep{Brunken2024b}, and in the gas phase toward protostars \citep{Smith2015,Jorgensen2016,Jorgensen2018}. However, the detections of $^{13}$CO and corresponding isotopic ratio does not guarantee that $^{12}$CO is optically thin, since selective UV photodissociation of CO isotopologues can be at play and deplete $^{13}$CO in the gas phase, as it was suggested in the JWST observations of protostars of \citet{Rubinstein2024}.


\subsection{Other molecular lines in the SC source}
\label{sec:mol_em_other_species}

\begin{figure*}[!tbh]
\centering
\includegraphics[scale=0.53,clip,trim= 2.5cm 1.5cm 3cm 3cm]{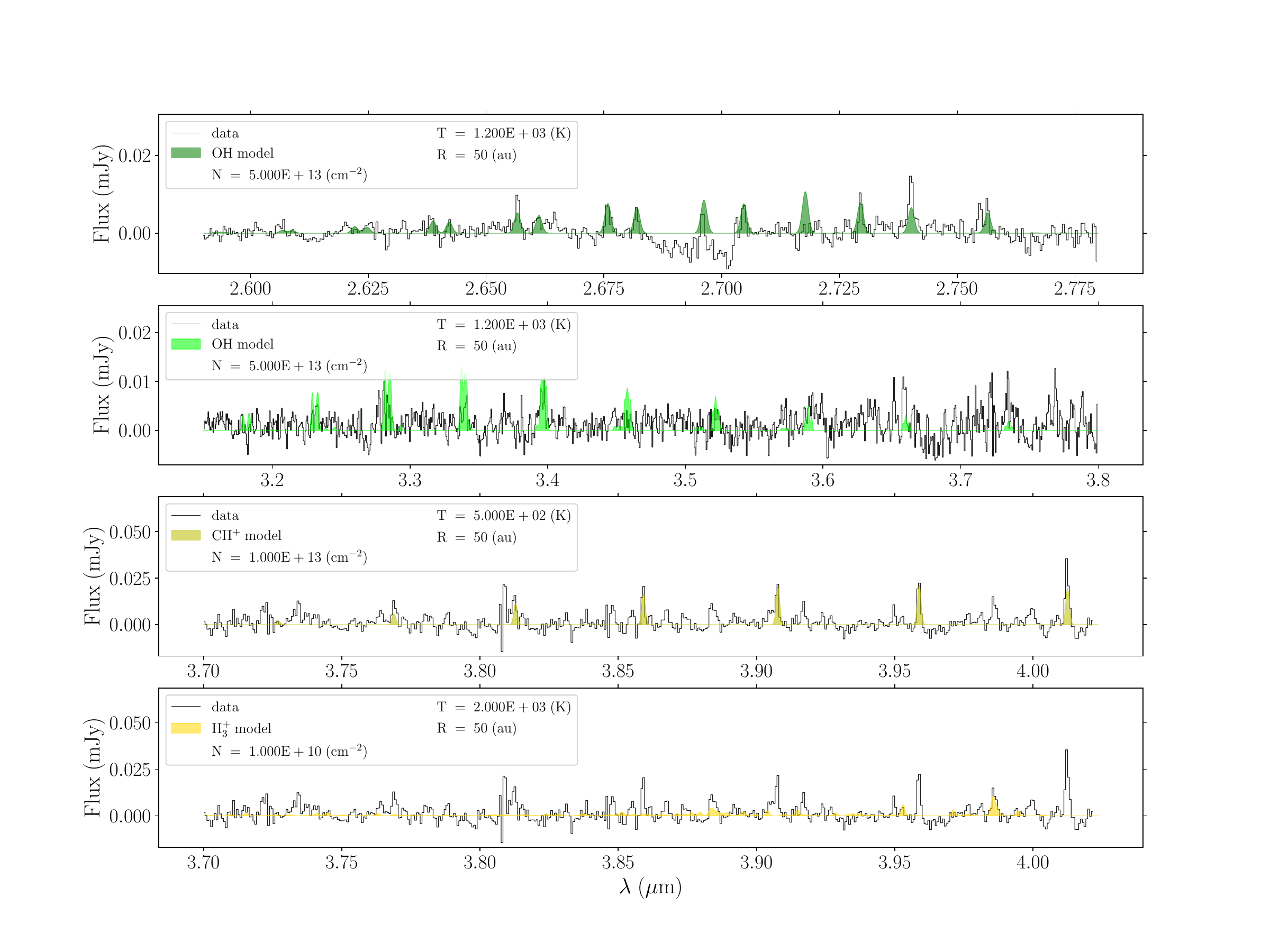}

\vspace{-0.3cm}
\caption{\small OH, CH$^+$, and H$_{3}^{+}$ detections in the SC source in the 2.6--4 $\mu$m wavelength range.
Solid black spectra were extracted from the aperture centered on the SC source, with \Ht lines fitted and removed and the continuum subtracted.
Slab models of OH, CH$^+$, and H$_{3}^{+}$ are shown in color assuming LTE conditions. The models are extinction corrected using the extinction value of the SC source.  The LTE parameters have been chosen manually in order to identify the spectral features.}
\label{fig:OH_CH+_detections}
\end{figure*}

We identified two spectral regions around 2.7 and 3.8 $\mu$m where several spectral features remain unidentified in our analysis of the SC source molecular wind; they are not associated with H$_2$, CO, [\ion{Fe}{2}], or \ion{H}{1} lines. We searched the HITRAN \citep{Gordon2022} and ExoMol databases for molecules with ro-vibrational emission consistent with these spectral features. 
We clearly detected OH and CH$^+$ lines, and potentially detect H$_{3}^{+}$.
Figure \ref{fig:OH_CH+_detections} shows observed spectra (with H$_2$ lines removed and the continuum baseline subtracted) along with 1D-slab models for these three molecules where the LTE parameters have been set manually to fit spectral features). 

Using the HITRAN database line list, we detect OH lines between 2.6 and 2.8 $\mu$m, and tentatively between 3.2 and 3.5 $\mu$m. Eight lines pass our detection criteria around 2.6 $\mu$m, and only one at 3.4 $\mu$m. 
The strong water ice absorption precludes us from detecting the brightest lines, which lie around 3 $\mu$m.
These OH ro-vibrational lines have been recently detected with $JWST$/NIRSpec toward the external FUV irradiated protoplanetary disk d203-506 in Orion \citep{Berne2024}. \citet{Zannese2024} attributed the origin of these lines to OH chemical excitation via the O+H$_2\,\rightarrow\,$OH+H formation pathway \citep{Veselinova2021} or UV radiative pumping. 

In addition, we detect the methylidyne cation CH$^+$ around 3.8 -- 4 $\mu$m, using values from the ExoMol database \citep{Pearce2024}. The brightest CH$^+$ lines in our NIRSpec wavelength coverage are expected to be between 3.8 and 4.2 $\mu$m. We clearly detect 6 CH$^+$ lines between 3.75 and 4.02 $\mu$m, and longer wavelength lines are impacted by the gap between NRS1 and NRS2 detectors longward of 4 $\mu$m. These lines were also observed toward the d203-506 disk by \citet{Berne2024}. CH$^+$ IR emission lines are commonly associated with dense UV-irradiated gas \citep{Godard2013,Neufeld2021}, and were extensively studied in the context of fast dissociative shocks developing self-irradiation by \citet{Godard2019,Lehmann2020,Lehmann2022}, but can also be produced in irradiated $C$-type shocks. Indeed, CH$^+$ is efficiently formed from C$^+$+H$_2$ in energetic shocks, and by photodissociation of CH$_{3}^{+}$. Its abundance is increased by several orders of magnitude when shocks are externally-, or self-UV irradiated. The modeling of these processes at the origin of the detected CH$^+$ lines will be the focus of a future work. However, we constrained the CH$^+$ column density using the same techniques and assumptions as for CO, and found a most probable range for $N_{\textrm{CH}^{+}}$ of $5\times10^{10}$---$5\times10^{12}$ cm$^{-2}$. This column density range is in good agreement with the estimations by \citet{Neufeld1989a,Lehmann2022} for fast and dense shocks.

Finally, we report a tentative detection of H$_{3}^{+}$ at 3.986 $\mu$m using the line list from the ExoMol database \citep{Mizus2017,Bowesman2023}. The numbers of OH, CH$^+$, and H$_{3}^{+}$ detected lines are not sufficient to constrain the rotational temperatures and column densities of the ro-vibrational levels of these molecules. However, we report that the CH$^+$ seems to be optically thin from exploring line fluxes with LTE models. Comparing observed integrated fluxes with the predictions from the shock code we use to fit the H$_2$ lines is beyond the scope of this paper and will be performed in a future work. We note that we do not detect any HD or H$_2$O lines.

\section{Protostellar photosphere of the NC source}
\label{sec:photosphere}

We now model the clearly detected absorption features of the NC source with photospheric models. We assume that the NIR spectra we extract is a sum of different contributions: the photosphere of the central nascent protostellar embryo, continuum excess from the warm dust emission coming from the inner disk (the association of the IR NC source with the small continuum dust emission peak seen in the submillimeter is clear; \citealt{LeGouellec2019a}), and a surrounding cold and dense circumstellar envelope responsible for foreground extinction. The NIR emission of this system is likely dominated by scattered light, which hinders us from evaluating the intrinsic flux of the protostellar embryo and inner disk.

\subsection{Modeling setup}
\label{sec:photosphere_model}

We model the NC's source spectrum in two separate steps.
Firstly, we model the observed flux between 1.9 and 2.45 $\mu$m with the following model (adopted from \citet{Greene2018}):
\begin{multline}
F_{\lambda}\,=\, [ F_{\star,\lambda} \left( T_{\star,{\textrm{eff}}}, {\textrm{log}}\,g, [{\textrm{Fe/H}}] \right) \Omega_{\star} + \\ 
\int_{R_{i,\textrm{disk}}}^{R_{\textrm{max}}}
{B_{\lambda}\left( T_{i,\textrm{disk}} \sqrt{\frac{R_{i,\textrm{disk}}}{r}} \right) d\Omega_{\textrm{disk}}(r)} ]
\times10^{-0.4 A_K \left( \frac{\lambda}{\lambda_K}\right)^{-\alpha}}\,\,,
\label{equ:photo_model}
\end{multline}
where $F_{\star,\lambda} \left( T_{\star,{\textrm{eff}}}, {\textrm{log}}\,g, [{\textrm{Fe/H}}] \right)$ is the flux from the protostellar photosphere of effective temperature $T_{\star,{\textrm{eff}}}$ surface gravity ${\textrm{log}}\,g$ and metallicity $[{\textrm{Fe/H}}]$ (that we restrict to the solar value), $\Omega_{\star}$ is the surface solid angle of the photosphere, 
$B_{\lambda}(T_{\textrm{disk}}(r))d\Omega_{\textrm{disk}}(r)$ is the disk temperature profile modeling the $K$-band continuum emission with $\Omega_{\textrm{disk}}(r)$ for the disk solid angle at the radius $r$ and $T_{\textrm{disk}}(r) = T_{i,\textrm{disk}} \sqrt{R_{i,\textrm{disk}}/r}$ the temperature profile as a function of radius, 
$A_K$ is the effective $K$-band extinction with wavelength dependence $(\lambda/\lambda_K)^{\alpha}$ (where $\lambda_K$ is 2.2 $\mu$m, the $K$-band effective wavelength). We fix $\alpha$ to $1.54$ to match the  extinction law from \citet{Pontoppidan2024a} used throughout this paper. 
We fix $R_{\textrm{max}}$ to 10 AU, but this parameter does not influence much the fitting.
\citet{Greene2018} performed a first model of this object, using however relative flux-calibrated Keck NIRSPEC data of spectral resolution $\simeq$1000. Our JWST NIRSpec data are flux calibrated, more sensitive, and of higher spectral resolution, which allows us to derive more physical parameters. We turn the star and disk solid angles into actual emitting radii, using the distance adopted for the core of 445 pc (assuming an emission of $R$ in radius at a distance $d$, we have $\Omega=\pi R^{2}/d^{2}$). 

We restrain this model exploration to the $1.9-2.45\mu$m spectral region in order to minimize any wavelength dependence in the extinction due to the scattering off grains in the cavity walls. This range also contains spectral features needed to determine the photospheric properties, \ie the different absorption features (Na, Ca, CO, for the deepest ones).
The photospheric spectral features longward of 2.5 $\mu$m becomes sub-dominant with respect to the disk contribution, and contains several ice absorption features. Expanding the range would dilute their contributions and increase the uncertainty of the photospheric fit. 
Within this spectral region, we remove the \Ht emission lines from the spectra using the emission line fitting procedure outlined in Section \ref{sec:obs_jwst_maps} to avoid biasing the covariance estimation. \HtonetoOSthree, \HtonetoOStwo, \HtonetoOSone,  \HtonetoOSO, \HttwotooneSone, \HttwotooneSthree, \HttwotooneStwo, are considered detected and their fit is subtracted from the spectra before performing the fitting.

We use the Starfish modeling framework from \citet{Czekala2015} and \citet{GullySantiago2017} and fit the parameters described in Equation \ref{equ:photo_model} via a Markov Chain Monte Carlo (MCMC) model exploration. A radial velocity (RV) parameter was also used to allow for shifting the wavelength array.
We detail the Starfish fitting and the MCMC processes in Appendix \ref{app:MCMC}.

Secondly, we model the full 1.9---5.3 $\mu$m continuum to improve our characterization of the disk thermal emission. 
The disk temperature profile we use in the first step described above largely overestimates the flux longward of 4 $\mu$m, as the 1.9---2.4 $\mu$m region we focused on is only sensitive to the hot dust.
The 1.9--2.4 $\mu$m slope of this disk component is also steeper compared to what is expected from a heated disk (\eg see the modeling of \citep{Chiang1997}), which is probably related to the role of scattered light and the unknown subsequent extinction law of the NC source.
Therefore, in order to better account for the hot dusty disk contribution, we construct a simple model for the full 1.9---5.3 $\mu$m continuum using both the photospheric parameters we just constrained (\ie $T_{\star,\textrm{eff}}$, log $g$, and $R_\star$), and a geometrically flat, optically thick disk with a temperature profile following $T_{\textrm{disk}}(r)\,=\,T_{\textrm{disk,i}}\times(R_{\textrm{disk,i}}/r)^{\beta}$, between an inner and outer disk radius $R_{\textrm{min}}$ and $R_{\textrm{max}}$. 
No inclination effects are taken into account since they are directly correlated with the flux-loss problem induced by the dominance of scattered light (see Section \ref{sec:photosphere_extinction_veiling}).
The fit only broadly reproduces the full spectra, but while this approach corresponds to a total flux density (integration of $\lambda F_{\lambda}$ on 1.9---5.3 $\mu$m) which agrees within 10\% with the observed 1.9---5.3 $\mu$m integrated flux density, the $K$-band only modeling overestimates it by a factor of $\sim$3. The posterior and adopted prior distributions for the photosphere ($T_{\star,\textrm{eff}}$, and log $g$) and circumstellar disk parameters ($T_{\textrm{disk}}(1\,AU)$, $R_{\textrm{min}}$, $R_{\textrm{max}}$, $\beta$) are listed in Table \ref{t.photosphere_MCMC_results}. We do not report on the stellar radius in this Table because its value is most likely an underestimate, since scattered light is likely the major contributor to the NIR light.

\subsection{Photospheric parameters}

The MCMC posterior probability distributions (derived median, 16\%, and 84\% confidence values) of the NC source $1.9-2.45 \mu$m spectra model parameters are listed in Table \ref{t.photosphere_MCMC_results}, and the best model (corresponding to the median values of the posterior probability distributions) is shown in Figure \ref{fig:best_model_photosphere}. The covariances between each of the other parameters are shown in Appendix \ref{app:MCMC}, Figure \ref{fig:corner_plot_photosphere}. 

The posterior distribution of foreground $K$-band extinction is correlated with the posterior distribution of star solid angle. This is explained by the small spectral region we use, within which the wavelength dependence of the extinction is small. De-reddening the spectra is thus similar to a uniform scaling of the spectra (\ie likewise scaling the photosphere spectra with the star solid angle). In spite of these (anti-)correlation between these two pairs of parameters, the width of the posterior distributions remains small such that we can interpret the results. The effective temperature and surface gravity of the best model photosphere, \ie 3113 K, and log $g$ of 1.95, respectively, correspond to a low mass photosphere.

The low log $g$ value likely represents the true state of the young protostar and is not produced by spurious effects. We highlight that log $g$ is constrained by relative depths of different absorption features of photospheric spectra. The deep CO bands and weak Na and Ca, contribute strongly to the log $g$ fit, more especially in the 2.2 — 2.3 $\mu$m region of the spectra, where the \ion{Ca}{1} and \ion{Na}{1} lines (around 2.206--2.209 $\mu$m and Ca 2.263--2.266 $\mu$m, respectively) are well reproduced by the model. These lines are close enough together that they are less affected by veiling, whose impact is much less shortward of 2 $\mu$m.
We tested the sensitivity of log $g$ to continuum level and shape, and found we retrieve the same log $g$ (and $T_{\star,\textrm{eff}}$) within 1 sigma. These experiments also showed that the surface gravity was sensitive to the ratio of CO and atomic (Na and Ca) equivalent widths, as expected for photospheres in this temperature range.
It is unlikely that cold foreground gas contributes to the measured CO absorption bands. The CO fundamental ($\lambda \sim 4.6\mu$m) absorption is sensitive to cold foreground gas, but it is detected only marginally in the NC source's spectrum. Therefore, cold foreground gas would be expected to produce weak CO overtone ($K$-band) absorption. The relatively strong overtone absorptions indicate the bulk of the CO gas responsible for the overtone absorption is hot.

The faint CO fundamental lines are not likely to be of photospheric origin according to our model, and can thus come from the wind or the disk. More especially we notice that while faint absorption features are seen on the aperture centered on the NC source, we also notice faint CO emission features toward the extended \Ht emission to the North of the NC source. Both of these emission and absorption components are hard to disentangle from one another because of their low S/N, and we thus do not attempt to derive their CO excitation properties.

\subsection{Extinction and veiling}
\label{sec:photosphere_extinction_veiling}

The foreground extinction we determine from our fitting of the NC source's spectrum does not take into account the dominant contribution of scattered light to the photospheric and disk fluxes. Indeed, the highly embedded inner regions of protostars are often modeled with extremely high  absorption in direct lines-of-sight\citep[e.g., $A_v \sim 1000$ mag;][]{Andre1993}. Therefore scattered light escaping through the blueshifted cavity must be the main contributor to the observed NIR light.
The extinction reported in Table \ref{t.photosphere_MCMC_results} thus measures the effective extinction to the last main scattering surface. This causes the measured photospheric and circumstellar disk emitting radii to be underestimated.

Several observational facts support the idea that the NIR continuum is dominated by scattered light.
The negative velocity shift of the photospheric spectra constrained in the MCMC of Section \ref{sec:photosphere} is consistent with that of the \Ht lines (see Figure \ref{fig:line_H2_CO_vel_shift}). The extinction derived from the photospheric model and the extinction derived from the ratio of \Ht lines are consistent with each other, suggesting that the environment of the photosphere and the medium onto which its light is scattered are similarly embedded. We also notice a small shift ($\sim 0\farcs06$) of the IR peak location from 2 to 5 $\mu$m, suggesting that the fraction of scattered light to the total flux evolves with wavelength. Finally, the sub-millimeter peak lies at $\sim 0\farcs15$ (\ie $\sim$ 67 AU) from the 2 $\mu$m peak.

If we assume that the light we receive from the inner disk and the photosphere components are equally affected by scattering (\ie scattered by the same material, with the same efficiency), their relative ratio can be considered as the intrinsic disk to star contribution ratios. Using the fitting of the 1.9---5.3 $\mu$m spectrum performed with a combination of the photosphere model and the disk temperature profile, we can quantify the relative energy of each component. Defining the $K$-band veiling as $r_K \simeq \int^{R_{\textrm{max}}}_{R_{\textrm{min}}}{B_{\lambda_{K}}(T_{\textrm{disk},r}) \, d\Omega_{\textrm{disk}}(r)} / B_{\lambda_{K}}(T_{\textrm{eff},\star})\Omega_{\star}$, we obtain $r_K = 1.25$. Integrating this over wavelength yields the a ratio of disk to star luminosity of $L_{\textrm{disk}}/L_\star = 0.86$.


Finally, we note that our results slightly differ from the model fitting by \citet{Greene2018} slit-based Keck NIRSpec ($R\sim1000$) spectra taken in 2014 to derive the photosphere parameters. 
Specifically, while the estimated effective temperatures are consistent, we find a lower $K$-band extinction (by $\sim$\,1 mag) and a lower surface gravity (by $\sim\,0.4$). We also find a higher $K$-band veiling value than \citet{Greene2018}, who estimated $r_K\,\sim\,0.10$.
However, \citet{Greene2018} used non absolute flux calibrated data, their spectral resolution was three times poorer, and the SNR was $\sim$ one order of magnitude poorer at 2.4 $\mu$m than our JWST spectra. We thus cannot conclude about the time variability of the SC source veiling.
Thanks to its improved capabilities (IFU, higher spectral resolution and sensitivity) the JWST/NIRSpec observations provide much more precise constraints on the properties of this young star-disk system.

\begin{table}
\small
\caption[]{Best NC's source parameters and prior ranges}
\label{t.photosphere_MCMC_results}
\setlength{\tabcolsep}{0.2em} 
\begin{tabular}{p{0.22\linewidth}cccc}
\hline \hline \noalign{\smallskip}
Parameters & S68N NC results$^{a}$  &  priors & units  \\
\noalign{\smallskip}  \hline \noalign{\smallskip}
$T_{\star,\textrm{eff}}$ & $3113^{+23}_{-26}$ & 2600 -- 5550  &  K \\  
log $g$ & $1.95^{+0.16}_{-0.15}$ & 0.5 -- 5 & cm\,s$^{-2}$ \\  
$T_{\textrm{disk}}(1\,\textrm{AU})$ & $157^{+4}_{-4}$ & 10 -- 2000 &  K   \\  
$R_{\textrm{min}}$$^{b}$ & $3$ & 3 -- 10 & $R_\star$  \\ 
$R_{\textrm{max}}$ & $1.5^{+0.1}_{-0.1}$ & 2e-5 -- 100 & AU  \\ 
$\beta$ & $0.50^{+0.01}_{-0.01}$ & -2 -- 2 &   \\ 
$A_K$$^{c}$ & $4.74^{+0.15}_{-0.15}$ & 0 -- 14 & mag  \\  
$v_{\textrm{shift}}$$^{d}$ & $-25.2^{+0.9}_{-1.0}$ & -30 -- 30 & km\,s$^{-1}$  \\   
\noalign{\smallskip} 
\hline
\noalign{\smallskip}
\end{tabular}
\vspace{0.03cm}
\tablecomments{\small 
$^{a}$Posterior parameter distribution median $\pm$ 68\% confidence widths. 
$^{b}$The $R_{\textrm{min}}$ parameter converges toward the low prior boundary which makes its posterior distribution small, which does not reflect the true uncertainty; we thus do not report uncertainty for this parameter.
$^{c}$
The extinction $A_K$ measures the effective extinction to the last scattering surface within the inner regions of the NC source. The intrinsic combination of scattering and absorption is not constrained, and we thus underestimate the intrinsic photospheric and disk fluxes (scaled by their respective solid angle, which in turn are underestimated). An accurate determination of the total flux loss should take into account the dominance of scattered light as the main source of NIR light.
$^{d}$Corrected to LSR. 
The rotational broadening posterior distribution is consistent with it being unresolved (see Figure \ref{fig:corner_plot_photosphere}).
}
\end{table}

\begin{figure*}[!tbh]
\centering
\includegraphics[scale=0.475,clip,trim= 0cm 0.0cm 0cm 0cm]{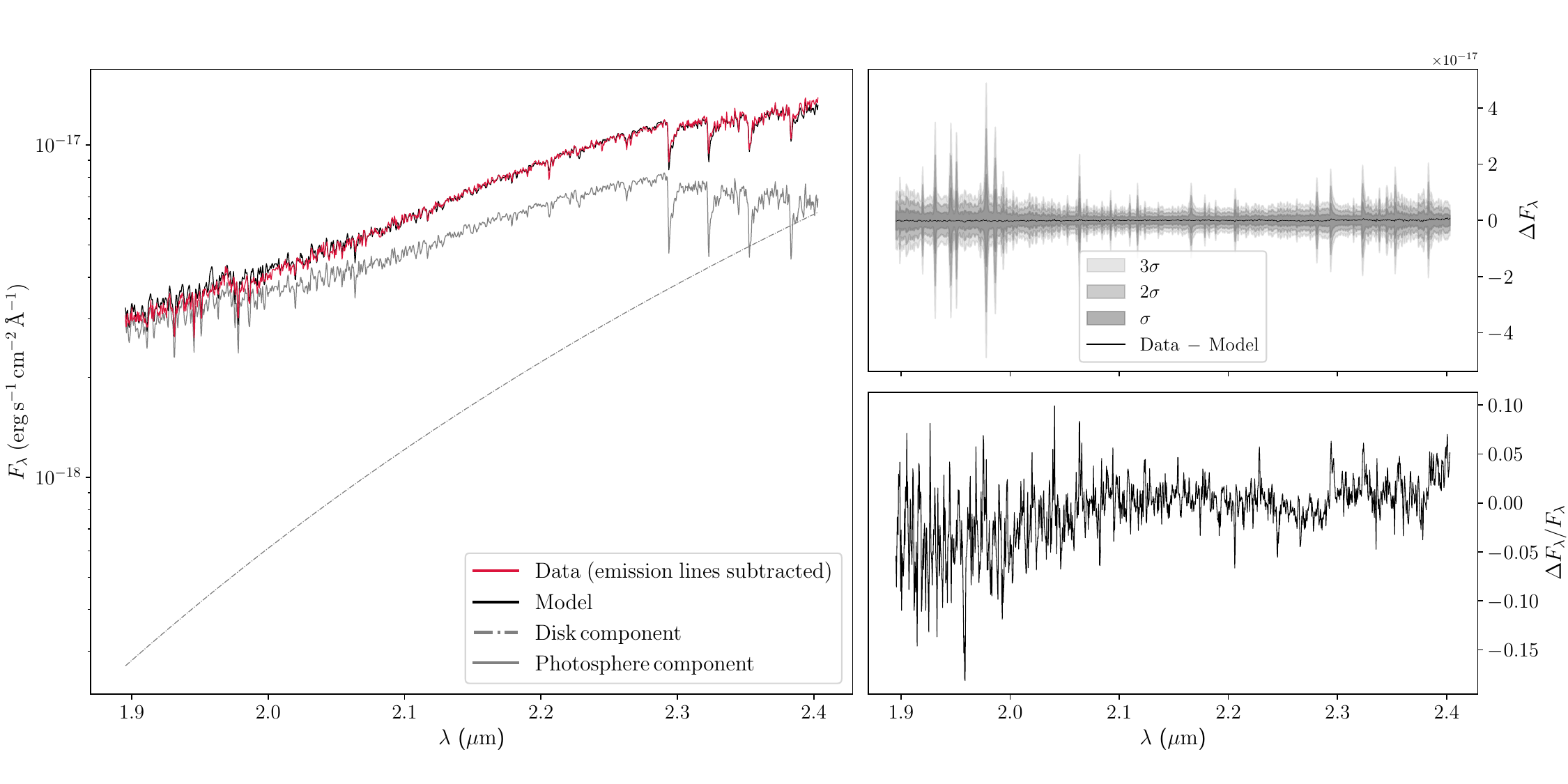}

\vspace{-0.5cm}
\caption{\small Best model from the Starfish MCMC exploration of the NC source's $1.9-2.45\,\mu$m spectrum. The left panel shows the S68N NC source's $1.9-2.45 \mu$m spectrum from the NIRSpec observations in red (where \Ht emission lines were fit and subtracted), and the best model (see Equation \ref{equ:photo_model}) in black, which is also decomposed between the disk temperature profile (gray dot-dashed line; corresponding to the result from the first modeling step of Section \ref{sec:photosphere_model}) and the photospheric component (gray solid line). The final disk temperature profile is then constrained using the full 1.9---5.3 $\mu$m spectrum.
The residuals along with the diagonal of the covariance matrix are shown as $\sigma$ contours in the top-right panel, and the relative errors (residuals to data flux ratios) in the bottom-right panel.}
\label{fig:best_model_photosphere}
\end{figure*}

\section{Principal Component Analysis and Spatial distributions of source components}
\label{sec:PCA}

In order to interpret further the spatial distribution of the different features contributing to the NIRSpec IFU data, we employ a Principal Component Analysis (PCA) tomography to the G235H and G395H data. The PCA method transforms the data into a new coordinate system of orthogonal principal components, chosen to describe the highest variance. It allows the extraction of information from higher dimensions in the data. When applied to data cubes that contain two spatial dimensions and one spectral dimension, the PCA reduction technique allows the decomposition of the data along eigenvectors (in this case eigenspectra), using the information contained in the spatial axes. Therefore, spectral signals that correlate with some spatial distribution that may be hidden in the current dataset can be unveiled. We can thus isolate the different contributors to the original dataset that have a self-consistent spatial and spectral signal.

While we develop the method and results in Appendix \ref{app:PCA}, we aim to outline here the main results of this PCA analysis. In both the G235H and G395H data, the nature of the first three principal components (PC) seem to be similar. The first PC exhibits mostly the contribution from the continuum emission, \ie the photosphere component of the NC source shortward of 3 $\mu$m, and the continuum from both the NC and SC sources longward of 3 $\mu$m. Then, while the second PC clearly highlights the extended \Ht emission lines of both the SC and NC source's outflows in the G235H data, it mostly consists of the bright CO molecular wind of the SC source in the G395H data. PC3 shows an interesting gradient (whose direction is consistent with the outflow direction, as suggested by the morphology of the \Ht emission lines) in the eigenvalues toward the NC source in both the the G235H and G395H data, with a clear slope in the eigenspectra. We interpret this as a potential scattering contribution to the NC source continuum spectra. PC4 and PC5 of the G395H show complex spatial morphology in the eigenvalues toward the SC source outflow cavity associated with faint CO fundamental emission and absorption features, shifted in wavelength relative to one another, most likely highlighting the complexity in the spatial variation of the CO excitation properties.

Such analysis opens a promising avenue to further interpret the different contributors of NIRSpec IFU datasets of protostellar cores. The spatial distribution of the continuum and its scattered-light counterpart, the extended molecular outflows and its different velocity components can ultimately be identified and studied separately.

\section{Discussion}
\label{sec:discussion}

\subsection{The evolutionary status of the S68N protostellar system}
\label{sec:discussion_accretion_S68N}

Spectroscopic observations performed with the JWST toward protostellar systems opens a new avenue to explore their evolutionary status, \ie the accretion and ejection status and properties of the disk and protostellar embryo, of the youngest and most embedded protostars. 
NIRSpec provides a complementary view of protostars to that provided by ALMA.
We can discuss the physics and chemistry at play in the innermost regions of protostars from the analysis of the photosphere, continuum, and molecular emission lines from NIRSpec, whereas ALMA constrains the properties of the cold gas and warm dusty emission in the envelope, disk, and outflow cavities.
More vigorous accretion is expected in the Class 0 stage compared to the Class I stage \citep{LeGouellec2024a}. However, it remains challenging to reliably estimate the mass accretion rate of the most embedded systems, as the accretion mechanisms and inner disk physical properties at this stage are complex and may be of different nature than in more evolved objects, of which the different accretion-related scaling laws and assumptions may not be applicable to Class 0s. The two S68N protostellar sources we study in this work bring new quantitative discussions about the NIR properties of Class 0 protostars.

\subsubsection{Status of the NC source} 
\label{sec:discussion_accretion_S68N_NC} 



The low $K$-band veiling we found for the NC source would suggest a small accretion rate for a Class 0 protostar, given the typical veiling of Class I protostars \citep{Doppmann2005,Connelley2010}, which are expected to be weaker accretors. However, considering the apparent low luminosity of the NC source young photosphere, the $K$-band veiling we constrain may be consistent with a significant accretion rate. 
In order to broadly estimate the mass accretion rate of the NC source, we use the constraints we obtain on the disk and photosphere. The idea is to determine whether the disk temperature we obtain corresponds to the radiative flux expected from the re-emission of the central object, or if an additional accretion-related term is needed. Indeed, at the radii of interest here (\ie $\leq\,1$ AU), the dust opacity is expected to be high enough for the disk to be optically thick,  both to stellar radiation and to its own re-emitted radiation. Therefore, the radial temperature profile of a geometrically thin optically thick disk heated solely by starlight of the central star has an analytical solution of the following form, using the prescription of \citet{Adams1986}:
\begin{equation}
\label{equ:Tdisk_from_Tstar}
T_{\textrm{disk}}(r)\,=\,T_{\star} \times \left(\frac{2}{3 \pi}\right)^{1/4} \times \left(\frac{r}{R_{\star}}\right)^{-3/4}\,\,.
\end{equation}
Integrating Equation \ref{equ:Tdisk_from_Tstar} for all radii of an optically thick disk reprocessing the energy from the center yields $L_{\textrm{disk}}/L_{\star} = 1/4$, while we found a higher ratio of 0.92.
This temperature profile underestimates by a factor of 3---5 the disk temperature profile we constrained in Section \ref{sec:photosphere}, suggesting that another source of heating on top of the central photosphere, accretion-related, can be invoked to explain the IR excess toward the inner disk of the NC source. 

The viscous disk formalism is usually used in Class II disks to relate the dissipation of gravitational potential energy associated to accretion with the turbulent transport of angular momentum in the disk, which induces viscous heating (\eg see the formalism by \citealt{Shakura1973,LyndenBellPringle1974,Pringle1981}, and the seminal modeling of \citealt{Calvet1991,Meyer1997}). 
Class 0 disks may have different properties than more evolved disks because they are still forming and dynamically coupled to the collapsing envelope. While several caveats come along with this approach, we attempt to reproduce the NC source's IR excess with this formalism in Appendix \ref{app:NC_accretion}. 
The mass accretion rate is found to be significant, \ie $\dot{M} \gtrsim$ $3\times10^{-6}$ $M_\odot$\,yr$^{-1}$.
However, it is not possible to conclude firmly because of the scattered light problem hindering us to precisely constrain the stellar parameters.

Surprisingly, despite this IR excess of the hot inner disk, almost no \ion{H}{1} lines are detected, although such line emission is commonly seen toward more evolved accreting objects.
Only a faint Pa$\alpha$ emission is detected; this \ion{H}{1} transition was, however, not previously calibrated with the accretion luminosity because it cannot be easily observed from the ground (see, however, \citealt{Rogers2024a}).
Using the extinction we derived for the NC source, we derive an upper limit for the \Brgamma line luminosity of $L_{\textrm{Br}_\gamma}\,\leq\,2.3 \times 10^{-5}\,L_{\odot}$.
This value corresponds to an accretion luminosity upper limit of 1.7 $\times$ 10$^{-4}$ L$_\odot$, where we used the empirical relation of \citet{Alcala2017} relating $L_{\textrm{acc}}$ to $L_{\textrm{Br}\gamma}$ in T-Tauri stars.
Even considering the full range of parameters explored in Appendix \ref{app:NC_accretion} 
, the corresponding accretion luminosity that can reproduce the IR excess lies $\geq$ 3 orders of magnitude above the accretion luminosity upper limit obtained via the non-detection of the \Brgamma line.
This raises the question whether the accretion mechanism at play at the Class 0 stage is similar to the one occurring in more evolved objects, \ie magnetospheric accretion, where the correlation between $L_{\textrm{\ion{H}{1}}}$ and the accretion luminosity is clearly established \citep{Najita1996b,Muzerolle1998,Alcala2017}.
Another possibility is that much material loads up in the disk, but does not transfer to the central protostellar embryo in a steady way. The high mass accretion rate we estimate suggests a short dynamical timescale for the disk accreting material which is usually associated with thermal-viscous instabilities in the disk (\eg \citealt{Martin2011}). However, no significant variability in the IR or millimeter has been noticed during the last decade of WISE/NEOWISE, JCMT, and CARMA/ALMA monitoring of the S68N core \citep{Francis2019,LeeYH2021,Park2021}.

The NC source is thus a peculiar object, harboring a hot disk but without exhibiting the \ion{H}{1} lines typical related to the accretion column. Despite these findings, there is much evidence that this object is a Class 0 protostar. Its extremely high extinction and embedded nature revealed by both our JWST ($A_K$ measurements and presence of ices in absorption) and ALMA observations \citep{Hull2017b,LeGouellec2019a}, support the fact that the NC source is embedded in a dense circumstellar envelope. The low log $g$ value we constrain is more likely to be encountered toward young accreting objects rather than PMS stars \citep{Baraffe2017}. The extended blueshifted \Ht emission also suggests shocks within a nascent outflow structure. Finally, given its proximity with the SC source and its connected ALMA dust continuum emission, it does not appear probable to have a Class II object (age of 1-5 Myr) inside a Class 0 protostellar core. This rather suggests co-evolution.






\subsubsection{Status of the SC source}
\label{sec:discussion_accretion_S68N_SC}

The prevalence of emission lines in the SC source is similar to many Class 0s in the Keck survey by \citet{LeGouellec2024a} (see also \citealt{Yoon2022}), but the lack of $K$-band continuum emission is suggestive of the most embedded evolutionary state, similar to a few Class 0s of the Keck survey exhibiting only emission lines tracing the outflow base (\Ht and [\ion{Fe}{2}] lines). The totally obscured inner regions preclude us from analyzing potential accretion related \ion{H}{1} lines or CO overtone bands, but given the powerful ejected molecular jet we detect, this source is likely more heavily accreting compared to the NC source. The dynamics of the outflow suggest a significant mass ejection rate ($\sim\,10^{-7}$---$10^{-6}$ $M_\odot$ yr$^{-1}$), with strong shocks (fast with velocity $\geq$ 30 km s$^{-1}$, and high pre-shock density of $\geq$ $10^7$ cm$^{-3}$), and potential irradiation from the accretion shock-driven radiative flux and/or from shock self-irradiation. This latter point is suggested by the \Ht shock modeling results, the bright and cavity-shaped UV-sensitive C$_2$H emission from \citet{LeGouellec2023a}, the detection of the CH$^{+}$ cation, and the CO non-LTE excitation showing hints of UV-pumping. Further work will be needed to constrain precisely the origin and intensity of the irradiation field, which may be the most straightforward way to quantify the accretion in this source, if one can differentiate the external irradiation field affecting the molecular gas excitation from the potential self-irradiation of shocks at the outflow base. 
Indeed, in the case of fast (\ie $\geq\,$30 km s$^{-1}$) weakly magnetized, J-type molecular shocks, self-generated UV radiation is eventually produced by the shock front, in which case Ly$\alpha$ is emitted locally (see \citealt{Lehmann2020} for a thorough analysis of the treatment of the Ly$\alpha$ photons self-absorption). 
The mass ejection rate, which itself remains challenging to estimate given the uncertainty of the inclination angle, will only provide an approximate idea of the associated mass accretion rate. 

Although larger than the NC source, the outflow of the SC source also exhibits a short dynamical timescales of $\sim 10^{3}$ yr \citep{Tychoniec2019}. The NIRCAM mapping of the entire region presented by \citet{Green2024} shows  that the bipolar outflow does not extend beyond 5000 AU. Dust emission is also detected at the apex of the bipolar lobes (see Figure \ref{fig:RGB_alma_image}) suggestive of entrained dust during the recent propagation of the outflow throughout the dusty envelope. While the SC source appears older than the NC source from the spatial extents of their \Ht emissions, it still appears to be a Class 0 protostar with an accreting object still deeply embedded in its dense envelope as mapped in dust emission with ALMA \citep{Aso2019}. The SC source envelope is likely more massive than the NC source envelope given their ALMA 870 $\mu$m fluxes \citep{LeGouellec2019a}. 

Interestingly, no [\ion{Fe}{2}] lines are seen in the NIRSpec data of the SC source. This can be related to the amount of dust in the shocked material (and thus to the launching radius of the molecular jet), which can in turn have important implications for the chemistry of the jet \citep{Tabone2020}. Other tracers of dust within the outflow shocks of the SC source include the SiO ($5\rightarrow4$), SO ($5_6\rightarrow4_5$) transitions \citep{Tychoniec2019,Podio2021}, tracers of dust destruction in shocks, and complex organic molecules (COMs) attributed to ice sputtering from dust grains \citep{Tychoniec2021}. However these emission lines are seen 2000 AU away from the SC source in its blueshifted outflow; the corresponding dust may originate from the outer envelope being shocked by the apex of the outflow. \citet{Narang2024} has observed an ionized jet in a young Class 0 source, but at a low accretion stage. This suggests that the strong accretion of the SC source leads to powerful ejection of a dense jet that converts everything to molecular gas \citep{Ray2023}. We note that faint [\ion{Fe}{2}] lines are detected in the MIRI/MRS of S68N (priv. communication with the JOYS MIRI GTO team), only with the lines of lowest excitation energy (\ie $\lesssim$ 2500 K of upper level excitation temperature, at 5.34, 17.92, 24.50, and 24.99 $\mu$m). 





\subsection{On the nature of NIR characteristics of the youngest protostars}
\label{sec:discussion_C0_NIR}

\subsubsection{Young photospheres and inner accretion disks}

Only a few photospheres of Class 0 protostars have been observed so far \citep{Greene2018,LeGouellec2024a}. These objects are deeply embedded ($A_K$ of $\sim$ 4.8 mag for the NC source) and thus require long integrations. While the consensus is that these young protostars go through the most intense mass accretion activity of their lifetime during the Class 0 phase, in S68N and the four other photospheres observed in the survey of \citet{LeGouellec2024a} none of the \ion{H}{1} accretion tracer lines are detected. 
In the surveys of more evolved Class I and Flat Spectrum (FS) objects from \citet{Doppmann2005,Connelley2010,Connelley2014}, several sources are devoid of \ion{H}{1} emission, but yet are often associated with high veiling, suggestive that an accretion-based heating process is at play.
These \ion{H}{1} emission-less Class 0 spectra could thus be associated with some level of accretion activity.

\citet{LiuH2022} explored models of viscously heated young protostellar disks and determined that such disks can outshine the central star in the case of low-mass stellar embryos that did not accrete enough mass yet to dominate the NIR emission (\eg see also \citealt{Yoon2021}). Such systems, dominated by their viscously heated disk contribution, are usually attributed to FU Ori-type objects (FUors), which are decades-long outbursts of accretion onto young stars, producing prototypical emission free NIR spectra. One can question whether the disk of the S68N NC source is undergoing such an episode of high accretion activity. 
However, as mentioned above we note that the viscous disk model may not be applicable to the Class 0 inner regions. We also detect clear Na and Ca lines which are as deep as the CO absorption bands, a good sign that the absorption features originate from the photosphere of the protostellar embryo. Indeed, the disk photosphere would then produce much deeper CO overtone absorption bands (see \citealt{Connelley2018} and their Figure 9), which is not observed in S68N \citep{LeGouellec2024a}. In addition, the temperature of the disk component we constrain appears to be lower than the typical hot dust emission of FUor objects (\eg \citealt{Zhu2009c}), and we note that Paschen $\alpha$ is expected to be seen in
absorption in FUors \citet{Connelley2018} while a faint emission is detected in the spectra of the NC source. 
An active viscous accretion disk would also produce deep CO fundamental and water ro-vibrational absorption features due to the strong temperature gradient induced by the mid-plane-viscous heating toward the dense inner disk. This was observed in several NIR and MIR spectra of FUors (\eg the Spitzer IRS spectra of FU Orionis \citealt{Green2006,Zhu2009a}, and recent SOFIA/EXES observations of FUors, C. DeWitt private communication, \citealt{DeWitt2023AAS}). However, no water ro-vibrational absorption features are seen in the JWST MIRI spectra of \citet{vanGelder2024b}, and only marginal CO fundamental absorption is seen in S68N NC.

The low value of the surface gravity we constrain for the NC source is low compared to models of young low-mass PMS stars of similar temperature. For example, \citep{Baraffe2015} predicts a log $g$ of $\sim\,3.3$ for a 0.5 Myr object at an effective temperature of $\sim\,3100$ K. 
However, it is not inconceivable that the increase of internal energy of the protostellar embryo induced by a recent episode of strong accretion would cause the radius to increase, and thus lower the surface gravity. \citet{Baraffe2017} showed for example that accretion bursts onto T Tauri stars can increase their luminosities by over 0.5 dex compared to their non-accreting state at the same effective temperature.
An alternative explanation would be that the disk contributes to the absorption features.
Recent models by \citet{Carvalho2023a,Carvalho2024a} successfully fitted the optical and NIR absorption lines of FUors with disk models implementing PHOENIX models of low surface gravity values of $\sim\,1$ at each annuli.
However, the log $g$ value we obtain is one dex higher and recent high spatial resolution numerical simulations of collapsing protostellar cores have shown that the disk mid-plane of young Class 0 objects are nearly dead to turbulence, and that most of the infalling material transits through the disk upper layers before being accreted by the central stellar embryo \citep{Kuffmeier2018,LeeYN2021,Ahmad2024}. The viscous disk model that describes the turbulent transport in evolved disks may thus not be appropriate to embedded disks, but this remains to be explored in more detailed modeling of such JWST NIRSpec spectra of Class 0 disks.

The non-detections of \ion{H}{1} in several Class 0 objects remain puzzling given that these objects are expected to be strong accretors, as suggested by the Class 0 exhibiting bright \Brgamma and CO overtone emission in \citet{Laos2021,LeGouellec2024a}. While accretion time-variability can often be invoked in the Class 0 phase \citep{Yoon2021, Yoon2022, Guo2020, LeeYH2021, Park2021, Zakri2022,LeeSieun2024}, we can also question the status of the magnetosphere at this early evolutionary stage. In more evolved objects, the \ion{H}{1} is thought to originate from the accretion column of material guided by the central star magnetic field lines. The morphology of the accretion column may be different in nature in the Class 0 phase if the nascent stellar embryo has not yet developed a magnetosphere and a strong enough gravitational potential to develop accretion columns and their associated shocks at the origin of the hot plasma, where \ion{H}{1} emission lines are then produced. Numerical simulations resolving the collapse of the second Larson core have suggested that turbulent motions could initiate a dynamo process early on during the protostellar phase \citep{Bhandare2020,Ahmad2023}. Velocity broadening measurements of Zeeman-sensitive photospheric absorption lines revealed significant large-scale magnetic field strengths in Class I and FS objects of $\sim 1$--2 kG \citep{JohnsKrull2009,Donati2024,Flores2024}. 
Such observations remain challenging for current instruments when looking at Class 0 objects, which are fainter than $\sim 14$ mag in the $K$-band.

\subsubsection{Young outflow/jet systems}

Accreting protostars are known to generate a significant amount of UV radiative flux propagating in outflow cavities, induced by the accretion and/or outflow shocks \citep{Spaans1995,vanKempen2009,Visser2012,Yildiz2012,Kristensen2013b,Benz2016}. JWST enables the thorough study of the excitation of ro-vibrational lines and the photo-chemistry at play in the outflow cavities of protostars. \citet{Tabone2021} showed that measurements of OH rotational line intensities and water vapor column density can constrain the UV field responsible for photodissociation of water into OH. To this end, OH high-$J$ rotational lines in the MIR were recently analyzed in the energetic shocks of the HOPS 370 protostellar outflow \citep{Neufeld2024} (see also work by \citealt{Zannese2023} for an externally irradiated disk). 
However, water vapor is not detected longward of 5$\mu$m in our NIRSpec observations of the SC source.
Ultimately, MIRI MRS will be needed to quantify the UV field and confirm the constraints we obtain from the \Ht and CO fundamental lines analysis.


Our shock modeling results must still be confirmed by additional shock models implementing self-generated UV radiation fields in order to ascertain the shock parameters on the SC source.
Specifically, one still needs to investigate what transverse magnetic field strength would allow the generation of fast UV-irradiated shocks in the nested dense envelope. Our current results suggest that the magnetic precursor is strong enough to not be fully dissociative. Our best models can account well for the excitation and column densities of the \Ht molecular gas. The CO excitation and reported detections of OH and CH$^+$ lines argue toward a UV-irradiated wind. Indeed, the UV field, whether internally-generated in the shocked environment or accretion-generated within the central regions, is a key component enhancing the production and excitation of CH$^+$ in shocks.


The non-detections of NIR [\ion{Fe}{2}] lines toward the energetic shocks of the SC source molecular jet suggest that a significant amount of refractory elements is not liberated into the gas phase. Fast magnetized $C$-type shocks are usually efficient at eroding dust grains, as the magnetic precursor causes a velocity drift between charged and neutral species which enhances collisions \citep{McCoey2004}.
However, we note that the stratification into a separated central atomic jet (exhibiting NIR [\ion{Fe}{2}] and [\ion{Ni}{2}] lines, for example) and a surrounding lower velocity \Ht wind was highlighted in several recent IR observations of Class 0/I protostars \citep{Harsono2023,Assani2024,CarattioGaratti2024,Delabrosse2024,Dionatos2014,Hodapp2024,Federman2024,Narang2024,Nisini2024b,Tychoniec2024}. Such stratification may operate later than the current protostellar evolutionary stage of S68N. This can also be a matter of accretion rate/mechanism that enables the ejection of material whose velocity is more organized, \ie axisymmetric, and more homogeneously distributed as a function of launching radius. In the \citet{LeGouellec2024a} Keck survey of Class 0 protostars, 20\% of sources had [\ion{Fe}{2}] in their MOSFIRE $K$-band spectra. However, among all the sources with a heavily obscured faint continuum level (\ie $\geq$ 16 mag for their $K$-band magnitude, thus suggestive of a relatively young age based on the embededness of the central source) that exhibited at the same time bright \Ht lines (similar to the SC source), none of them had [\ion{Fe}{2}] line detections. 
In the work of \citet{Federman2024}, 4 out of their 5 Class 0 protostars showed [\ion{Fe}{2}] jets in the NIR. The only exception was the most luminous source in the sample, which only showed knots of emission. The emergence of central high-velocity atomic and bright NIR [\ion{Fe}{2}] jets might take place in the Class 0 phase, and these ionized jets may currently be inactive in several of these Class 0 objects.



The molecular jet of the S68N SC source appears collimated but does not exhibit a variation of excitation or velocity properties across its minor axis. The IR molecular lines are also not resolved, and we measure the width of the molecular jet to be $\sim$ 100 AU at 800 AU from the driving source using the \Ht lines (using an average PSF FWHM of 0.26$^{\prime\prime}$ at 5 $\mu$m). Comparing with the ALMA molecular gas data (see Figure \ref{fig:JWST_ALMA_images}), we note that the blueshifted CO (2$\rightarrow$1) emission line has a ``V-shape'' morphology embracing the more collimated IR molecular jet.
Deeper and higher angular resolution ALMA data would be needed to perform a detailed analysis of the ALMA CO outflow to discuss its wind origin, similar to what has been done in Class I objects \citep{Bjerkeli2016,Louvet2018,deValon2020,deValon2022}.
The spatial and spectral resolution of these JWST observations also remain limited to evaluate the driving mechanism of this young jet (the source is at 445 pc), \ie whether we can differentiate between a wide-angle X-wind model morphology \citep{Shu1995,Matzner1999}, from a jet driving outflow through successive bowshocks \citep{Masson1993,Raga1993}. Recent (magneto-)hydrodynamic models of protostellar outflows provide valuable insights to discuss and analyze the morphology and kinematics of young outflows with respect to these launching mechanisms \citep{Rabenanahary2022,Shang2023a,Shang2023b}. 

The transition from the hot SC molecular jet to the dense and colder cavity walls (n = 10$^{6}$--10$^{8}$ cm$^{-3}$ from the shock modeling) appears sharp in Figure \ref{fig:JWST_ALMA_images}. The shock interface between the jet and the surrounding envelope may carry along envelope material as described in the wind model scenario, where a radially expanding swept-up cavity grows from the interaction of an inner wide-angle wind with an outer stratified environment \citep{Shu1991}. This model predicts that the shell expansion velocity increases with distance from the driving source (so called ``Hubble law''). Figures \ref{fig:line_H2_CO_vel_shift} and \ref{fig:H2_param_jet} show that the LOS velocity shift of the \Ht and CO lines increase along the jet. This is also the case with the ALMA CO outflow (with a gradient of -5 km s$^{-1}$ per arcsec in the blueshifted cavity). Such a model would deplete the envelope in too short of a timescales compared to the lifetime of the a protostellar envelope if shocks were not confined toward narrow enough regions, and active during short enough timescales \citep{Kristensen2013b}. Alternatively, the approach by \citet{Liang2020} alleviates this problem by accounting for a constant speed turbulent mixing layer that may develop between the shocked cavity walls and the infalling envelope.

More insights are needed about the magnetic field morphology inside young molecular jets. Magnetic fields are a key component in the launching and propagation of winds, and observing their orientation in outflows cavities would be a valuable asset to compare with the kinematics of the molecular gas and the shock conditions (\ie what is the role of the transverse magnetic field in setting the shock type and properties). In this context, the Goldreich-Kylafis effect \citep{Goldreich1981} can polarize emission lines of certain outflow-tracer molecular gas species in outflow cavities of protostars \citep{LeeCF2018Nat,Barnes2023}, but the observability of this phenomenon remains challenging \citep{Lankhaar2020}. One can still utilize ALMA dust polarization observations of outflow cavity walls \citep{Maury2018,LeGouellec2019a,Hull2020a} alongside \Ht-based JWST shock parameter diagnostics to discuss the magnetization of outflow shocks propagating in young outflow cavities, which are still nested in a dense envelope that could have accumulated significant magnetic field during collapse. In this case, the magnetic precursor constrained by shock models can be related to the total magnetic field via the POS magnetic field mapped with the polarized dust emission.


\section{Conclusions and summary}
\label{sec:ccl}

We present new JWST NIRSpec 1.7---5.3 $\mu$m observations of the S68N protostellar core, located in the Serpens Main star forming region. While four fragments are identified in the ALMA sub-millimeter high angular resolution observations of this core, our NIRSpec observations cover the two central sources, the North Central (NC), and the South Central (SC) sources (separated by $\sim$ 1.4$\farcs$, \ie $\sim$ 600 AU), whose NIR properties are surprisingly different. Our conclusions and results concerning these two protostellar sources are summarized as follows:

\begin{itemize}
    \item The NC source exhibits continuum emission whose peak location is consistent with the sub-millimeter source. We attributed the origin of the 1.8--2.4$\mu$m continuum to a young, low-mass photosphere ($T_{\star,\textrm{eff}} = 3113^{+23}_{-26}$, log $g = 1.95^{+0.16}_{-0.15}$) accompanied by a warm continuum emission coming from a circumstellar disk. 
    It is challenging to quantify the fraction of near-IR scattered light originating from the photosphere and inner disk that escapes via the outflow cavity. This makes it difficult to constrain its stellar luminosity and mass.
    The surprisingly low surface gravity is discussed along with the relatively young age and strong accreting nature of Class 0 objects, for which standard evolutionary models don't apply.
    
    \item The NC source disk heating suggests that significant accretion onto the central protostellar embryo is on-going. Estimating the accretion luminosity from this IR excess also depends on the fraction of intrinsic photospheric and disk Near-IR light that scatter out the cavity. The prototypical \ion{H}{1} line accretion tracers are not detected, suggesting that a different accretion mechanism than the magnetospheric accretion model adopted in more evolved objects may be at play, or that the accretion onto the central object is unsteady or episodic. 
    
    \item The short dynamical timescale ($\leq$ 100 yr) of the \Ht blueshifted outflow, which does not extend beyond 500 AU, also suggests that the NC protostar is very young.

    \item The SC source's continuum is totally obscured shortward of 3.5 $\mu$m. A faint continuum is detected between 3.5 and 5.3 $\mu$m. The IR source is shifted toward the blueshifted outflow cavity of the sub-millimeter source, suggesting the continuum is dominated by scattered light.
    
    \item $\sim$70 \Ht lines, forming a collimated molecular jet, are detected toward the SC's molecular jet. We model them with the shock model grid presented by \citet{Kristensen2023}, and establish that energetic $C$-type shocks provide the best match with the observations. More specifically, fast shocks ($\geq$ 30 km s$^{-1}$), high pre-shock density ($\geq$ $10^7$ cm$^{-3}$), strong magnetic field ($B \sim $ 10--100 mG), and a non-negligible irradiation field ($G_{\textrm{0}}$ not consistent with 0) best match the data. This energetic molecular jet is suggestive of vigorous mass ejection, for which the accretion-related origin and strength of the UV radiation field can be quantified through further modeling.
    
    \item A bright line forest of CO fundamental transitions is also detected in the wind of the SC source, with specifically the detections of the CO $v =  1\rightarrow0$, $2\rightarrow1$, and $3\rightarrow2$ bandheads. The excitation analysis of the CO lines reveals hints of significant non-thermal excitation, likely IR- or UV-pumping. Furthermore, several $^{13}\textrm{CO} (v=1\rightarrow0)$ lines are detected, yielding a $^{12}\textrm{CO}/^{13}\textrm{CO}$ consistent with the ISM value. 
    
    \item For the SC source OH (at $\sim$ 2.7 and 3.4 $\mu$m), CH$^+$ (at $\sim$ 3.9 $\mu$m), and potentially H$^{+}_{3}$ lines are detected at the base of the outflow, also suggestive of radiative pumping.
    
    \item No [\ion{Fe}{2}] lines are detected in the NIRSpec data of the SC source molecular jet.
\end{itemize}
 

This study reveals the interesting NIR properties of the S68N fragmented core that hosts at least two young Class 0 protostars. This opens a promising avenue to understand and constrain the first steps of the star formation processes, as young photospheres and energetic irradiated outflows start to be observed. JWST thus enables unprecedented diagnostics of the spatial location of these emissions and the extinction of the gas. 

\begin{acknowledgments}

The authors thank C. Boersma, O. Berné, K. Kaplan, M. G. Navarro for fruitful discussions.
V. J. M. L.'s research is supported by an appointment to the NASA Postdoctoral Program at the NASA Ames Research Center, administered by Oak Ridge Associated Universities under contract with NASA.
V. J. M. L. gratefully acknowledges support from the ALMA Ambassadors Program, operated by the North American ALMA Science Center. The National Radio Astronomy Observatory is a facility of the National Science Foundation operated under cooperative agreement by Associated Universities, Inc. 
T. P. G. and B. W. P. L. acknowledge support from the NASA Next Generation Space Telescope Flight Investigations program (now JWST) via WBS 411672.07.05.05.03.02.
A. Gusdorf gratefully acknowledges the support of the Programme National "Physique et Chimie du Milieu Interstellaire" (PCMI) of CNRS/INSU with INC/INP co-funded by CEA and CNES.
DJ is supported by NRC Canada and by an NSERC Discovery Grant.
EvD, MvG, and LF, acknowledge support from ERC Advanced grant 101019751 MOLDISK, TOP-1 grant 614.001.751 from the Dutch Research Council (NWO), the Netherlands Research School for Astronomy (NOVA), the Danish National Research Foundation through the Center of Excellence “InterCat” (DNRF150), and DFG- grant 325594231, FOR 2634/2.
This work is based on observations made with the NASA/ESA/CSA James Webb Space Telescope. 
We thank Marcia Rieke for providing the observing time for this program (PID 1186) from the JWST NIRCam GTO allocation.
The data were obtained from the Mikulski Archive for Space Telescopes at the Space Telescope Science Institute, which is operated by the Association of Universities for Research in Astronomy, Inc., under NASA contract NAS 5-03127 for JWST.
This paper makes use of the following ALMA data: ADS/JAO.ALMA\# 2013.1.00726.
ALMA is a partnership of ESO (representing its member states), NSF (USA) and NINS (Japan), together with NRC (Canada), MOST and ASIAA (Taiwan), and KASI (Republic of Korea), in co-operation with the Republic of Chile. The Joint ALMA Observatory is operated by ESO, AUI/NRAO and NAOJ. Astrochemistry in Leiden is supported by the Netherlands Research School for Astronomy (NOVA). This research has made use of NASA’s Astrophysics Data System Bibliographic Services.
The JWST data presented in this article were obtained from the Mikulski Archive for Space Telescopes (MAST) at the Space Telescope Science Institute. The specific observations analyzed can be accessed via \dataset[doi: 10.17909/phr8-cz37]{https://doi.org/10.17909/phr8-cz37}.
\end{acknowledgments}

\textit{Facilities:} JWST/NIRSpec, ALMA

\textit{Software:}  Astropy \citep{Astropy2013,Astropy2018,Astropy2022}, Specutils, matplotlib \citep{Hunter2007}, numpy \citep{vanderWalt2011,Harris2020}, panda, scipy \citep{Jones2001}, emcee \citep{ForemanMackey2013}, Spectres \citep{Carnall2017} python packages. Starfish \citet{Czekala2015}, DS9 \citep{Joye2003}, CDS, Vizier, Simbdad softwares.

\bibliography{ms}
\bibliographystyle{apj}

\newpage
\clearpage
\newpage


\appendix
\addcontentsline{toc}{section}{Appendix}
\renewcommand{\thesection}{\Alph{section}}

\section{\normalfont{Modeling of the photosphere of the NC embedded source}}
\label{app:MCMC}

Starfish provides a rapid estimation of the goodness of a model parameter fit thanks to an efficient computation of the likelihood function during the Markov Chain Monte Carlo (MCMC) process. This is enabled by the pre-computation of a model grid performed via a principal component analysis (PCA) of the grid. During the MCMC, the photospheric flux is directly computed from the PCA eigen basis which allows for a much faster retrieval of the likelihood estimation than if the entire stellar model was fully retrieved from the initial grid each time. We use the BTSettl model grid adapted to low-mass objects \citep{Allard2012} as the model grid for Starfish to compute the PCA on, allowing for effective stellar temperature and surface gravity values of $[2600-5500]$, and $[0.5,5.0]$, respectively. We use the emcee python package \citep{ForemanMackey2013} to perform the MCMC using this Starfish likelihood estimation.

In addition to the photospheric flux from the BTSettl model grid, we also incorporate the other parameters described in Equation \ref{equ:photo_model} in the MCMC model exploration, \ie the stellar radius, disk temperature and emitting radius, and the $K$-band extinction. We broaden the model spectrum with a rotational broadening kernel parametrized by the $v \textrm{sin} i$ parameter (from \citealt{Gray2008}) and also convolve the resulting spectra with a Gaussian kernel following the NIRSpec IFU spectral resolution of $R = 2700$. The model spectrum is then resampled to the wavelength data array of the NIRSpec observations (we first consider an over-sampled initial grid of BTSettl spectra with respect to the NIRSpec data). Finally, two additional parameters are used, \ie the scale $ls_{\textrm{cov}}$ and amplitude $A_{\textrm{cov}}$, to build a kernel that is be used to characterize the covariance between pixel residuals in the definition of the global covariance matrix (see Section 2.3.1 of \citealt{Czekala2015}).  Starfish converged within the first 5,000 steps and was run for 2,000 additional steps after convergence.

\begin{figure}[!tbh]
\centering
\includegraphics[scale=0.32,clip,trim= 0cm 0.0cm 0cm 0cm]{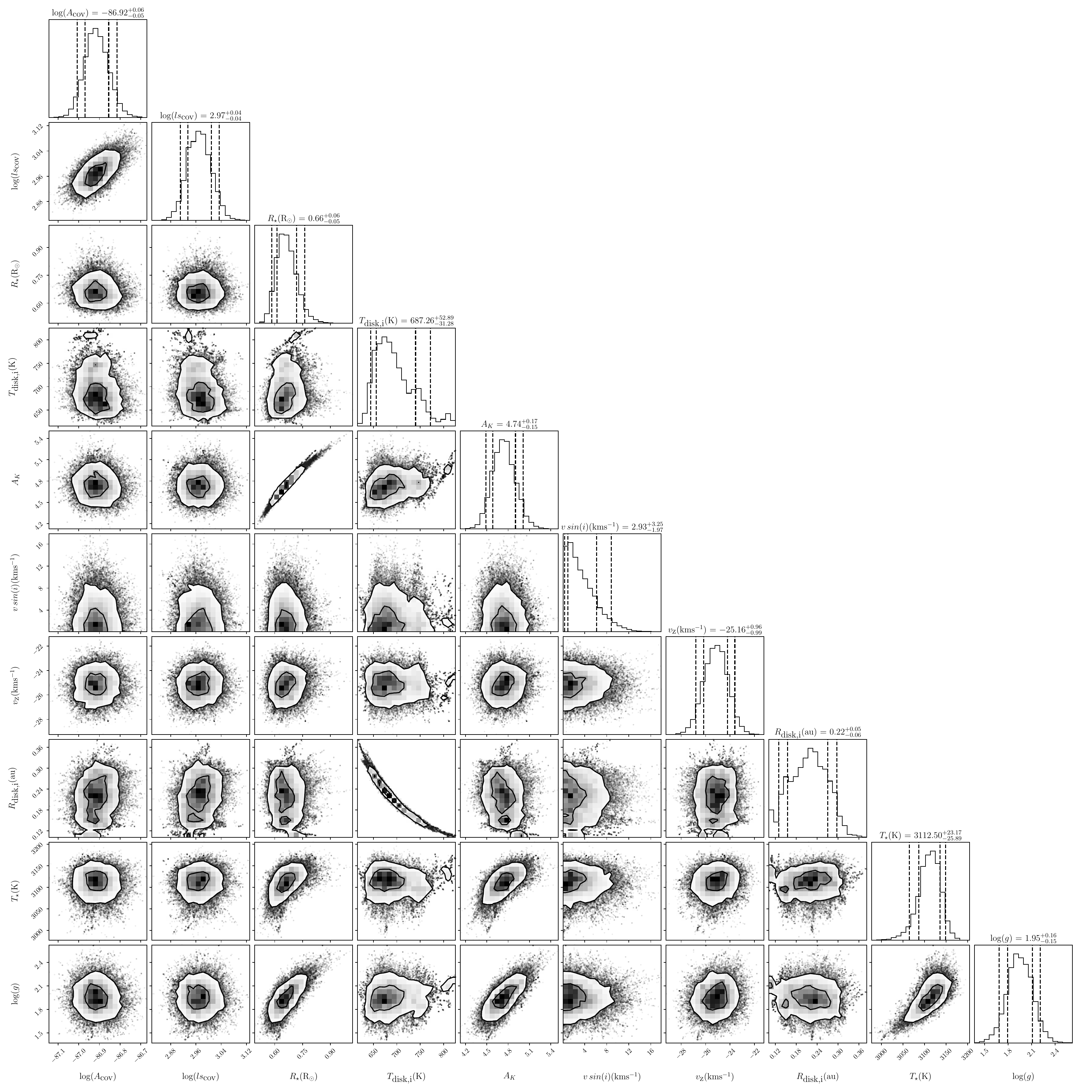}
\caption{\small Corner plots of the MCMC posterior probability distributions and covariances
of the model parameters of the NC source's $1.9-2.45 \mu$m spectrum. Values of the last 4000 steps of the Starfish run are shown. The grayscale plots show the 2D correlations between the parameters listed along the vertical and horizontal axes. Plots of the posterior probability distributions run along the top diagonal above the 2D correlation plots. The median values, 16\%, and 84\% confidence values of the each parameter are given above each posterior probability distribution plot.}
\label{fig:corner_plot_photosphere}
\end{figure}

Figure \ref{fig:corner_plot_photosphere} presents the corner plots of the MCMC posterior probability distributions and covariances of the model parameters of the $1.9-2.45 \mu$m spectrum of the NC source. We note that in the line broadening we apply in the MCMC, which is made of a rotational broadening (via the $v\,\textrm{sin} i$ parameter) due to the stellar rotation and a Gaussian broadening to take into account the NIRSpec IFU spectral resolution, the $v\,\textrm{sin} i$ parameter converges toward 0. The effect of the stellar rotational broadening is thus not constrained. The best-fit model from the Starfish MCMC of the photosphere is shown in Figure \ref{fig:best_model_photosphere}. As explained in Section \ref{sec:photosphere_model}, we consider in a second step a disk temperature profile (see their parameters in Table \ref{t.photosphere_MCMC_results}) in order to better reproduce the 1.9---5.3 $\mu$m continuum, using the best-fit photospheric parameters (\ie $R_\star$, log $g$, and $T_{\star,\textrm{eff}}$). 


\section{\normalfont{Modeling of the accretion of the NC source from its NIR excess}}
\label{app:NC_accretion}

In this Appendix, our goal is to attempt to quantify the accretion energy that reproduces the disk heating (or IR excess) measured in the NC source. In order to account for the additional heating caused by accretion, we use the simple approach of adopting the formalism of the steady-state viscous disk, that assumes that the dissipation of gravitational potential energy associated with accretion occurs via viscous heating (\eg see the formalism by \citealt{Shakura1973,LyndenBellPringle1974,Pringle1981}, and the seminal modeling of \citealt{Calvet1991,Meyer1997}). Ignoring external irradiation flux, the effective disk temperature radial profile is given by :
\begin{equation}
T^{4}_{\textrm{visc}}(r)\,=\, \frac{3 G M_{\star} \dot{M}}{8 \pi \sigma r^{3}} \left(1 - \sqrt{\frac{R_{\star}}{r}} \right)\,\,,
\end{equation}
where $\dot{M}$ is the mass accretion rate, $\sigma$ the Stefan-Boltzmann constant, and $G$ the gravitational constant.
Then, we designate $f_{\textrm{acc}}$ the fraction of the accretion luminosity $L_{\textrm{acc}}$ radiated away, such that $(f_{\textrm{acc}} L_{\textrm{acc}} + L_{\star})/4$ is the energy re-emitted by the disk. In the protostellar phase the accretion luminosity is actually the dominant source of external disk heating (\eg see \citealt{Hennebelle2020b,LeeYN2021,Lebreuilly2024a}). 
$L_{\textrm{acc}}$ is expressed using the following relation of \citet{Gullbring1998}:
\begin{equation}
L_{\textrm{acc}} = (1 - R_\star/R_{\textrm{min}}) G M_\star \dot{M} / R_\star\,,
\label{equ:disk_energy_budget}
\end{equation}
that links the mass-accretion-rate-related release of gravitational potential to the accretion luminosity.
We saw above that a geometrically thin and optically thick disk would re-emit 1/4 of the central luminosity. This yields the following energy budget equation:
\begin{equation}
L_{\textrm{disc}} = L_{\textrm{visc}} + (f_{\textrm{acc}} L_{\textrm{acc}} + L_{\star})/4\,,
\end{equation}
where $L_{\textrm{disc}}$ is the disk energy radiated away (\ie $\int_{R_{\textrm{min}}}^{R_{\textrm{max}}}{\sigma T^{4}_{\textrm{disk}}(r) \pi rdr}$), and $L_{\textrm{visc}}$ is the radiated energy due to the disk viscous heating (obtained integrating $\sigma T^{4}_{\textrm{visc}}(r) \pi r dr$ between $R_{\textrm{min}}$ and $R_{\textrm{max}}$). 

As stated in Section \ref{sec:photosphere}, the $R_\star$ we constrain is lower limit because of the flux-loss induced by scattering in this highly embedded system. We thus introduce here a correction factor $\epsilon$, that corresponds to the fraction of intrinsic photospheric/inner disk flux that is scattered initially onto the cavity walls, prior to extinction by the foreground. 
The total flux correction we propose would thus be $10^{-0.4\times A_\lambda} / \epsilon$, \ie the effective extinction down to the last-scattering surface we measure with Starfish, and the fraction of light that is scattered out by the inner regions, $\epsilon$.
$R_\star$ is initially determined in Section \ref{sec:photosphere} where we applied the extinction $A_K$ to correct for the 2 $\mu$m flux, and we now correct it $1/\sqrt{\epsilon}$, since the total flux scales with ${R_\star}^2$.
$M_\star$ is obtained with the following relation: $\textrm{log}\,g = \textrm{log}(g_\odot \times M_\star \times \epsilon/R_\star^{2})$.
Since we assume $L_{\textrm{visc}}/L_{\star}$ does not depend on the flux loss induced by scattering in the inner regions, we divide Equation \ref{equ:disk_energy_budget} by $L_{\star}$ and deal with the flux loss via the correction of $R_\star$ and $M_\star$ by $\epsilon$.

Future work could quantify this flux loss that we model by $\epsilon$ in this work via radiative transfer modeling. Indeed, while the NIRCAM images and NIRSpec \Ht lines can constrain the cavity geometry, the ALMA data can constrain the dust mass in the disk and inner envelope. As a consequence, these S68N NC source characteristics could be used to estimate more precisely this fraction of scattered light $\epsilon$, and thus the true extinction affecting the NIR flux originating form the inner regions.

\begin{figure}[!tbh]
\centering
\includegraphics[scale=0.5,clip,trim= 0cm 0.0cm 0cm 0cm]{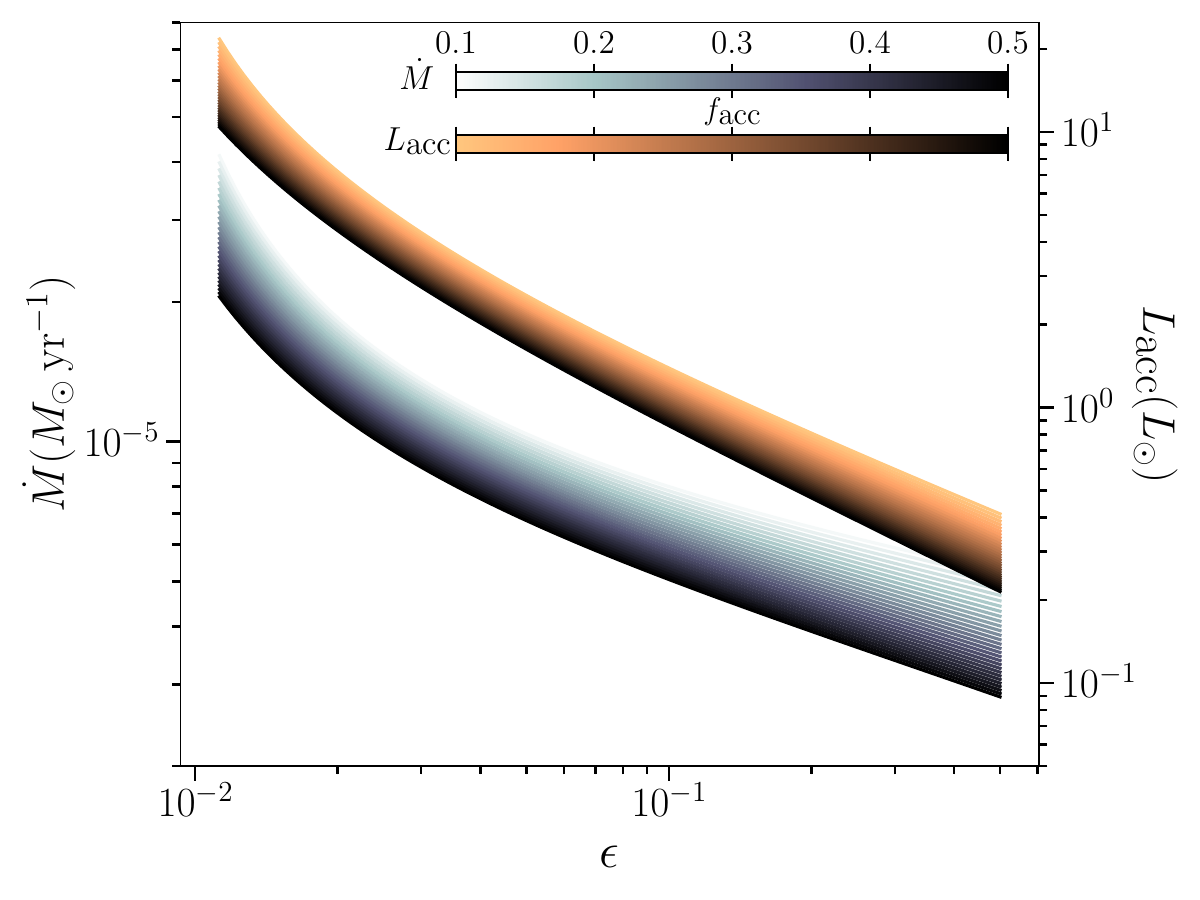}

\vspace{-0.5cm}
\caption{\small Evolution of the estimated mass accretion rate $\dot{M}$ (gray color scale) and accretion luminosity $L_{\textrm{acc}}$ (orange color scale) of the NC source as a function of the fraction of the accretion luminosity radiated away $f_{\textrm{acc}}$, and the fraction of intrinsic photospheric/inner disk flux that is initially scattered in the cavity prior to extinction by the envelope foreground $\epsilon$ (used to model the fraction of scattered light escaping in the cavity). $R_\star$ is thus corrected by a factor of $1/\sqrt{\epsilon}$, and $M_\star$ by a factor $1/\epsilon$ (propagating the corrected $R_\star$ in the equation relating $R_\star$, $M_\star$, and log $g$).
As a result, if we underestimate the flux of the photosphere by a factor of 10 (\ie $\epsilon =$ 0.1), the disk heating we model (via the direct heating from the accretion shock onto the disk, and the disk viscous heating) gives an estimated mass accretion rate of 6---9$\times\,10^{-6}\,M_\odot\,\textrm{yr}^{-1}$, and an accretion luminosity of $\sim$ 0.6---1 $L_{\odot}$, for $f_{\textrm{acc}} =$ 0.1---0.5.}
\label{fig:Mdot_vs_facc_epsilon}
\end{figure}

Equation \ref{equ:disk_energy_budget} thus yields a relation of the mass accretion rate as a function of $\epsilon$, $f_{\textrm{acc}}$. Such parameter exploration is shown in Figure \ref{fig:Mdot_vs_facc_epsilon}. 
The dependence of both $R_\star$ and $M_\star$ with the flux loss fraction $\epsilon$ explains the decrease of the mass accretion rate with increasing values of $\epsilon$.
We don't observationally constrain the $\epsilon$ parameter in this work, as it depends on the yet unexplored properties of dust grains located in the inner regions of the NC source's blueshifted outflow cavity, as well as on the cavity geometry and inclination.
We employ the conservative range of 1---50\% and 0.1---0.5 for $\epsilon$ and $f_{\textrm{acc}}$, respectively, which encompass the typical values of grain albedo at 2---3 $\mu$m, and expected values of $f_{\textrm{acc}}$ from theory \citep{Baraffe2009,Baraffe2012}.
We constrain a range for the mass accretion rate of $\dot{M} \sim$ $3\times10^{-6}$---$4\times10^{-5} M_\odot \textrm{yr}^{-1}$. This mass accretion rate is considerably higher (by $\sim$ 1-2 dex) than the typical mass accretion rate encountered in the more-evolved Class I protostars \citep{Fiorellino2021,Fiorellino2023}.

We use this simple formulation to posit that the disk radiative flux must account for an additional source heating, which in the case of a viscously heated disk, would suggest that the NC source is heavily accreting.
However, these calculations are only first order approximations and rely on assumptions not necessarily true for Class 0 disks. For example, the evacuation of angular momentum to allow accretion on the central object can also occur via disk winds (\eg see the seminal paper by \citealt{Blandford1982}, and \citealt{Lesur2021,Lesur2023} for reference works), in which case no energy dissipation in the form of heat is required for accretion.
Another caveat is that the vertical distribution of the viscous heating is not well known, and thus the vertical energy transport occurring via radiative diffusion affects the relation between the mass accretion rate and the corresponding disk radiated flux captured from the disk upper layers. Specifically, if the disk is optically thick in the vertical direction the mass accretion rate is then underestimated.

\section{\normalfont{Shock modeling parameter exploration}}
\label{app:shock_models}

\begin{figure}[!tbh]
\centering
\includegraphics[scale=0.55,clip,trim= 0cm 0.0cm 0cm 0cm]{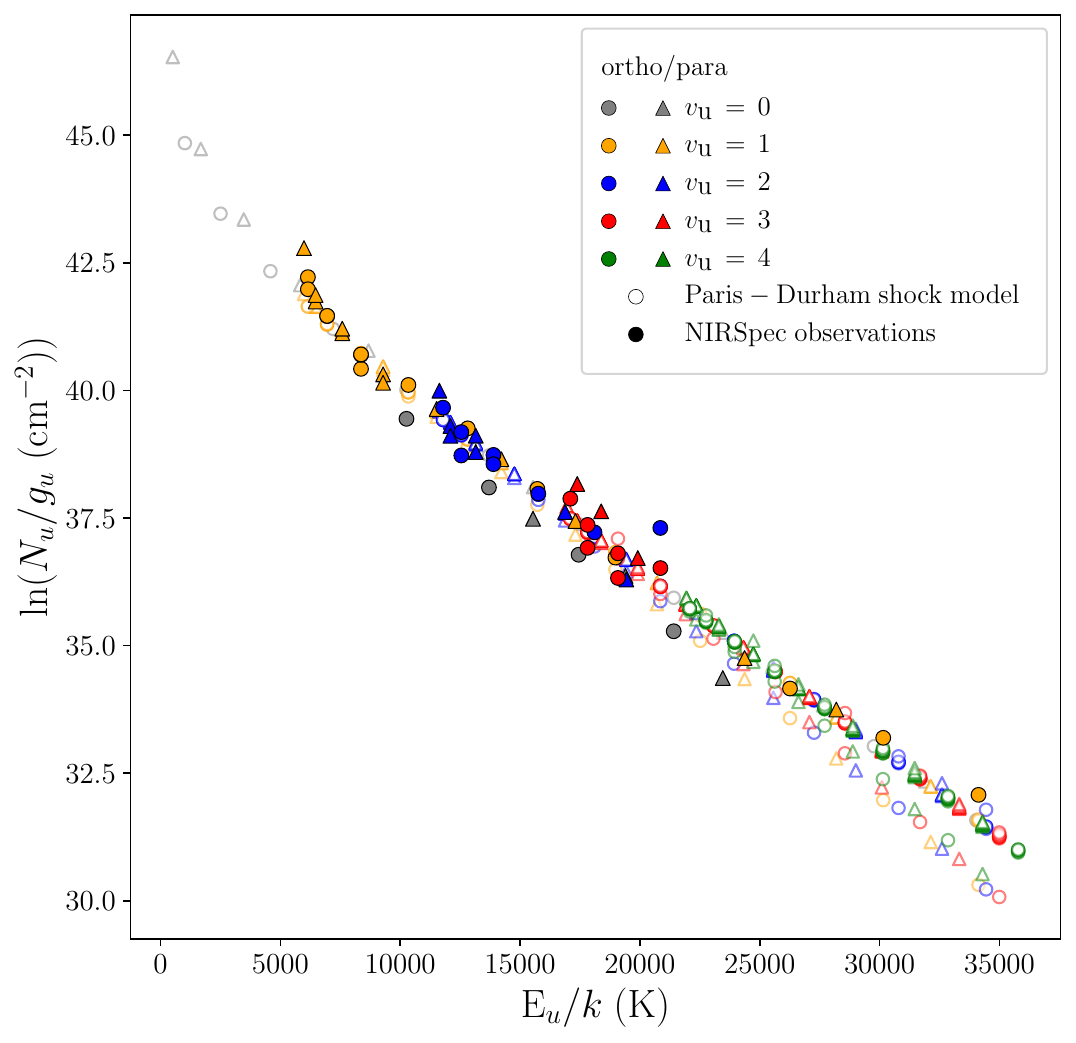}

\vspace{-0.3cm}
\caption{\small \Ht excitation diagram obtained from the spectra extracted at the first spatial location along the jet of the SC source. The \Ht excitation diagram of the best model for this location (a $C$-type shock, whose parameters are \nH$ = 10^{8}$ cm$^{-3}$, \vs$ = 40$ km s$^{-1}$, \b$ = $10, \Go$ = $10, \CIR$ = 10^{-16}$ s$^{-1}$, \XPAH$ = 10^{-6}$) is shown with open symbols, while plain symbols correspond to the \Ht excitation diagram constructed from the NIRSpec data.}
\label{fig:H2_diagram_comp_model}
\end{figure}

This Section contains further details on our comparisons between the \Ht excitation extracted from the data and the shock model grid of \citealt{Kristensen2023}. For every location along the molecular jet of the SC source, the observed excitation of \Ht lines is compared with the grid in order to constrain the evolution of the shock parameters along the jet (see Section \ref{sec:mol_em_jet_south_H2shocks}). As an example, Figure \ref{fig:H2_diagram_comp_model} shows the \Ht excitation diagram obtained from the spectra extracted at the first spatial location along the jet of the SC source, alongside the \Ht excitation diagram of the best model fit for this location.

\begin{figure}[!tbh]
\centering
\includegraphics[scale=0.5,clip,trim= 2cm 2cm 1cm 2cm]{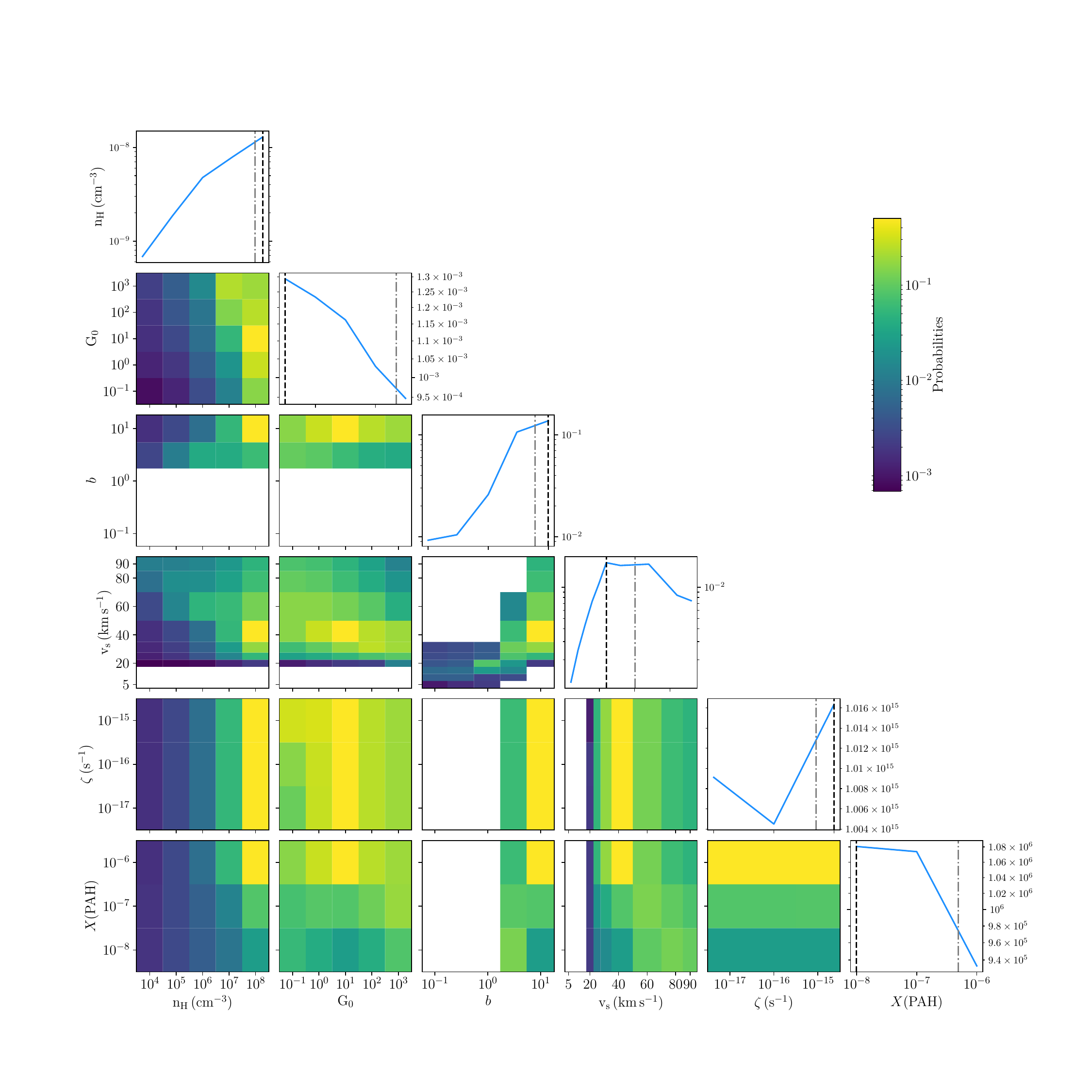}

\vspace{-0.2cm}
\caption{\small Shock model grid parameter exploration for the \Ht lines extracted at the first spatial location along the jet of the SC source. The 2D diagrams off the diagonal show the different likelihood values of each pair of parameters, using the best shock model as a reference. The 1D diagrams on the diagonal show the probability distribution functions of each parameter, with the mean and maximum values indicated by the vertical grey and black lines.}
\label{fig:shock_param_corner_plot}
\end{figure}

\begin{figure}[!tbh]
\centering
\includegraphics[scale=0.55,clip,trim= 0cm 0.0cm 0cm 0cm]{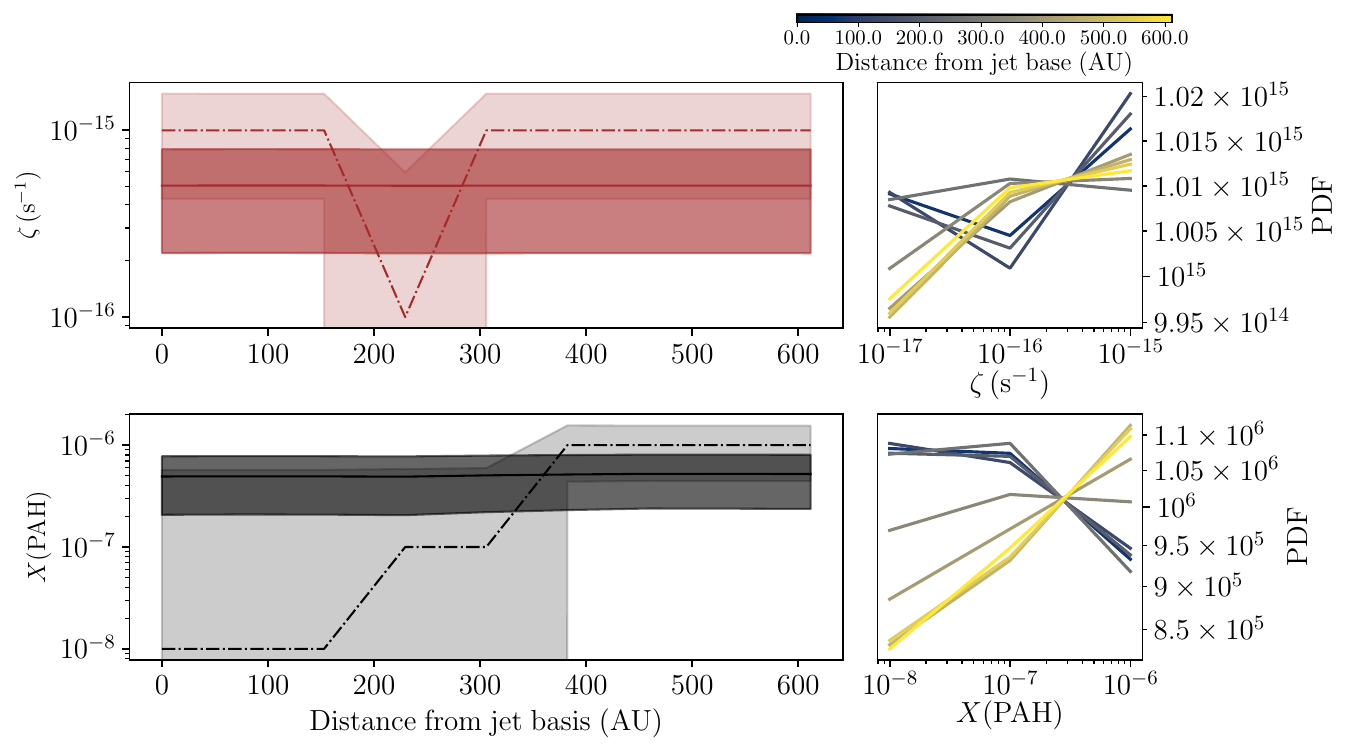}

\vspace{-0.3cm}
\caption{\small Same as Figure \ref{fig:H2_shock_modeling}, for the last two parameters of the shock model grid of \citet{Kristensen2023}, \ie the \Ht cosmic-ray ionization rate \CIR, and the PAH abundance fraction \XPAH. The sampling of these two parameters is not precise enough, such that we do not constrain these two parameters.}
\label{fig:H2_shock_modeling_irrelevant_params}
\end{figure}

We adopted the same technique as \citet{Poorta2023} to build likelihood distributions
for each modeled parameter of the non-uniformly sampled shock model grid (see the method explained in their appendix C). We quantify the goodness of fit of every shock model of the grid to the data with a $\chi^2$ parameter calculated taking the logarithm of the data-to-model ratios $\sum \textrm{log}_{10} (F_{\textrm{H}_{2},\,\textrm{data}}/F_{\textrm{H}_{2},\,\textrm{model}})^{2}$, where $F_{\textrm{H}_{2},\,\textrm{data}}$ and $F_{\textrm{H}_{2},\,\textrm{model}}$ are the extinction corrected H$_2$ fluxes of the observed data and shock model, respectively. This gives a uniform weight to all detected H$_2$ lines, regardless of the different level populations. We convert all $\chi^2$ values to a likelihood taking $1/\chi^2$. For a given shock parameter, the likelihood of a given parameter value is obtained by summing over the likelihood of all other parameter value combinations. These likelihood functions are normalized by the area under the curve to obtain the probability distribution functions, whose mean and standard deviation are calculated. Figure \ref{fig:shock_param_corner_plot} shows, for the first spatial location along the jet, a corner plot showing the different likelihood values for each pair of parameters in a 2D diagram. The best shock model serves as a reference for each 2D diagram. Along the diagonal, 1D plots show the probability distribution functions of each parameter, whose mean, maximum, and respective standard deviation are shown in Figure \ref{fig:H2_shock_modeling}, for \nH, \b, \vs, and \Go. Figure \ref{fig:H2_shock_modeling_irrelevant_params} presents the probability distribution functions for the \CIR and \XPAH parameters. However, the sampling of the shock model grid for these two parameters is not adequate with respect to the shocks we observe, such that the dependence of the best model on these two parameters is very low. Typically, the \citet{Kristensen2023} grid only implements three values for each of these two parameters, spanning a range of 2 orders of magnitude. We thus cannot make any conclusions on the most likely values of \CIR and \XPAH in the shocks of the SC source's outflow.

\section{\normalfont{LTE parameter exploration of the CO fundamental line forest}}
\label{app:CO_models}

\begin{figure}[!tbh]
\centering
\includegraphics[scale=0.45,clip,trim= 0cm 0.0cm 0cm 0cm]{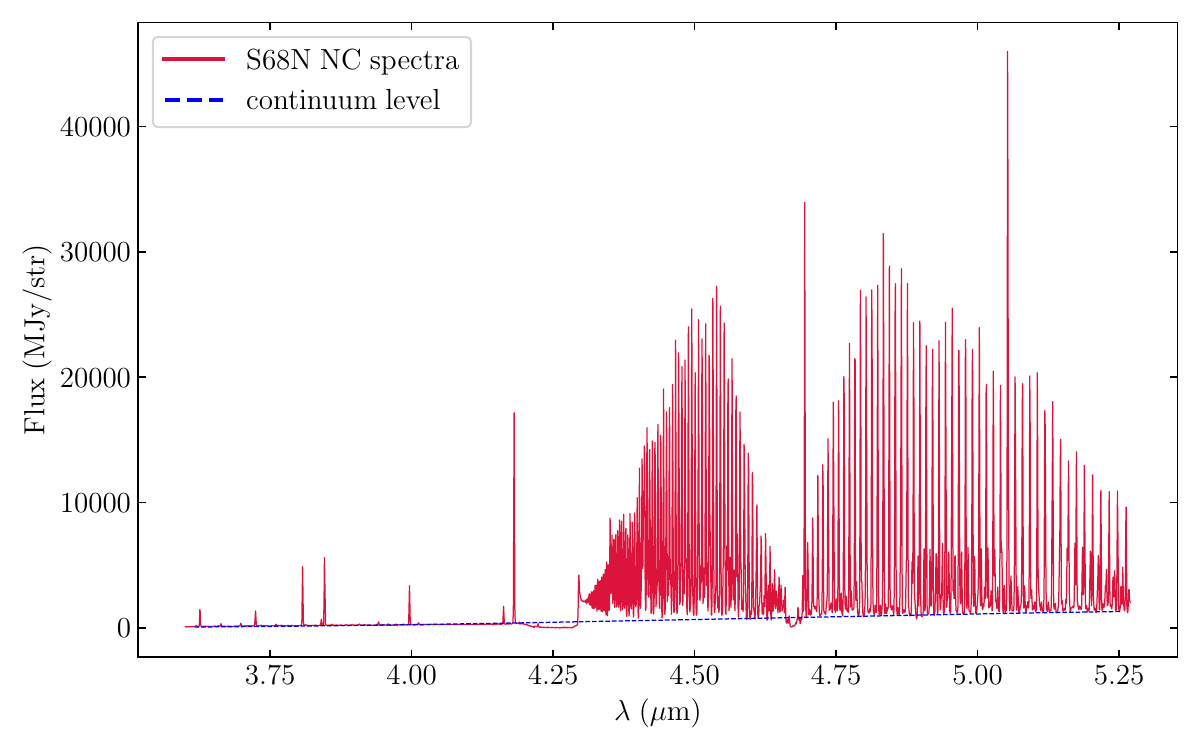}

\vspace{-0.4cm}
\caption{\small 3.6--5.3 $\mu$m spectra of the S68N NC source. The observed spectra is shown in red, and the continuum level used to analysis the CO fundamental line forest seen in emission is shown with the dashed blue line.}
\label{fig:CO_cont}
\end{figure}

\begin{figure*}[!tbh]
\centering
\includegraphics[scale=0.7,clip,trim= 0cm 0.0cm 0cm 0cm]{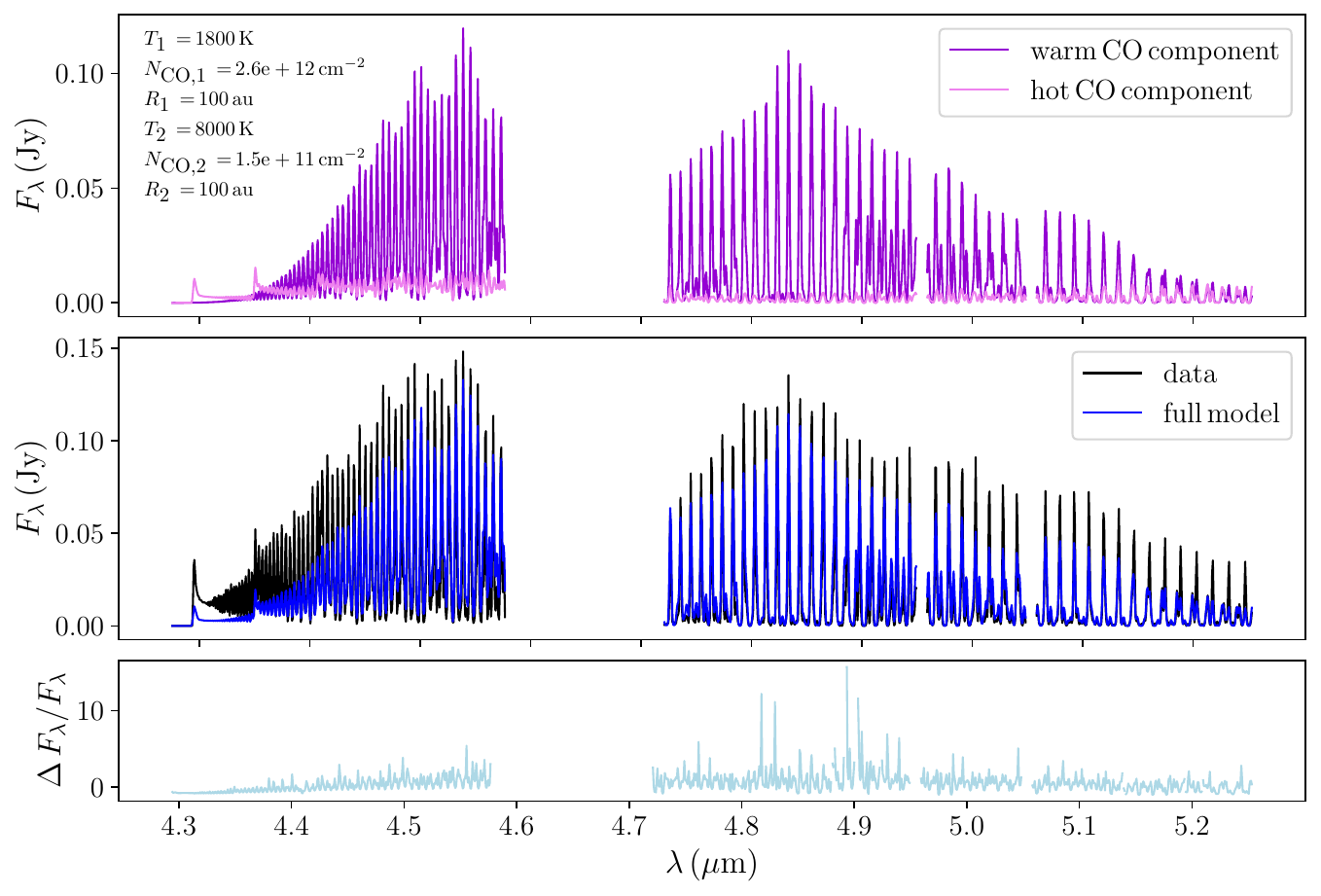}

\vspace{-0.4cm}
\caption{\small Example of a two-component LTE model of the SC source CO fundamental line forest. 
\textit{Top panel:} Two LTE 1D slab models are shown, a warm ($T_{\textrm{1}}\,=\,1800$\,K), and a hot ($T_{\textrm{1}}\,=\,8000$\,K) component. This model is not a precise fit, and the surface area and column densities (which cannot be disentangled, see the analysis of appendix \ref{app:CO_models}, Figure \ref{fig:CO_corner_plot}) are fixed by hand. For reference, these parameters correspond to the following total number of CO molecules: $\mathcal{N}_{\textrm{CO},1}\,=\,1.8\times\textrm{10}^{43}\,\textrm{cm}^{-2}$ and $\mathcal{N}_{\textrm{CO},2}\,=\,1.1\times\textrm{10}^{42}\,\textrm{cm}^{-2}$.
\textit{Middle panel:} The extinction corrected, continuum subtracted observed spectrum is shown in black, and the sum of the two LTE models is shown in blue. 
\textit{Bottom panel:} The fractional residual between the data and the model is shown.
The spectral regions covering the bright \Ht lines \Ht\,0-0\,S(9), \Ht\,0-0\,S(8), and \Ht\,1-1\,S(9), and the saturated $^{12}$CO ice absorption band (between 4.58 and 4.72 $\mu$m) are flagged and not used.
These LTE models do not fit the data well. Reproducing the flux of the detected $v = 1-0$, $v = 2-1$, and $v = 3-2$ bandheads would cause the model to largely overestimate the flux of the P-branch lines.
}
\label{fig:CO_data_model_spec}
\end{figure*}
 
We present in Figure \ref{fig:CO_cont} the continuum level used for the analysis of the CO emission line.
In this Appendix, we explore how multiple CO emitting gas populations at different temperatures and column densities could also contribute to the total CO excitation. Each component is characterized by a temperature $T$, column density $N_{\textrm{CO}}$ and surface area $R_{\textrm{area}}$; we follow \citealt{Goldsmith1999,Mangum2015}, with a similar approach as in \citealt{Tabone2023,Francis2024}. Indeed, if only one LTE component was at the origin of this CO line forest, reproducing the three first bandheads would largely overestimate the $^{12}$CO $v = 2-1$ and $^{12}$CO $v = 3-2$ P-branch lines compared to what is observed. 

Figure \ref{fig:CO_data_model_spec} presents an illustrative two-component LTE model. We add together two 1D-slab models of 1800\,K and 8000\,K, with a filled disk surface area of an outer radius of 100 AU, and then let the CO column densities of each component vary to minimize the $\chi^2$ with the observed spectra. We see that to reproduce the bandheads and the faint $^{12}$CO $v = 3\rightarrow2$, the hot component needs to be $\geq 3000$ K and one order of magnitude lower column density (or surface area) than the warm CO component, responsible for the bulk of the $^{12}$CO $v = 1\rightarrow0$ and $^{12}$CO $v = 2\rightarrow1$ $J_{\textrm{u}}\,\leq\,40$ lines. 

A hot CO component of $\geq $ 3000 K could thus be at the origin of the high-$J$ CO $v = 1-0$, $v = 2-1$, and $v = 3-2$ lines detected toward the bandheads. However, such high temperatures could also favor the hypothesis of non-LTE effects impacting the CO excitation at the base of the cavity, especially given the high vibrational temperatures we derive and since the $^{12}$CO $v = 1-0$ level population does not align with the $v_{\textrm{u}}\,\geq\,2$ level populations. 
This CO emission lines ($\Delta\,v\,=\,1$) analysis is limited by the difficulty of precisely determining the continuum baseline, the complexity of the CO excitation, and the assumed extinction law. 
Also, we stress that our analysis of the $^{12}$CO $v = 1-0$ lines can be affected by absorption, which is is usually seen in more evolved disks and winds \citep{Herczeg2011,Brown2013,Thi2013,Banzatti2022}. 
Further work including observations with $JWST$/MIRI MRS would constrain more precisely the conditions of the CO emitting gas.

\begin{figure}[!tbh]
\centering
\includegraphics[scale=0.5,clip,trim= 3cm 2.5cm 2cm 2cm]{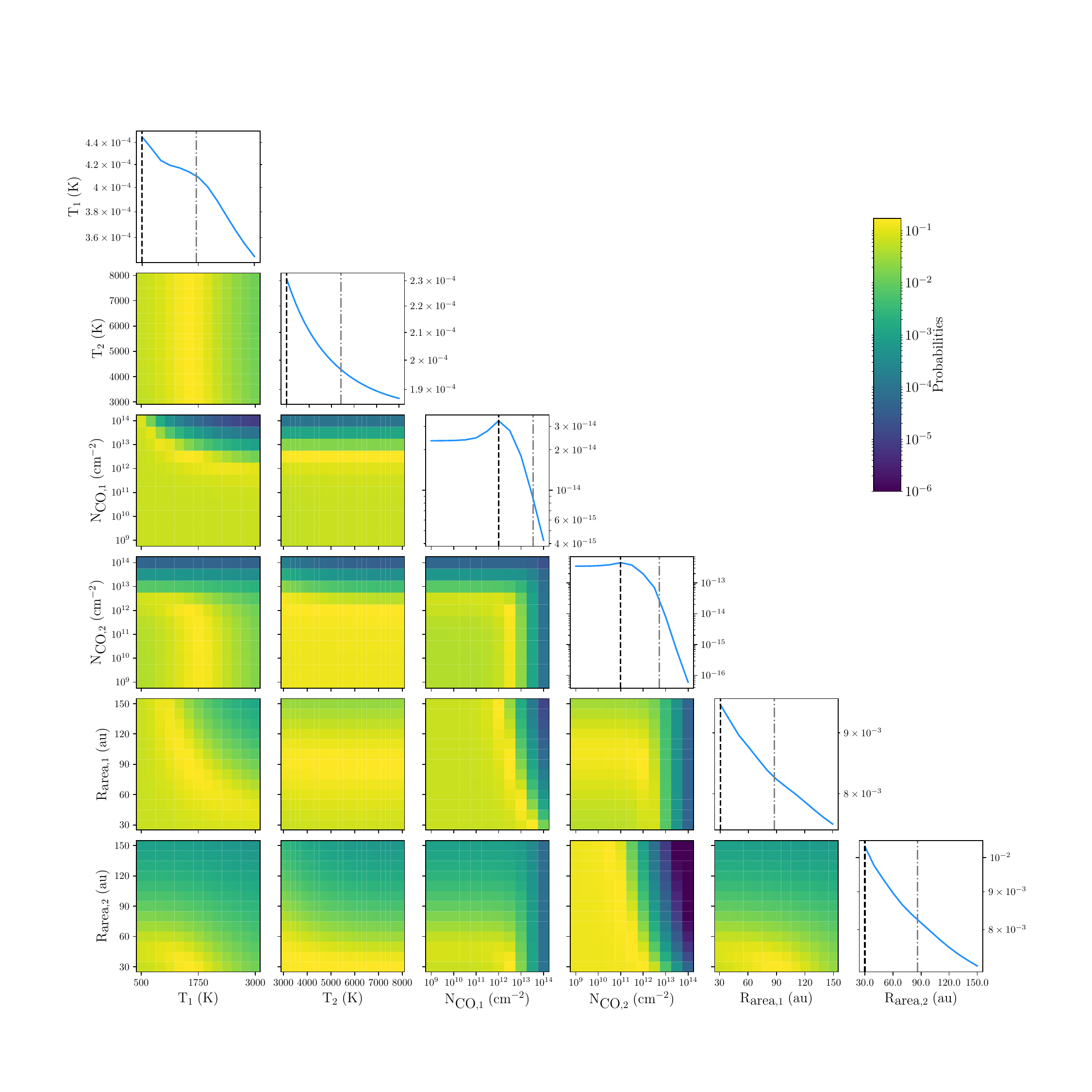}

\vspace{-0.2cm}
\caption{\small Exploration of the two LTE component parameters (temperature $T$, column density $N_{\textrm{CO}}$, and surface area radius $R_{\textrm{area}}$ for each component) modeling the CO fundamental line forest. The 2D diagrams off the diagonal show the different likelihood values of each pair of parameters, using the best shock model as a reference. The 1D diagrams on the diagonal show the probability distribution functions of each parameter, with the mean and maximum values indicated by the vertical grey and black lines.}
\label{fig:CO_corner_plot}
\end{figure}

We perform an analysis similar to that presented in Appendix \ref{app:shock_models} for H2 shock excitation, but this time to explore the hypothesis of the existence of hot CO gas in LTE at two different temperatures ($T_1$, $T_2$), each with its associated column density ($N_{\textrm{CO,1}}$ and $N_{\textrm{CO,2}}$) and emitting area ($R_{\textrm{area,1}}$, $R_{\textrm{area,2}}$). We compute the likelihood of each model from the $\chi^2$ that we obtain comparing the modeled and observed CO fundamental line forest (shown in Figure \ref{fig:CO_spectra}). We then construct the 2D distributions of likelihood values, and 1D probability distribution functions for each parameter as in Figure \ref{fig:shock_param_corner_plot}. These distributions are shown in Figure \ref{fig:CO_corner_plot}. 

We notice that the column density probability distribution function of each component rejects the possibility of $N_{\textrm{CO}}\,\geq\,10^{13}\,\textrm{cm}^{-2}$. 
The highest probabilities in the $N_{\textrm{CO}}$ versus $R_{\textrm{area,2}}$ 2D diagrams suggest that these parameters are correlated together for both of the two components, such that we cannot disentangle one from the other.
This is expected from optically thin emission, however, we cannot rule out optical depth effects, especially for the $^{12}$CO $v_{\textrm{u}}\leq2$ lines. 
While the temperature of the warm component shows a preference for $T\,\leq\,1800\,$K, the temperature of the hot component is not constrained. 
The model presented in Figure \ref{fig:CO_data_model_spec} is only illustrative, and the parameters have been picked by hand, following the probability distribution of Figure \ref{fig:CO_corner_plot}.
The non-LTE effects explain the difficulty of fitting the observed spectrum with this two LTE-component approach, especially as the $^{12}$CO$v_{\textrm{u}}\geq2$ lines show different excitation properties than the $^{12}$CO$v=1\rightarrow0$ lines.

\section{\normalfont{PCA analysis of the IFU cubes}}
\label{app:PCA}

PCA analysis is a powerful dimension reduction method that has been widely applied in various fields of astronomy, such as direct imaging of exoplanets \citep{WangJJ2015}, tomography of polycyclic aromatic hydrocarbons in Galaxies \citep{WangL2011,Donnan2024}, characterizing black hole Images and variability \citep{Medeiros2018}, supernova remnant shocks \citep{Neufeld2007}, analyzing the structure of molecular gas with respect to environmental conditions in star forming molecular clouds \citep{Gratier2017}, and prestellar and protostellar cores \citep{Spezzano2017,Okoda2020}. PCA analysis maps the multi-dimensional data to a coordinate system consisting of a set of new base axes. In the new coordinate system, the axes, which are also named eigenvectors or principal components (PC), are orthogonal to each other. The axes are ordered by their corresponding eigenvalues, which correspond to the data variances projected on the axes. The first axis or eigenvector has the highest eigenvalue and indicates the direction in which the data vary the most. With a PCA analysis, we can construct a set of axes ordered from high to low eigenvalues, which is useful for exploring the minimal number of axes to explain the variances in the dataset.

\begin{figure}
    \centering
    \includegraphics[scale=0.5,clip,trim= 0cm 0cm 0cm 0cm]{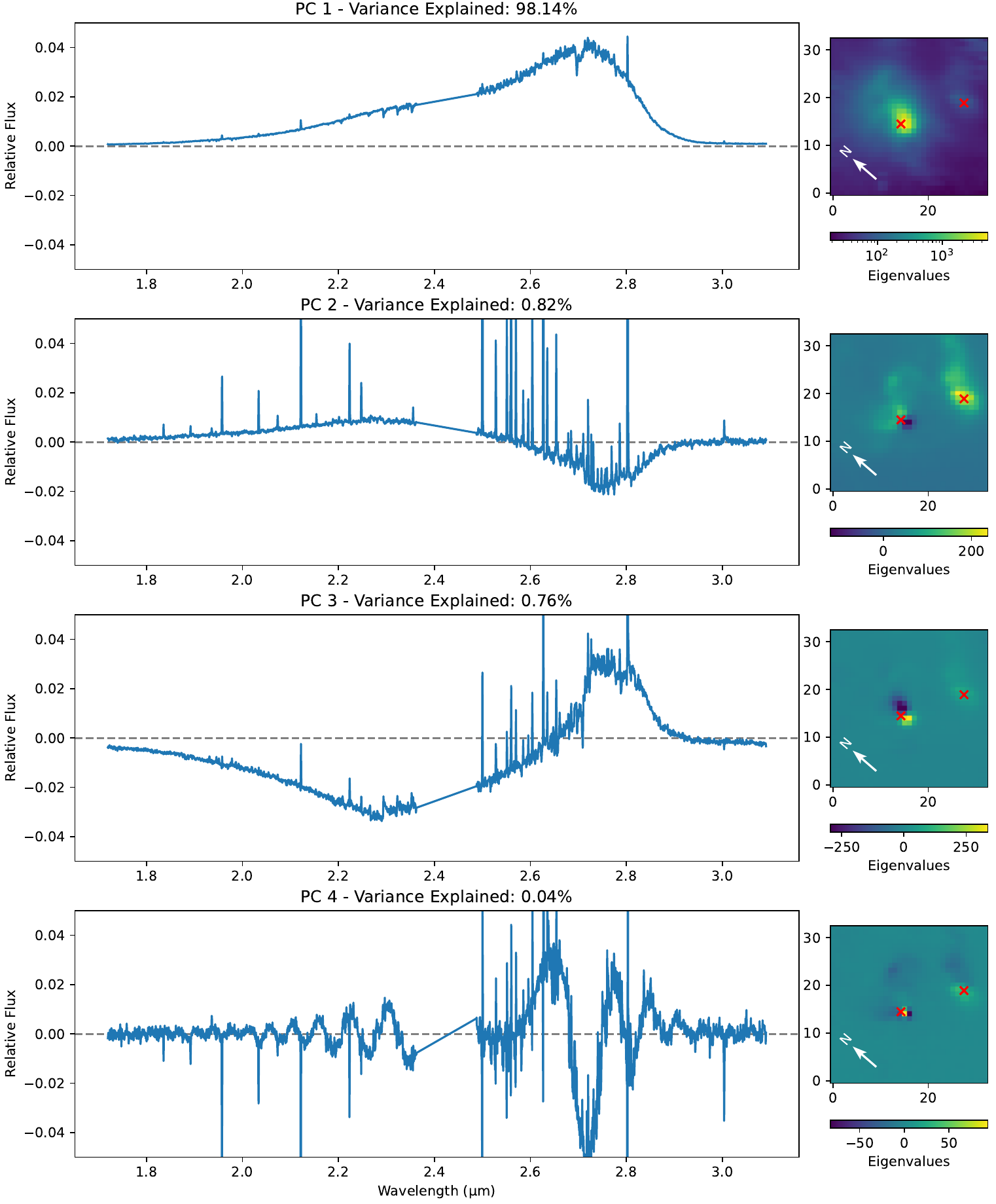}

    \vspace{-0.2cm}
    \caption{The principal components based on the PCA analysis performed on the G235H data. Left columns show the principal component eigenspectra while the right column of the same row shows the spatial distribution of the eigenvalues of each principal component. The red crosses indicate the locations of the NC and SC sources. The first principal component (top row) shows the main photospheric spectral feature of the NC source. The second principal component highlights the outflow region with strong $H_2$ emission lines toward both the NC and SC sources. The third principal component seems to indicate the non-uniform scattering around the NC source. The fourth principal component is likely systematic noise.}
    \label{fig:PCA_g235h}
\end{figure}

In the context of IFU data, PCA maps $n$ rows $\times$ $m$ columns $\times$ $N_{\lambda}$ spectral points to a new coordinates system with $q$ number of eigenspectra $P$. The first step of PCA analysis is to remove the mean spectral values from the IFU data and calculate the covariance matrix $X$ between the $n\times m$ spectra. By performing diagonalization of the covariance matrix, the observed spectra can now be reproduced by a set of eigenvectors $P_{k,\lambda}$ (in this case eigenspectra, as they are function of wavelength) for each principal component $k$ multiplied with their corresponding eigenvalues $W_{i,j,k}$:
\begin{equation}
F_{i,j,\lambda} = W_{i,j,k} \cdot P_{k,\lambda}\,\,,
\end{equation}
where $i\in [1:n]$, $j\in [1:m]$. With the eigenvalues $W_{i,j,k}$, also called tomograms, we can then study the corresponding variances explained by the eigenspectra and their corresponding spatial distribution.
In other words, the different eigenvalues describes the spatial correlation of its corresponding eigenspectrum, which in turn reveals how the signal is spatially correlated in the data. A negative eigenvalue suggests that the data projected on the eigenspectrum is in the negative direction, while a positive eigenvalue suggests the opposite. Thus, the eigenvalues inform us about the importance of an eigenspectrum to explain the observed IFU spaxel variations from the mean spectrum.
We perform PCA analysis on the G235H and G395H datasets and study the spatial distribution of the principal components. The results are shown in Figures \ref{fig:PCA_g235h} and \ref{fig:PCA_g395h}. The color bar shows the eigenvalues $W_{i,j,k}$ of each principal component $k$. We note that for this PCA analysis we kept the detector coordinate base and do not regrid to (RA,DEC) in order to maximize the spatial resolution and minimize the spatial correlation of the signal between spaxels.

\subsection{PCA analysis results of G235H IFU dataset}

The first principal component (PC1) has the highest eigenvalues in the NC source region. PC1 has a spectral profile similar to the best-fit Starfish model with clear CO overtone, Na and Ca absorption lines between 1.9 and 2.4 $\mu$m. 

The second PC (PC2) component shows strong \Ht emission line features, with their spatial distribution following the outflow morphology of both the NC and SC sources (\ie the jet structure of the SC source, and the two extended clumps of the NC source). The PC2 spectrum suggests a negative correlation between the flux below and above 2.6 $\mu$m. We interpret that this negative correlation is associated with the outflow emission and the absorption from the protostellar envelope each of which dominate the flux below and above 2.6 $\mu$m respectively.

The third PC (PC3) has an amplitude comparable with that of the PC2. There are clear positive and negative eigenvalues toward the NC source, suggesting spatial variation (whose gradient is aligned with outflow direction) with a spectral slope that becomes steeper in the northward direction. The amplitude of PC3 is about 1\% of the PC1 amplitude in the NC region. No contribution from the SC source is seen in PC3. While this will have to be confirmed with modeling, we propose that this strong continuum gradient toward the NC source in PC3 may correspond to the scattering contribution of the photosphere and/or disk continuum.

The amplitudes of the fourth principal component (PC4) is an order of magnitude lower than PC2 or PC3, and is at a level two orders of magnitude below the PC1 amplitude. We interpret the wavelike spectral feature of PC4 is likely correlated with the under-sampling alias features of the detector.

\subsection{PCA analysis results of G395H IFU dataset}

\begin{figure}
    \centering
    \includegraphics[scale=0.46,clip,trim= 0cm 0cm 0cm 0cm]{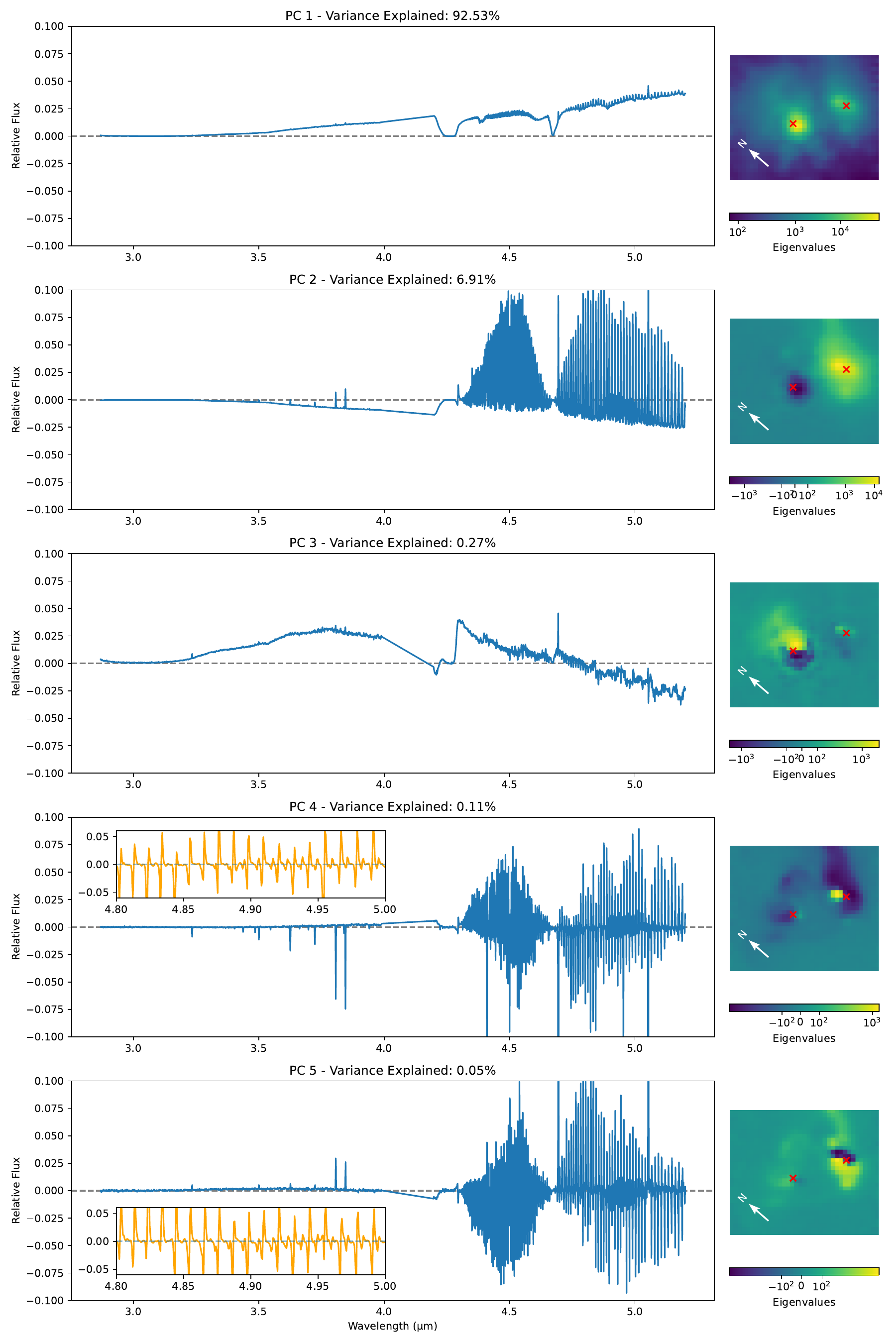}

    \vspace{-0.2cm}
    \caption{Same as Figure \ref{fig:PCA_g235h} for the G395H data. The first principal component (top row) has a rising spectral slope imprint with CO$_2$ and water ice absorption and weak CO emission lines. PC1 likely traces the cool envelope in the northern and SC source regions. The second component shows strong CO emission lines from 4.3 to 5.2 $\mu$m toward the SC source's outflow. The third component exhibits a broad continuum from 3 to 4.8 $\mu$m, implying an higher-than-mean emission from the west of the NC source. The fourth and fifth principal components feature positive and negative flux on the long and short wavelength side of CO emission lines respectively. These features seem to be distributed along the gradient in CO line velocity exhibited in Figure \ref{fig:line_H2_CO_vel_shift} toward the SC source.}
    \label{fig:PCA_g395h}
\end{figure}

The first principal component (PC1) of the G395H data features a gradually rising continuum slope, imprinted with a 3 $\mu$m wide water-ice band, 4.2 $\mu$m CO2 ice absorption, and weak 4.3--5.2 $\mu$m CO fundamental emission lines. The spectral slope peaks beyond 5 $\mu$m, indicating an envelope temperature cooler than 550 K, assuming the envelope emission follows a blackbody profile. Thus, we interpret PC1 as representing a cool ($<$550 K) envelope in both regions of the NC and SC sources. The eigenvalues map exhibits a strong similarity to the 5.1--5.3 $\mu$m continuum map extracted in Figure \ref{fig:cont_images}. This suggests there is a faint and spatially extended contribution of the CO fundamental emission lines.

PC2 highlights strong CO emission profiles, with the highest eigenvalue observed toward the SC source. The negative 3--4 $\mu$m features in PC2 indicate that the 3--4 $\mu$m emission is lower than the mean spectra. This lower-than-mean 3--4 $\mu$m continuum, combined with the bright CO emission lines, matches the observed spectral features in the SC source region, where the highest eigenvalue of PC2 is found.

PC3 features a strong continuum that peaks around 3.7 $\mu$m near the NC source. According to Wien's displacement law, a blackbody peaking at 3.7 $\mu$m corresponds to a temperature of around 800 K, which aligns with the fitted disk temperature constrained in Section \ref{sec:photosphere}. PC3 also shows weak CO absorptions, consistent with the interpretation of an 800 K warm envelope. The spectral feature turns negative beyond 4.8 $\mu$m, hinting at an additional astrophysical emission source contributing to the spectra. We especially note that a gradient is seen in the eigenvalues tomogram toward the NC source, which is similar to the PC3 of the G235H data (although it is fainter). This suggests there is a common physical origin behind these two PC3 components. Again, we point that it may correspond to the scattered-light contribution of the NC source continuum.

The maximum eigenvalues of PC4 and PC5 are on the order of a few percent level compared to that of PC1 in the SC source region. Both PC4 and PC5 show interesting pairs of sharp negative and positive features adjacent to each other. These features coincide with the location of CO emission lines, and the morphology of the CO line velocity shift map of Figure \ref{fig:line_H2_CO_vel_shift}. These features are consistent with a sharp negative flux followed by a sharp positive flux, that is redshifted from the mean CO emission line (see the subpanels of Figure \ref{fig:PCA_g395h}). Conversely, such a feature with a negative eigenvalue thus implies a blueshift offset relative to the mean CO emission lines. PC4 and PC5 both show amplitude gradients in the eigenvalue tomogram toward the SC source region, with a negative amplitude to the South of the SC source. This elongation follows the bright CO emission line region highlighted by PC1.

The major difference between PC4 and PC5 lies in the \Ht emission lines: the eigen spectra of PC4 features strong \Ht absorption lines while the eigen spectra of PC5 features strong H2 emission lines (e.g., lines at  5.0529, 4.694, 3.847, 3.807, 3.724, 3.626, 3.501, and 3.235 $\mu$m). Because the \Ht emission lines exist in the mean spectra, the negative (or positive) eigen spectra \Ht features associated with the spatial distribution of negative (or positive) eigenvalues suggest there are some variations in the relative contribution of \Ht emission lines throughout the outflow cavity of the SC source.

Beyond the fifth principal component, PC6, PC7, and PC8 show complex spatially extended features. Due to uncertainties about whether these features are intrinsic or systematic, we refrain from interpreting the higher-order principal components, and we do not show them in Figure \ref{fig:PCA_g395h}.

\end{document}